\pdfoutput=1
\documentclass[iop,apjl,tighten]{emulateapj}
\usepackage{graphicx,enumitem}
\usepackage[bottom]{footmisc}
\newcommand{\tabincell}[2]{\begin{tabular}{@{}#1@{}}#2\end{tabular}}
\usepackage{threeparttable}
\usepackage{multirow}
\usepackage{amsmath}
\usepackage{ulem}
\usepackage{pgffor}
\usepackage [caption=false]{subfig}

\usepackage{xcolor}
\usepackage{hyperref}
\hypersetup{
   colorlinks,
   linkcolor={blue!88!black!80},
   citecolor={blue!88!black!80},
   urlcolor={blue!88!black!80}
}

\newcommand{\lya}{Ly$\alpha$}

\newcommand{\OII}{[O\,\textsc{ii}]}

\newcommand{\OIII}{[O\,\textsc{iii}]}

\shorttitle{ $z\sim7.0$ LAEs}
\shortauthors{Hu W. et al.}

\begin{document}

\title{The L\lowercase{y}$\alpha$ Luminosity Function and Cosmic Reionization at  $\lowercase{z} \sim$ 7.0: \\a Tale of Two LAGER Fields}
\author{
Weida Hu\altaffilmark{1,6}, Junxian Wang\altaffilmark{1,6}, Zhen-Ya Zheng\altaffilmark{2}, Sangeeta Malhotra\altaffilmark{3}, James E. Rhoads\altaffilmark{3}, Leopoldo Infante\altaffilmark{4}, L. Felipe Barrientos\altaffilmark{5}, Huan Yang\altaffilmark{4}, Chunyan Jiang\altaffilmark{2}, Wenyong Kang\altaffilmark{1,6}, Lucia A. Perez\altaffilmark{7}, Isak Wold\altaffilmark{3}, Pascale Hibon\altaffilmark{9}, Linhua Jiang\altaffilmark{10}, Ali Ahmad Khostovan\altaffilmark{3}, Francisco Valdes\altaffilmark{11}, Alistair R. Walker\altaffilmark{8}, Gaspar Galaz\altaffilmark{5}, Alicia Coughlin\altaffilmark{7}, Santosh Harish\altaffilmark{7}, Xu Kong\altaffilmark{1,6}, John Pharo\altaffilmark{7}, XianZhong Zheng\altaffilmark{12}}

\affil{
$^1$CAS Key Laboratory for Research in Galaxies and Cosmology, Department of Astronomy, University of Science and Technology of China,
Hefei, Anhui 230026, China; urverda@mail.ustc.edu.cn, jxw@ustc.edu.cn\\
$^2$CAS Key Laboratory for Research in Galaxies and Cosmology, Shanghai Astronomical Observatory, Shanghai 200030, China; zhengzy@shao.ac.cn\\
$^3$Astrophysics Science Division, Goddard Space Flight Center, 8800 Greenbelt Road, Greenbelt,  Maryland 20771, USA; sangeeta.malhotra@nasa.gov, james.e.rhoads@nasa.gov\\
$^4$Las Campanas Observatory, Carnegie Institution of Washington, Casilla 601, La Serena, Chile;\\ linfante@carnegiescience.edu, hyang@carnegiescience.edu\\
$^5$Instituto de Astrof\'isica,  Facultad de F\'isica, Pontificia Universidad Cat\'olica de Chile, Santiago, Chile, barrientos@astro.puc.cl\\
$^6$School of Astronomy and Space Science, University of Science and Technology of China, Hefei 230026, China\\
$^7$School of Earth and Space Exploration, Arizona State University, Tempe, AZ 85287, USA\\
$^8$Cerro Tololo Inter-American Observatory, National Optical Astronomy Observatory, Casilla 603, La Serena,Chile\\
$^9$European Southern Observatory, Alonso de Cordova 3107, Casilla 19001, Santiago, Chile\\
$^{10}$The Kavli Institute for Astronomy and Astrophysics, Peking University, Beijing, 100871, China\\
$^{11}$National Optical Astronomy Observatory, 950 N. Cherry Ave, Tucson, AZ 85719\\
$^{12}$Purple Mountain Observatory, Chinese Academy of Sciences, Nanjing 210008, China\\
}

\begin{abstract}

We present the largest-ever sample of 79  Ly$\alpha$ emitters (LAEs) at $z\sim$ 7.0 selected in the COSMOS and CDFS fields of the LAGER project (the Lyman Alpha Galaxies in the Epoch of Reionization). Our newly amassed ultradeep narrowband exposure and deeper/wider broadband images have more than doubled the number of LAEs in COSMOS, and we have selected 30 LAEs in the second field CDFS. We detect two large-scale LAE-overdense regions in the COSMOS that are likely protoclusters at the highest redshift to date.
We perform injection and recovery simulations to derive the sample incompleteness. We show significant incompleteness comes from blending with foreground sources, which however has not been corrected in LAE luminosity functions in {the} literature. 
The bright end bump in the Ly$\alpha$ luminosity function in COSMOS  
is confirmed with 6 (2 newly selected) luminous LAEs (L$_{Ly\alpha}$ $>$ 10$^{43.3}$ erg s$^{-1}$). Interestingly, the bump is absent in CDFS, in which only one luminous LAE is detected. Meanwhile, the faint end luminosity functions from the two fields well agree with each other. 
The 6 luminous LAEs in COSMOS coincide with 2 LAE-overdense regions, while such regions are not seen in CDFS. 
The bright-end luminosity function bump could be attributed to ionized bubbles in a patchy reionization. It appears associated with cosmic overdensities, thus supports an inside-out reionization topology at $z$ $\sim$ 7.0, i.e., 
the high density peaks were ionized earlier compared to the voids. 
An average neutral hydrogen fraction of $x_{HI}$ $\sim$ 0.2 -- 0.4 
is derived at $z\sim$ 7.0 based on the cosmic evolution of  the Ly$\alpha$ luminosity function.

\end{abstract}

\section{Introduction} \label{introduction}
Cosmic reionization is a critical epoch in the {history of the} universe, during which most of the neutral hydrogen is ionized by the hard UV photons arising from the star forming galaxies and active galactic nuclei (AGNs).
Observations of the Gunn-Peterson troughs in the quasar spectra show that the epoch of reionization (EoR) ends at $z\sim 6$ \citep{Fan2006}.
Meanwhile, \cite{Planck2018VI} derived a mid-point reionization redshift $z\sim 7.7\pm0.7$ through measuring the Thompson scattering of CMB photons from free electrons.
High-$z$ gamma ray bursts (GRBs), quasars and galaxies are also probes to constrain the evolution of the neutral hydrogen fraction in the intergalactic medium {\citep[e.g.][]{Greiner2009, Mortlock2011, Banados2018, Finkelstein2015, Bouwens2015}};
however, the constraints are still poor up to date, especially at $z\gtrsim 7$. 

\lya\ emitters (LAEs) {are powerful probes to investigate cosmic reionization}, as \lya\ photons from galaxies in the early universe {are} resonantly scattered by the neutral hydrogen atoms in the intergalactic medium (IGM), {and} thus are sensitive to the neutral hydrogen fraction $x_{\mathrm{H\,\textsc{i}}}$ 
\citep[for a review see][]{Dijkstra2014}. 
High redshift LAEs can be effectively selected with narrowband imaging surveys \citep[e.g.][]{Malhotra2004, Hu2010, Ouchi2010, Tilvi2010, Hibon2010, Hibon2011, Hibon2012, Krug2012, Konno2014, Matthee2015, Santos2016, Konno2018}. 
In the past two decades, more than a thousand of LAE candidates have been selected at $z\sim5.7$, $6.5$, and $6.6$ \citep[e.g.][]{Konno2018, Jiang2017}. 
{However,  very small number of LAEs at $z\gtrsim 7$ had beed selected.
So far before this study, the largest samples of LAEs at $z\gtrsim 7$ include three ones at $z$ $\sim$ 7.0, i.e., 
the 23 candidates by \citet{Zheng2017}, 20 candidates by \citet{Ota2017} and 34 candidates by \citet{Itoh2018}.}

Lyman Alpha Galaxies in the Epoch of Reionization (LAGER) is an ongoing large area narrowband imaging survey for LAEs at $z\sim7.0$, using {the} Dark Energy Camera (DECam) 
installed on the Cerro Tololo Inter-American Observatory (CTIO) Blanco 4-m telescope.
DECam, with a red-sensitive 520 Megapixel camera and pixel scale of $0.27\arcsec$ and a superb field of view (FoV) of $\sim3$ deg$^2$,
is one of the best instruments in the world to conduct such surveys. 
A custom-made narrowband filter NB964{\footnote{Please find more information about the filter following: \url{http://www.ctio.noao.edu/noao/content/Properties-N964-filter}}} 
(with central wavelength $\sim$ 9642\AA\ and FWHM $\sim$ 92\AA, see Fig. \ref{fig:transmission})
was installed in the DECam system in December 2015 to search for LAEs at $z$ $\sim$ 7.0.
The narrowband filter bandpass was optimally designed to avoid strong sky OH emission lines and atmospheric absorption (see the transmission of NB964 filter and sky OH emission\footnote{Cerro Pachon sky emssion lines smoothed with a gaussian kernel of 4\AA\ for illustration, \url{http://www.gemini.edu/sciops/telescopes-and-sites/observing-condition-constraints/ir-background-spectra}} in Fig. \ref{fig:transmission}, for more details please see \citealt{Zheng2019}).

In the first LAGER field COSMOS, we selected 23 $z\sim7.0$ LAE candidates with 34 hours NB964 exposure \citep[in the central $2$ deg$^2$ region;][]{Zheng2017}.
A bright end bump in the \lya\ LF is revealed, suggesting the existence of ionized bubbles in a patchy reionization {process}. 
Six of the LAE candidates have been spectroscopically confirmed \citep{Hu2017}, including 3 luminous LAE with \lya\ luminosities of $\sim 10^{43.5}$ erg s$^{-1}$. 

In this paper, we present new results of $z\sim7.0$ LAEs selected in the deeper LAGER-COSMOS field and a second LAGER-CDFS field.
In \S \ref{s:data}, we describe the observations  and data reduction. 
We present the LAE selection in \S \ref{sc:rt}. The sample completeness  and the derived \lya\ LF are given in \S \ref{sc:lf}.
In section \ref{sc:ds}, we discuss the evolution of \lya\ LF and the cosmic reionization at $z\sim7.0$. 
Throughout this work, we adopt a flat $\Lambda$CDM cosmology with $\Omega_{\mathrm{m}}=0.3$, $\Omega_{\mathrm{\Lambda}}=0.7$ and H$_0=70$ km s$^{-1}$ Mpc$^{-1}$.

\section{Observations and Data Reduction} \label{s:data}

\subsection{Observations}

\begin{table*}
\caption{Summary of Imaging Observations}
\label{tab:seeing}
\centering
\begin{tabular}{l c c c c c l }
\hline
\hline
\tabincell{l}{Filter\\ \ } & \tabincell{c}{$\lambda_c$$^a$\\ (\AA)} & \tabincell{c}{$\Delta \lambda$$^b$\\ (\AA)}  & \tabincell{c}{Exp. Time\\ (s)} & \tabincell{c}{PSF Size \\ (arcsec)} & \tabincell{c}{5$\sigma$ Limiting Magnitude (AB) \\ 2\arcsec/1.35\arcsec (aperture diameter)} & \tabincell{l}{Observation Dates and Notes\\ \ } \\
\hline
\multicolumn{7}{c}{COSMOS} \\
\hline
\tabincell{l}{NB964\\ \ } & \tabincell{c}{9642.0\\ \ } & \tabincell{c}{92.0\\ \ } & \tabincell{c}{170,100\\ \ } & \tabincell{c}{0.90\\ \ } & \tabincell{c}{25.2/25.7\\ \ } & \tabincell{l}{2015 Dec 8, 2016 Feb 4-9, Mar 9-12, 2017 \\Dec 24-27}\\
SSC-$B$ & 4458.3 & 851.1 & $*$ & 0.61 & 27.7/28.1 & Public data\\
HSC-$g$ & 4816.1 & 1382.7 & 8,400 & 0.92 & 28.0/28.5 & Public data \citep{Aihara2018, Tanaka2017}\\
HSC-$r$ & 6234.1 & 1496.6 & 5,400 & 0.57 & 27.7/28.2 & Public data \citep{Aihara2018, Tanaka2017}\\
HSC-$r^{c}$ & 6234.1 & 1496.6 & $*$ & 0.63 & 26.0/26.5 & Public data \citep{Aihara2018}\\
HSC-$i$ & 7740.6 & 1522.2 & 21,600 & 0.63 & 27.5/27.9 & Public data \citep{Aihara2018, Tanaka2017}\\
HSC-$z$ & 9125.2 & 770.1 & 12,600 & 0.64 & 26.7/27.1 & Public data \citep{Aihara2018, Tanaka2017}\\
HSC-$y$ & 9779.9 & 740.5 & 34,200 & 0.81 & 26.2/26.6 & Public data \citep{Aihara2018, Tanaka2017}\\
 \hline
\multicolumn{7}{c}{CDFS} \\
 \hline
\tabincell{l}{NB964\\ \ } & \tabincell{c}{9642.0\\ \ } & \tabincell{c}{92.0\\ \ } & \tabincell{c}{121,200\\ \ } & \tabincell{c}{0.97\\ \ } & \tabincell{c}{25.0/25.5\\ \ } & \tabincell{l}{2015 Dec 7 \& 23, 2016 Mar 9-12, 2016 Nov 19-20, \\2017 Dec 24-27}\\
DECam-$g$ & 4734.0 & 1296.3 & 47,000 & 1.33 & 27.5/28.0 & Public data from NOAO Archive \\
DECam-$r$ & 6345.2 & 1483.8 & 161,750 & 1.20 & 27.6/28.0 & Public data from NOAO Archive\\
DECam-$i$ & 7749.6 & 1480.6 & 105,450 & 1.10 & 27.3/27.8 & Public data from NOAO Archive\\
DECam-$z$ & 9138.2 & 1478.7 & 274,680 & 1.02 & 27.0/27.5 & Public data from NOAO Archive\\
\hline
\end{tabular}
\begin{tablenotes}
\item{$^a$} { Central wavelength of the filter.}
\item{$^b$} { Effective width (FWHM) of the filter.}
\item{$^c$} { From HSC SSP deep survey. Note other HSC images are from HSC SSP ultra-deep survey.}
\end{tablenotes}
\end{table*}

Our DECam NB964 exposures of two LAGER fields were obtained between Dec. 2015 to Dec. 2017.
The total exposure time is 47.25 hrs in COSMOS and 32.9 hrs in CDFS. 
The NB964 data were scientifically reduced and calibrated by DECam Community Pipeline \citep{Valdes2014},
and the individual DECam frames were stacked with our customized pipeline (see \S\ref{ss:dr}).

\begin{figure}
\centering
\includegraphics[width=3.5in]{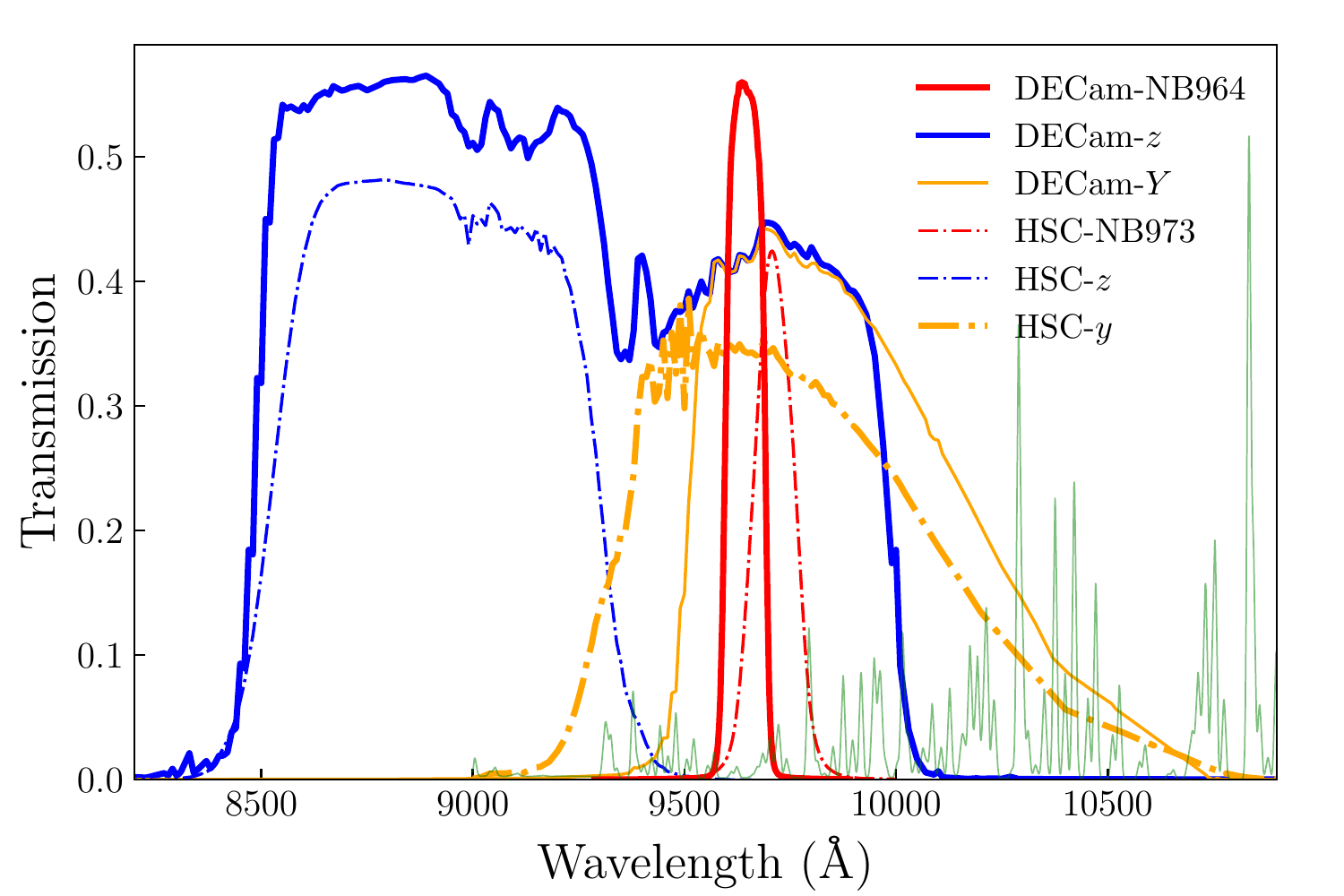}
\caption{\label{fig:transmission}{Total transmission curves of DECam filters and HSC filters, including the full system response from atmosphere (at airmass of $1.2$) to detector. 
The underlying broad bands (DECam-$z$ for CDFS and HSC-$y$ for COSMOS) adopted in this work are shown as bolded lines. 
The HSC-NB973 \citep{Itoh2018} is over-plotted for direct comparison with NB964 on DECam. 
The sky OH emission lines are also plotted (green line).  }}
\end{figure}

In COSMOS, the recently released {Hyper Suprime-Cam Subaru Strategic Program (HSC-SSP)} ultra deep broadband images \citep[$grizy$,][]{Tanaka2017}
are considerably deeper than the public DECam broadband images and the { Subaru Suprime-Cam (SSC)} images we used in \citet{Zheng2017}.
In this work we use the ultradeep HSC-SSP broadband images for LAE selections, and the deep SSC-$B$ band image is kept to extend the blue wavelength coverage down to 3500\AA.
The merged HSC observations of COSMOS by SSP team and University of Hawaii  \citep[][]{Aihara2018, Tanaka2017} were downloaded from HSC-SSP archive. 
HSC-$y$ band is selected as the underlying broadband (see Fig. \ref{fig:transmission} for the total transmission curve{\footnote{\url{https://hsc-release.mtk.nao.ac.jp/doc/index.php/survey/}}}) 
for comparison with NB964 narrowband for emission line selection.

In CDFS, ultradeep DECam broadband exposures ($griz$, together with much shallower $u$ and $Y$ exposures) are available and downloaded from 
National Optical Astronomy Observatory (NOAO) Science Archive\footnote{\url{http://archive.noao.edu/}}.
DECam-$Y$ image is however too shallow. We opt to {intend} use DECam-$z$ as the underlying broadband.
Its bandpass, unlike that of HSC-$z$, does overlap with NB964 (see Fig. \ref{fig:transmission} for the total transmission curve{\footnote {\url{http://www.ctio.noao.edu/noao/node/13140}}}).
The broadband DECam exposures were also stacked as described in \S\ref{ss:dr}.
Note the total transmission curves of DECam filters in Fig. \ref{fig:transmission} are higher than those of HSC filters. This is mainly because the CCD detector of DECam \citep{Diehl2008} has better quantum efficiency in the near-infrared than that of HSC\footnote{\url{https://www.naoj.org/Observing/Instruments/HSC/sensitivity.html}}.

All the broadband and narrowband images used in this paper, {including the 5$\sigma$ limiting magnitudes (for 2\arcsec\ and 1.35\arcsec\ diameter aperture respectively)}, are listed in Tab. \ref{tab:seeing}.

\subsection{Image stacking \label{ss:dr}}
In this section, we describe our optimal weighted-stacking approach following \citet{Annis2014} and \citet{Jiang2014}.
Briefly, for each individual DECam frame, we obtain the PSF, atmospheric transmission, {and} exposure time {to} generate a weight mask using those parameters and weight map provided by DECam Community Pipeline. Below are details of the approach. 

Firstly, we use PSFEx \citep{Bertin2011} to extract the PSF of each image and run SExtractor \citep{Bertin1996} to detect objects in the image.
To perform relative photometric calibration, we take one photometric frame for each band with low PSF FWHM as a standard image, 
and select a set of bright, unsaturated point-like sources as standard stars.
 {We obtain the zero-point of each frame relative to the standard image through cross-matching the standard stars in the images with a matching radius of $1\arcsec$.
We use these zero-point offsets to normalize the images to the same flux level.}

We utilize a 4$\sigma$-clipping method to reject artifacts in each frame (i.e., satellite trails, meteors, etc) which 
have not been masked out by the bad pixels masks provided by DECam Community Pipeline.
Since PSF varies in different images which will affect the clipping, we allow a fraction of 30\% flux variation per pixel during the clipping.

We assign each exposure a weight based on their exposure time $t_i$, PSF $\mathrm{FWHM}_i$, atmospheric transmission $T_i$, and background variance $\sigma_i^2$:
\begin{equation}
w_i = \frac{T^2_i t^2_i}{\mathrm{FWHM}^2_i \sigma^2_i}.
\end{equation}
This is similar to inverse variance weighting to minimize the variance of stacked image.
Here, the background variance $\sigma^2_i$ is given by DECam Community Pipeline, named {\textit{wtmap}}, which is the inverse variance of the local background. The atmospheric transmission $T_i$  is calculated with the relative zero-point of each individual frame aforementioned. 

Finally, we use SWarp \citep{Bertin2002} to resample and stack flux-normalized images with weight masks, and we obtain a stacked science image and a composite weight map. 

\subsection{Photometric Calibration}
\label{photometrycal}
We use SExtractor dual-image mode to extract sources from the images and {measure} photometry. 
The magnitude zero-points of broadband images in CDFS  and COSMOS  are calibrated using DES DR1 catalog \citep{Abbott2018} and COSMOS/UltraVISTA catalog \citep{Muzzin2013}, respectively.
The NB964 images are photometrically calibrated with $\sim 900$ A and B type stars in each field.
More specifically, we use Python package SED Fitter \citep{Robitaille2007} to perform spectral energy distribution fitting to the broadband photometries of stars with Castelli \& Kurucz (2004) models \citep{Castelli2004}, {and} then convolve the spectra with NB964 transmission curve to calculate the magnitudes of these stars in NB964 images. 
\section{LAE candidates} \label{sc:rt}
\begin{figure*}
\centering
\includegraphics[width=6.5in]{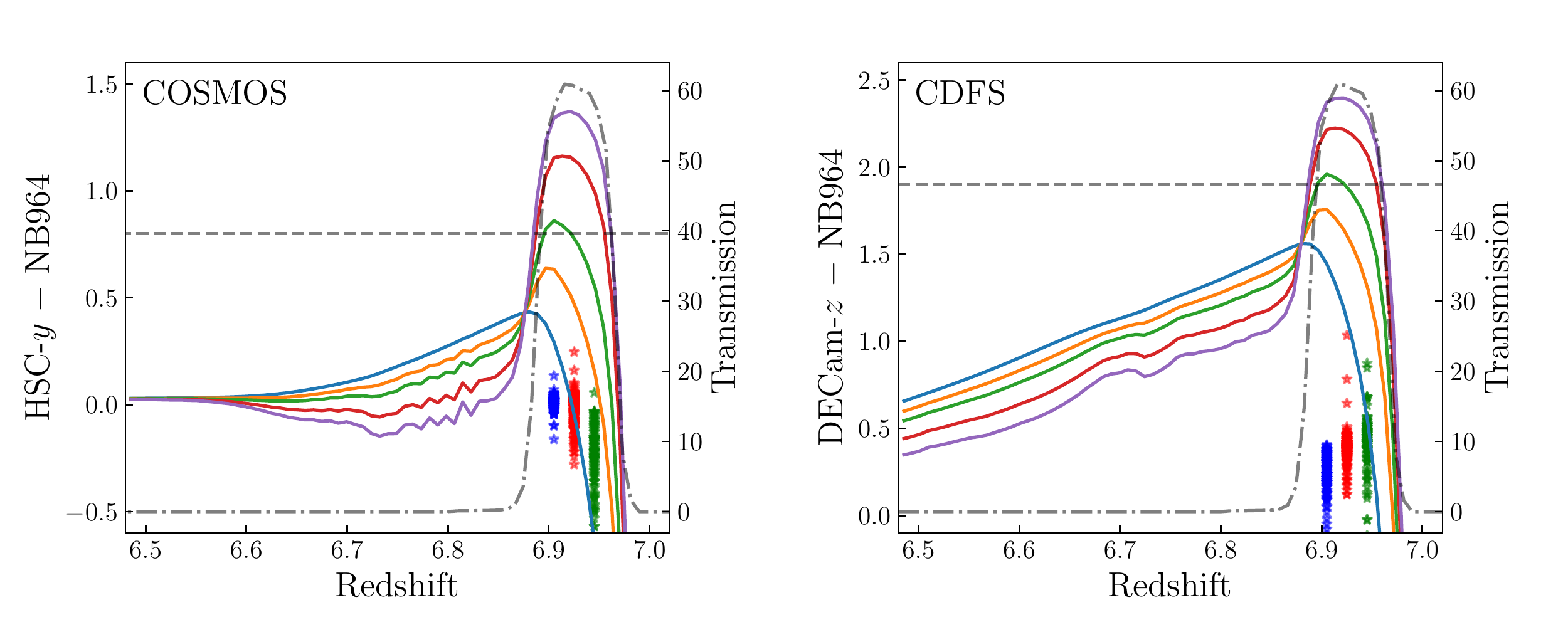}
\caption{\label{fig:color} Expected underlying BB $-$ NB964 color for LAEs from $z\sim6.5$ to $7.0$. Blue, orange, green, red, and purple lines present the colors of LAEs with rest frame \lya\ EWs of 0, 5, 10, 20, 30\AA, respectively. 
Grey dash-dotted lines show the profile of our NB964 filter. The grey dashed lines show the color cut we used for selecting LAEs, 1.9 for CDFS field and 0.8 for COSMOS field. 
{ We also plot the expected colors of the M/L/T dwarfs (in blue, red and green stars) using the spectra from SpeX Prism Library.
Note that the $x$-axis values for dwarfs are artificial.}}
\end{figure*}

\subsection{Selection Criteria{\label{ss:cs}}}

Our selection criteria of $z$ $\sim$ 7.0 LAEs consist of {three} components: 1) significant detection  in NB964 image; 2) color-excess of NB964 relative to the underlying broadband; and 3) non-detection in the bluer broadband (veto band) to filter out foreground galaxies.

We require our LAE candidates to be detected in NB964 image with { signal-to-noise ratio (SNR)} $>$ 5 in 2\arcsec\ diameter aperture. 
In order to rule out ``diffuse" artificial signals in the images, we find it is useful to  further apply a cut of {SNR} $>$ 5 in {a} 1.35\arcsec\ aperture. 
The completeness of such detection criteria is more complex than a single aperture photometry cut, but can still be estimated with injection and recovery simulations (see \S\ref{ssc:compp}). 

NB964 selected LAEs at $z$ $\sim$ 7.0 would exhibit flux excesses between the narrowband and the underlying broad band images
(HSC-$y$ for COSMOS, and DECam-$z$ for CDFS). 
To estimate the color excess, we simulate the photometric properties of $z\sim7$ LAEs from model spectrum following \citet{Itoh2018}. 
We assume a $\delta$-function-like \lya\ line profile and { power-law UV continuum ($f_{\lambda} \propto \lambda^{\beta}$ with $\beta=-2$).}
The UV spectra are attenuated by the neutral IGM with the model from \citet{Madau1995}.
We convolve the model spectrum with filter transmission (convolved with {instrument response} and atmospheric transmission, hereafter the same) to calculate the expected color excess for $z$ $\sim$ 6.5 -- 7.0 LAEs with various Ly$\alpha$ line equivalent widths. 
As shown in Fig. \ref{fig:color}, we adopt color cuts of BB -- NB $>$ $1.9$ and $0.8$ for DECam-$z$ and HSC-$y$, respectively, corresponding to rest-frame EW$_0$ of \lya\ line $\geqslant 10\mathrm{\AA}$. 
Following \citet{Ota2017}, we also plot the expected colors of the M/L/T dwarfs using the spectra from SpeX Prism Libarary\footnote{\url{http://pono.ucsd.edu/~adam/browndwarfs/spexprism}} to examine whether dwarfs could be selected with our color cuts.
Clearly, none of these dwarfs satisfies our criteria.

For COSMOS, we select LAEs using NB964 and Subaru HSC-$g,r,i,z,y$ Ultra-deep plus SSC-$B$ images with the following criteria: 
\begin{align}
\begin{split}
& \mathrm{SNR_{2\arcsec}(NB964)>5\; \&\ SNR_{1.35\arcsec}(NB964)>5}; \\
& \&\ \mathrm{SNR_{1.35\arcsec}(}B,g,r,i,z\mathrm{)<3}; \\
& \&\ \mathrm{[(}y \mathrm{- NB964 > 0.8\ \&\ SNR_{1.35\arcsec}(}y\mathrm{)>3)\ or}\\ 
& \mathrm{SNR_{1.35\arcsec}(}y\mathrm{)<3]},
\end{split}
\end{align}
{ Here we utilize SExtractor AUTO magnitudes (measured with dual imaging model on NB and BB) to calculate the color excess,
as it is known the Ly$\alpha$ emission in LAEs is more extended than the UV continuum \citep[e.g.][]{Momose2016,Yang2017,Leclercq2017}. 
In such case, the AUTO magnitudes, measured within regions defined by the narrowband image, could better recover the intrinsic color comparing with the common approach 
using aperture magnitudes (PSF-matched) to measure the color (see 
Appendix \ref{appd:a} for detailed comparison).}
 
{ We note that sources with SNR$_{1.35\arcsec}(y)$ $<$ 3 
automatically satisfy the color excess criterion ($y$ - NB964 $>$ 0.8), as the the underlying broadband image is much deeper than the narrowband image (see also Fig. \ref{fig:2d}).}

{ After masking out regions with significant CCD artifacts and bright stellar halos}, 
we select LAEs in the central region of NB964 exposure for which the NB964 image is covered with HSC-$y$ ultradeep exposure (with a total effective sky area  of 1.90 deg$^2$). 
For a small region of 0.45 deg$^2$ with no coverage of HSC-$r$ Ultra-deep exposure, we employ HSC-$r$ deep data from \cite{Aihara2018}. 
All HSC and SSC images are resampled to match DECam pixel scale.

Similarly, for CDFS, we select LAEs using NB964 and DECam-$u,g,r,i,z$ band with selection criteria:
\begin{align}
\begin{split}
& \mathrm{SNR_{2\arcsec}(NB964)>5\ \&\ SNR_{1.35\arcsec}(NB964)>5};\\
& \&\ \mathrm{SNR_{2\arcsec}(}u,g,r,i\mathrm{)<3};\\
& \&\ \mathrm{[(}z \mathrm{- NB964 > 1.9\ \&\ SNR_{2\arcsec}(}z\mathrm{)>3)\ or}\\ 
& \mathrm{SNR_{2\arcsec}(}z\mathrm{)<3]},
\end{split}
\end{align}
{ Again, sources with SNR$_{2\arcsec}(y)$ $<$ 3 
automatically satisfy the color excess criterion ($y$ - NB964 $>$ 1.9, see also Fig. \ref{fig:2d}).}
Since DECam broadband images were obtained without significant dithering, we lost a significant portion of sky coverage due to {CCD} gaps.
The final selection was performed in a total effective area of 2.14 deg$^2$ with both deep broad and narrowband coverage. 

Note we adopt 2\arcsec\ aperture for veto band photometry in CDFS, but 1.35\arcsec\ aperture {for} COSMOS. This is because
COSMOS broadband images generally have better seeing than those in CDFS. 
We find the 2\arcsec\ aperture veto band photometry of some good candidates would be contaminated by nearby foreground sources.
{Measuring} photometry with 1.35\arcsec\ aperture {avoids the loss of such candidates}.    
We note blending with foreground galaxies in the veto bands {can} still yield significant incompleteness in the final selected LAE sample,
{which we calculate with injection and recovery simulations in \S\ref{ssc:compp} and correct in the calculation of our luminosity function.}

\subsection {Selected Candidates{\label{sec:selection}}}

\begin{figure*}
   \centering
   \includegraphics[width=6.5in]{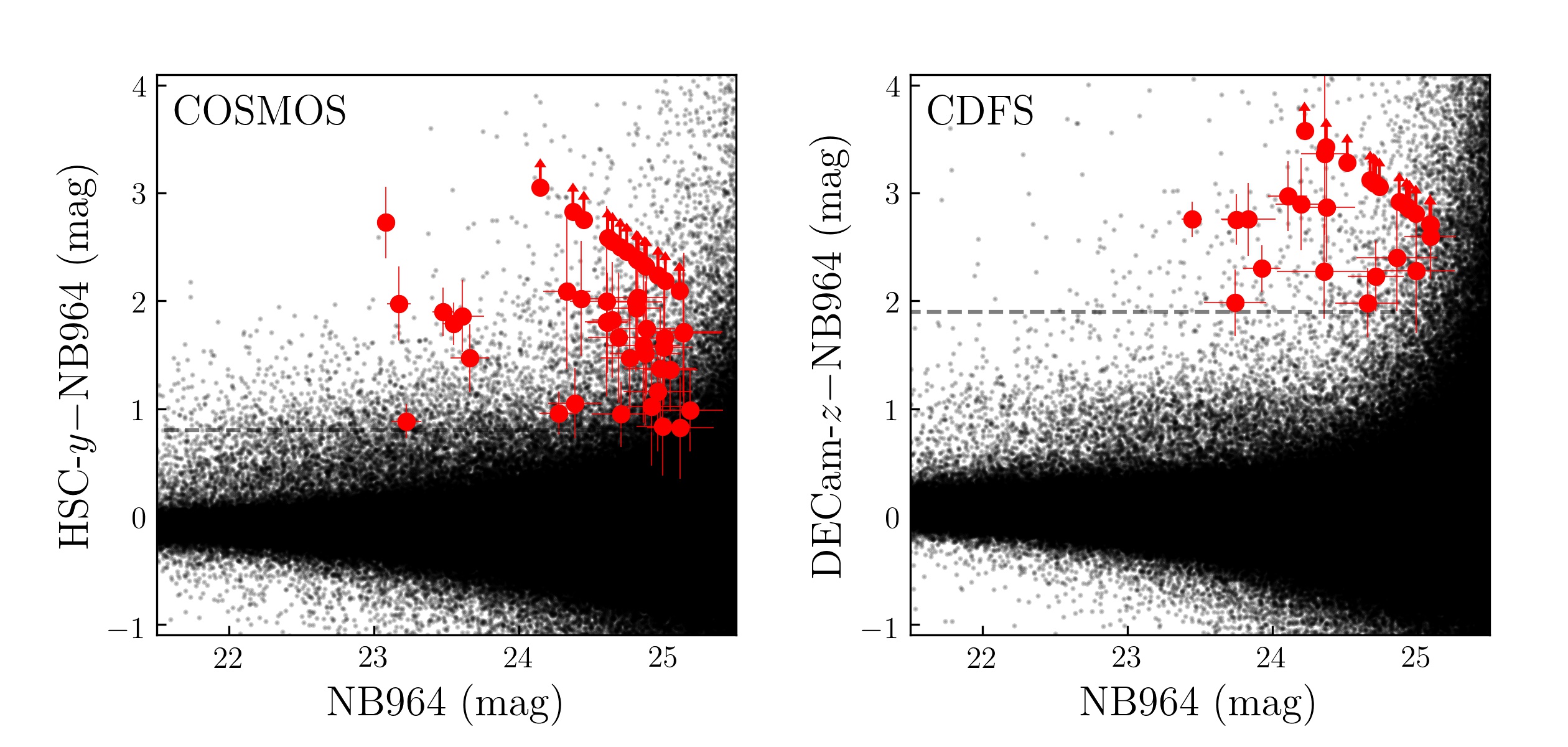}
   \includegraphics[width=6.5in]{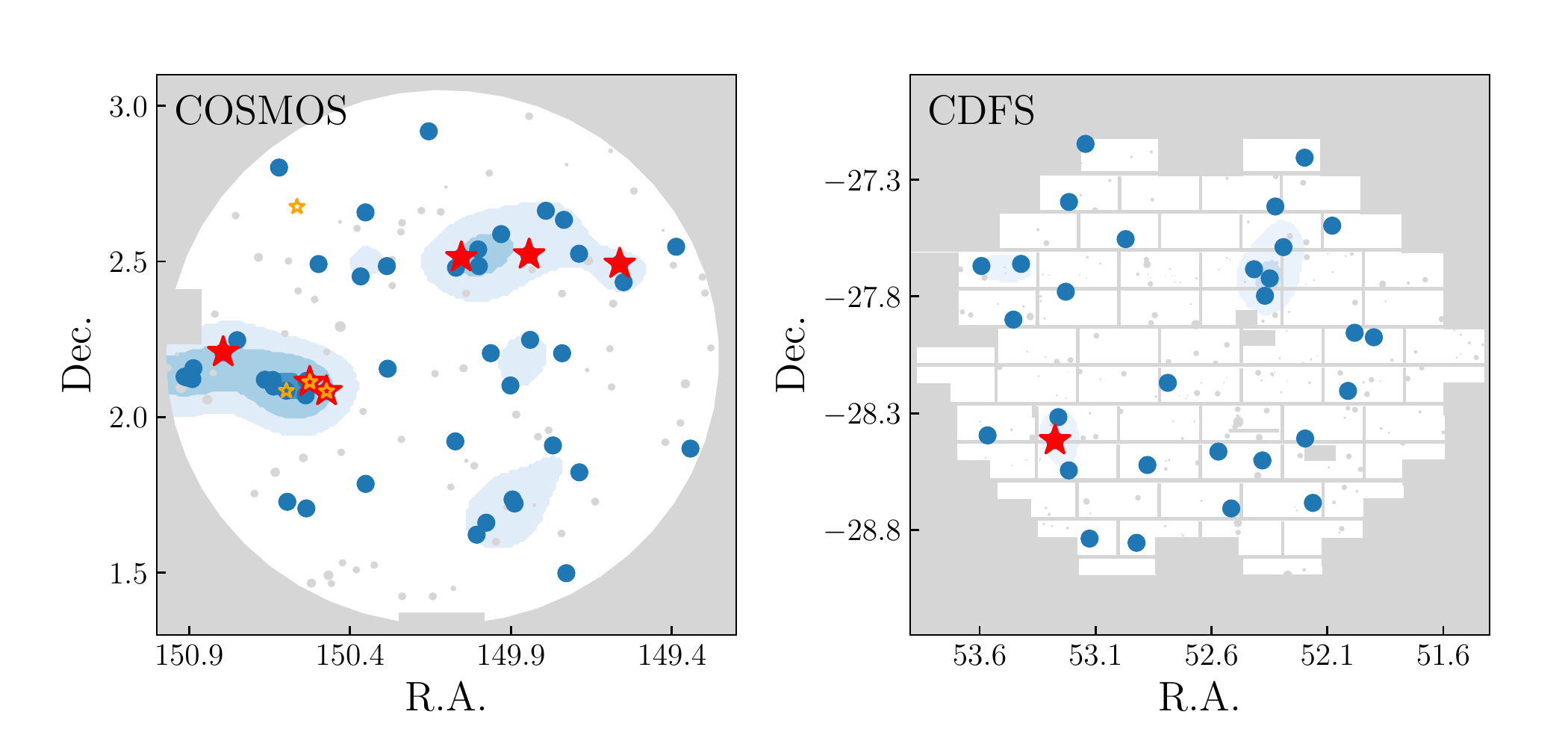}
   \caption{\label{fig:2d}
   { Upper panel: the color-magnitude diagrams (HSC-$y-$NB964 vs. NB964 for COSMOS and DECam-$z-$NB964 vs. NB964 for CDFS) of the LAE candidates (in red) and all other objects (in black). The dashed lines are the color criteria we used to select LAEs. 
   }
   Bottom panel: the spatial distribution of the selected LAEs in COSMOS and CDFS (blue dots). 
   The luminous LAEs with $\log_{10} \mathrm{L} > 43.3$ are marked with red stars.
   The four luminous LAEs selected by \cite{Itoh2018} are over-plotted with orange stars. 
   {The blue shadow contours show the local number densities of LAEs smoothed with a Gaussian Kernel of $\sigma=0.1 \deg$ to highlight the overdense regions, with the contour levels respect to  [1,2,3]$\times$ average number density of LAEs in the COSMOS and CDFS field.}
   {The regions we masked out when we perform candidate selections, completeness analyses and LF calculations are marked with grey shades. Note in CDFS as the archive broadband data were obtained without dithering, we exclude regions in CCD gaps.}
   Spectroscopic confirmations of 6 LAES in the COSMOS field have beed reported by \citet{Hu2017}, including the 3 luminous ones 
   at the left side. 
   More spectroscopic confirmations will be presented in future works. 
   }
\end{figure*}

A sample of 75 and 50 likely candidates were selected in COSMOS and CDFS respectively after excluding tens that were identified as obviously artificial due to 
CCD artifacts, stellar halos, bleeding trails, and significant veto band signals in quick visual examinations. 
3 transients were further excluded in COSMOS using NB964 exposures obtained at different epochs. 
We then perform a careful visual inspection of the candidates.
Although we adopt 3$\sigma$ rejection in veto bands, 11 and 12 candidates in COSMOS and CDFS show weak counterparts in at least one of the veto broadbands and are excluded.
These sources are more likely foreground transients or extreme emission line galaxies (see also \S\ref{ELG}).
We also remove 10 and 8 candidates in COSMOS and CDFS whose NB964 signals appear more like weak CCD artifacts {or} noise spikes, or {are} too close to adjacent bright objects.
Though such steps are somehow subjective, inspections from various team members often yield consistent classifications, and slight inconsistencies from various inspectors {do not} significantly alter the scientific results presented in this work. 
As foreground emission line galaxies (ELGs) may trace the overdense regions \citep[e.g.][]{Hayashi2018}, we further reject 2 candidates in COSMOS field which are adjacent to foreground ELGs (those NB964 excess sources with veto  band detections) within $3\arcsec$.

{ We then examine the stacked veto band images (which is deeper than a single veto band) of the remaining candidates,
and none of them show signal to noise ratio $>$ 2 in the stacked veto band.} 
Finally, we obtain a clean sample of 49 and 30 LAE candidates in COSMOS and CDFS field respectively,
{the thumbnail images of which are presented in Appendix \ref{appd}, and the catalog will be released in a future work, together with candidates to be selected in upcoming new LAGER fields.}
In Fig. \ref{fig:2d} we plot {the color-magnitude diagrams and the} spatial distribution of our selected LAEs in COSMOS and CDFS, in which 
two { possible} elongated over dense regions are seen in COSMOS. Each has a 2 dimensional  scale of $\sim$ 75 $\times$ 40 cMpc$^2$,  and contains $\sim$ 12 LAEs, including 3 of which are luminous with L$_{Ly\alpha}$ $>$ 10$^{43.3}$ erg s$^{-1}$.  
 Such large scale structures of high redshift LAEs probe the protoclusters in the early universe, and have been reported at redshift of 5.7 and 6.6 \citep{Wang2005, Ouchi2005, Jiang2018, Harikane2019} 
and smaller \citep[][]{Shimasaku2003,Zheng2016}. 
We note that 5 members of the large scale structures have been spectroscopically confirmed using \textit{Magellan}/IMACS \citep[][]{Hu2017},
with a remarkably high success rate of 2/3,
This indicates these two structures are physically real.  
Spectroscopic followup of the remaining members is ongoing to further secure their identifications, and will be presented in future work. 

Using 34 hr NB964 exposure, overlapping UltraVISTA Y band image, and deep Subaru SSC broadband images in the central 2 deg$^2$ region, 
 \citet{Zheng2017} selected 23 LAEs at $z$ $\sim$ 7.0 with EW $\geqslant10\mathrm{\AA}$.
Among them, 21 were recovered in this work using 47.25 hr NB964 exposure and considerably deeper broadband images. 
One of the {other} two was detected in the 47.25 hr NB964 image with S/N $<$ 5.0. It is a transient source, as the NB964 signal disappears in the latest 13.25 hr exposure. 
The {final} one passed the selection criteria, but was also rejected as a variable source with new NB964 data.

The total LAEs selected in COSMOS in this work is 49, more than doubling the number of \citet{Zheng2017}. 
Among the 28 new LAE candidates selected in this work but not in \citet{Zheng2017}, 
10 had NB964 S/N $<$ 5 in the 34hr exposure image;
7 had no SSC broadband coverage; 6 had too shallow or noisy broadband coverage; 1 was contaminated by a nearby source in veto band in 2\arcsec\ aperture;
2 were identified as possible noise spikes by visual examination of the faint NB signal (but re-classified as good candidates in this work with deeper NB964 exposure); and 2 were rejected due to visually identified marginal signal in one of the Subaru SSC veto bands. These two were re-classified as good candidates in this work using new and deeper HSC veto band images. 
The marginal veto band signals previously seen for those 2 sources {were} due to data processing flaw and disappear in re-processed images.

The \lya\ line fluxes are calculated using {the} NB and the underlying BB photometry by solving the following equation (Jiang et al. in prep):
\begin{align}
\begin{split}
\bar{f}_{\nu,NB/BB} &= \frac{\int (f_{\lambda,line}+f_{\lambda,con})T_{\lambda,NB/BB}d\lambda}{\int T_{\lambda,NB/BB}d\lambda} \times \frac{\bar{\lambda}^2_{NB/BB}}{c},
 \end{split}
\end{align}
where ${\bar{f}_{\nu, NB}}$, $\bar{f}_{\nu, BB}$ are the detected flux densities in the NB964 and the underlying broadband;
$T_{\lambda, NB}$, $T_{\lambda, BB}$ the corresponding filter transmission;
$f_{\lambda, line}$, $f_{\lambda, con}$ the \lya\ line and UV continuum flux at wavelength $\lambda$;
and $\bar\lambda_{NB}$, $\bar\lambda_{BB}$ the central wavelengths of the NB964 and broadband filter, respectively. 
{During the calculation, we assume a \lya\ line profile resembling a $\delta$-function at the center of the NB964 filter, }
and a power-law UV continuum with slope of $-2$ { and suffered neutral IGM attenuation with the model from \citet{Madau1995}.} 

For non-detections in the underlying broadband, we choose to calculate their BB flux densities using  2$\sigma$ limiting magnitudes. 
Note while this approach provides a conservative estimation of  \lya\ flux, it would 
systematically under-estimate the line flux if the underlying broadband is not sufficiently deep (see Section \ref{ssc:bbd} for further discussion).
If the output continuum flux from the equation is 0 or negative, we fix the the continuum flux to 0, which means the NB964 flux is completely contributed by \lya\ line.
Although several LAEs have been spectroscopically confirmed, we still use photometric fluxes to obtain their \lya\ fluxes due to the considerably large uncertainties in spectroscopic flux calibration. 

\subsection{Foreground Contaminant Emission Line Galaxies{\label{ELG}}}

The \lya\ emission line is often the only detectable feature of high-$z$ LAEs with optical/IR spectroscopic followup observations \citep[e.g.][]{Wang2009}.
Particularly, in many cases the spectral quality is limited, and the line profile is unresolvable. 
Can we safely identify such single line detections as high-$z$ LAEs?
Foreground ELGs are potential contaminants in such cases, especially the extreme emission line galaxies (EELGs) which have relative faint continua \citep[e.g.][]{Huang2015}. 
Below we estimate the number of expected contaminant foreground EELGs in our sample. 

Possible contaminant lines are \OII, \OIII, and H$\alpha$ emission lines at $z\sim1.59$, $0.93$, and $0.47$, respectively. 
With \textit{Hubble Space Telescope} (HST) slitless grism spectroscopic data, \citet{Pirzkal2013} obtained the LFs and rest frame EW$_0$ distributions of \OII\ line emitters at $z$ $\sim$ 0.5 -- 1.6, \OIII\ emitters at $z$ $\sim$ 0.1 -- 0.9, and H$\alpha$ emitters at $z\sim$ 0 -- 0.5. 
Assuming no strong evolution in these redshift bins and luminosity-independent EW distributions, we build artificial 
samples of \OII, \OIII, and H$\alpha$ emitters at  $z\sim1.59$, $0.93$, and $0.47$, respectively, utilizing the luminosity functions and EW distributions of \citet{Pirzkal2013}.
For each artificial ELG with assigned line luminosity and EW, we generate its mock spectrum by shifting 
the composite ELG spectrum from \citet{Zhu2015} to place the correspondent line at the central wavelength of NB964,
adjust the strength of the line relative to continuum to match the assigned line EW, and further normalize the spectrum to match the assigned line luminosity. 
Conservatively, the EW of other lines are fixed to values in the composite spectrum. 
Note that the \OIII\ doublet {was} unresolved by \citet{Pirzkal2013} but only one of {of the lines is} covered by our NB image. We adopt a line ratio of \OIII\ $\lambda4959/$\OIII\ $\lambda5007=0.40$ based on the ELG composite spectrum.

We convolve the mock spectra with {the} transmission curves of the narrow and broadband filters to calculate the expected magnitudes. 
We then apply our LAE selection criteria to the artificial ELG samples. 
The selection incompleteness described in Section \ref{ssc:compp} {is} also considered in the calculation.
The estimated numbers of \OII, \OIII, and H$\alpha$ emitters in our LAE samples are 0.14, 0.52, and 0.06 in COSMOS  and 0.24, 0.83, and 0.35 in CDFS, respectively.
In total, we predict the number of contaminant ELGs to be 0.72 in COSMOS and 1.42 in CDFS. 
The expected contamination in CDFS is higher, mainly because in this field the broadband images are slightly shallower (see Tab. \ref{tab:seeing}) comparing with COSMOS.

We note that only extreme ELGs can possibly contaminate our LAE sample. Such EELGs have continua steeper than the ELG composite spectrum and the rest emission lines are also stronger than those in the ELG composite spectrum \citep{Forrest2017}. These factors would elevate the veto broadband flux densities we estimated above.  Therefore, the number of contaminant foreground ELGs in the LAE sample we presented {above} have been conservatively overestimated. On the other hand, if the luminosity function of ELGs strongly evolves with redshift (e.g., the density of \OII\ emitters at $z$ $\sim$ 1.59 is higher than the average value at $z$ $\sim$ 0.5 -- 1.6), we would expect slightly more contaminants than expected above.  We finally note that some of such contaminants may have been excluded with our visual examination (\S\ref{sec:selection}).
Overall, we expect negligible foreground emission line contaminants in our LAE sample, thanks to the ultra-deep veto band images available. 

\begin{figure*}
   \centering
   \includegraphics[width=7in]{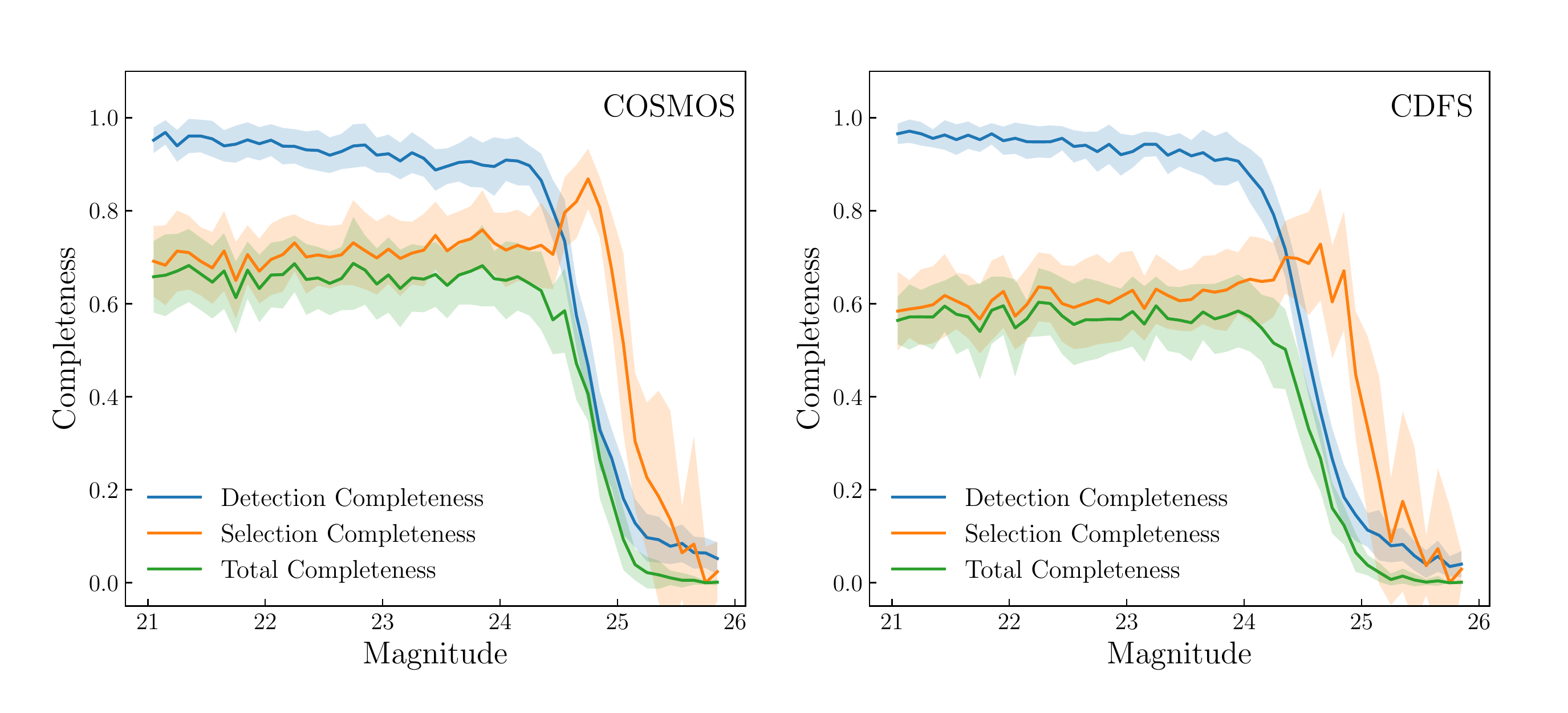}
   \caption{\label{fig:sltcomp} The LAE sample completeness (total completeness; green line) as a function of input NB964 total magnitude
   derived through injection and recovery simulations. 
   The detection completeness (blue line) plots the fraction of injected pseudo LAEs which can be detected in the corresponding narrowband image.
   The selection completeness (orange line) draws the fraction of NB964 detected pseudo LAEs which pass our LAE selection criteria.
{In each simulation trial we inject $\sim$ 3000 pseudo galaxies, and repeat 100 trials to estimate the average and 1$\sigma$ scatter (shaded regions) of the completeness measurements.}}
\end{figure*}

\section{\lya\ luminosity function} \label{sc:lf}

\subsection{Sample incompleteness \label{ssc:compp}}
It is essential to correct the sample incompleteness for the calculation of the luminosity function. 
Such incompleteness can be estimated through injection and recovery simulations, as described below.

We first run the Python package Balrog \citep{Suchyta2016} which utilizes GALSIM \citep{Rowe2014}, to simulate
pseudo LAEs, apply PSF convolution, and randomly insert the galaxies into the NB964 images.
The pseudo galaxies have a S\'ersic profile with a S\'ersic index $n$ of 1.5 and half-light radius of 0.9 kpc , corresponding to 0.17\arcsec\ at $z\sim7$.
The adopted S\'ersic index and half-light radius are similar to the recent UV continuum profile measurements of high redshift LAE and LBG galaxies in the EoR \citep[e.g., {0.5 -- 0.7 kpc for narrowband selected LAEs, and  0.9 -- 1.0 kpc for broadband selected LBGs,} ][]{Jiang2013, Allen2017, Shibuya2019}.
The magnitudes of pseudo galaxies in the narrowband are randomly given in the range of 21 to 26.
We then run SExtractor on the NB964 images after injections, with the identical configuration we used for detecting true LAEs. 

{Note the Ly$\alpha$ emission in LAEs could be more extended than the UV continuum \citep[e.g.][]{Finkelstein2011, Momose2016,Yang2017,Leclercq2017}.}
{For instance, with HST narrowband imaging data, \citet{Finkelstein2011} reported 3 $z\sim4.4$ LAEs have an averaged \lya\ emission half-light radius of 1.1 kpc, larger than that of the UV continuum (0.7 kpc).}
{\citet{Leclercq2017} reported the detection of extended \lya\ halos around $z\sim$ 3--6 LAEs,  observed with the Multi-Unit Spectroscopic Explorer (MUSE) at ESO-VLT.
Through two-component (continuum-like and halo) decomposition, they reported  an exponential scale length of $3.8\pm1.3$ kpc for the halo and $0.3\pm0.1$ kpc for the core in their highest redshift bin ($z\sim$ 5--6). Directly taking their measured \lya\ profiles, we calculate the effective half-light radius of the total \lya\ emission (core plus halo) for each LAE, and find a medium value of 1.5 kpc at $z\sim$ 5--6.
To address such effect, we also simulate pseudo LAEs with larger half-light radius of 1.2 kpc and 1.5 kpc, and find negligible difference in the completeness measurements. 
Thus, in this work, we adopt a half-light radius of 0.9 kpc to be consistent with previous works \citep[e.g.][]{Zheng2017,Konno2018}\footnote{A further note is that the effect of the extended halo 
relies on both its size and relative brightness, which are yet unknown for $z\sim7.0$ LAEs.  Our simulations could be insufficient for large \lya\ blobs or LAEs with strong and extended halos.}.}

We plot the fraction of the pseudo galaxies that are detected (with S/N $>$ 5 in both 2\arcsec\ and 1.35\arcsec\ aperture) as a function of magnitude in Fig. \ref{fig:sltcomp} (detection completeness hereafter).
{ Though masking out the regions around the bright stellar halos and CCD artifacts is a common approach for LAE selection \citep[e.g.][]{Ota2017, Itoh2018}, the detection completeness is still slightly less than unity even for the bright pseudo galaxies.
This is because they might still blend with bright foreground sources, making them undetectable by SExtractor in NB964 images.}
The detection completeness gradually drops with decreasing pseudo galaxy brightness before reaching the limiting magnitudes, as blending with foreground galaxies could hinder their detections. 

However, not all the NB964-detected pseudo LAEs passed our LAE selection criteria, since many of them were blended with foreground sources that were not bright enough to block them from NB964 detections,
but sufficiently bright to make them fail the requirement of veto band non-detection and/or color-excess. 
We apply our LAE selection criteria on the NB964-detected pseudo galaxies, and plot the recovery fraction (relative to the number of NB964 detected pseudo objects, hereafter ``selection completeness'') in Fig. \ref{fig:sltcomp}. 
{Note we insert the pseudo galaxies only into NB964 image and assume their underlying broadband fluxes to be zero. 
The effects if we  insert also pseudo underlying broadband fluxes based on \lya\ line EW are rather complicated \citep[see][]{Zheng2014}. 
Basically, while high EW pseudo LAEs can be easily recovered by the selection procedure, some low EW sources could be missed by our selection due to photometric fluctuations. 
However, a contrary effect is that some pseudo LAEs with intrinsic line EWs below our EW limit could also have their line EWs boosted by photometric fluctuations, thus be picked up. 
The net effect  is an Eddington type bias, and  can be quantitatively estimated with an accurate EW distribution, which is yet unavailable at $z\sim$ 7.0. 
Nevertheless, as shown by \cite{Zheng2014} , such bias is rather weak, as long as the underlying broadband image is $>$ 0.5 -- 1.0 mag deeper than the NB image, a condition that well satisfied by our datasets.  Thus in this work we do not consider broadband fluxes of pseudo galaxies in the simulations.}

Clearly, the effect of ``selection incompleteness" is remarkable and shall not be neglected.
Such incompleteness is mainly due to foreground contamination in the veto broadband photometry.
The effect of blending with foreground sources in the veto band can also be roughly estimated with random aperture photometry.
For example,  only $74.9\%$ of the $1.35\arcsec$ randomly placed apertures in the COSMOS HSC-$g$ band image
yield S/N $<3\sigma$.
The fraction further decreases to $68.3\%$ and $59.6\%$ if the aperture diameter increases to 2\arcsec\ and 3\arcsec, respectively.
Therefore, deriving veto band photometry using larger aperture would yield stronger sample incompleteness due to foreground contaminations.  
The incompleteness is also sensitive to the depth and PSF of the veto broadband images, i.e., 
the foreground contamination to the veto broadband would be more severe for deeper images, or those with poorer PSF.

We further note that the {``}selection completeness" is not constant, but magnitude-dependent (Fig. \ref{fig:sltcomp}). 
{The ``selection completeness" gradually increases with increasing magnitude. This is because a fainter pseudo LAE, if blended with foreground source(s), would more likely be treated as part of the adjacent object(s) and simply not detected in the narrowband image by SExtractor.}
Such effect, which is more important toward fainter magnitude, would suppress the detection completeness. {Consequently, those non-detected pseudo LAEs would be pre-excluded from the calculation of ``selection completeness", which in turn gets boosted (as seen in Fig. \ref{fig:sltcomp})}.
Near the detection limit where the detection completeness sharply drops, the effect is so strong {that most of the detected pseudo LAEs are located in sparse regions, i.e., free from contaminations, and } the selection completeness even exhibits a significant peak.
Meanwhile the ``selection completeness" significantly drops at faintest magnitudes, as
most of the faintest pseudo LAEs could be detected solely because they were injected by coincidence on top of foreground sources. 

The total sample completeness (the product of detection completeness and selection completeness, i.e, the fraction of injections that can be recovered as LAEs) is plotted in Fig. \ref{fig:sltcomp}.
We note that the ``detection incompleteness" in the narrowband images has usually been corrected for the Ly$\alpha$ luminosity function reported in the literature. Unfortunately, the ``selection incompleteness", which is indeed more prominent as we demonstrated above, has not been considered in previous studies.

\subsection{\lya\ Luminosity Function at $z\sim7$ \label{ssc:lf}}
Following \citet{Zheng2017},
we calculate the $z\sim7$ LAE luminosity function using the formula:
\begin{align}\label{equation_lf}
\Phi(L)dL=\sum_{L_i\in [L-\Delta L/2, L+\Delta L/2]}\frac{1}{V_{\mathrm{eff}} f_{\mathrm{comp}}(NB_i)} dL,
\end{align}
where $V_{\mathrm{eff}}$ is the effective volume of the survey which is calculated from sky coverage and redshift coverage, and $f_{\mathrm{comp}}$ the completeness described in Section \ref{ssc:compp} for each LAE with NB964 magnitude $NB_i$.
The effective volume is $1.29\times 10^6$ cMpc$^3$ and $1.14\times 10^6$ cMpc$^3$ for CDFS and COSMOS field, respectively, { with bad regions, such as CCD artifacts and bright stellar halos, removed}.
{ We do not take the contamination into account, since we expect only a few foreground ELGs can be included in our LAE sample (see Sec. \ref{ELG}).}

\begin{figure*}
\centering
\includegraphics[width=5.5in]{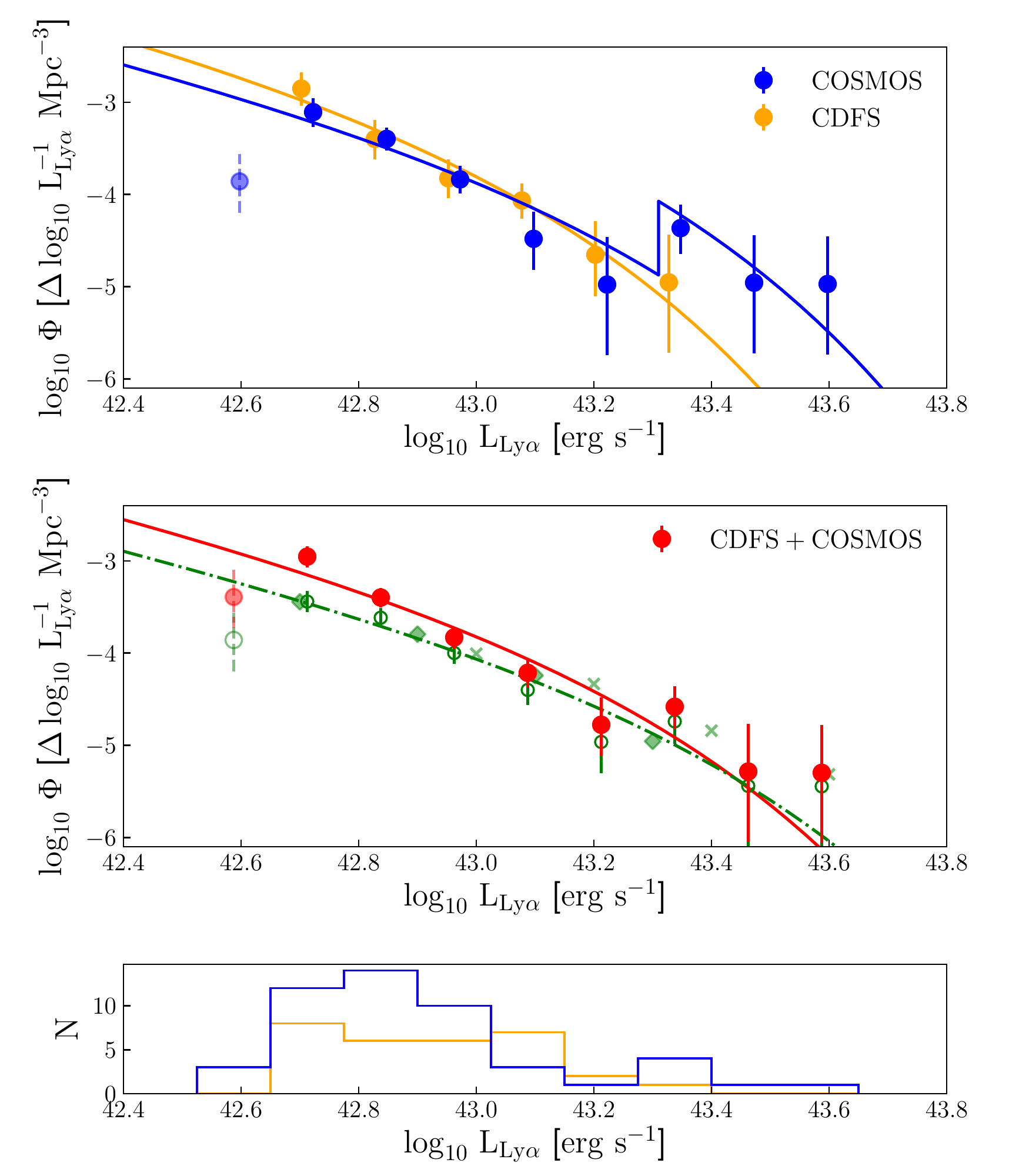}
\caption{\label{fig:lf_z7} Ly$\alpha$ LFs for our LAEs (selection incompleteness corrected). Upper panel:  the LFs from COSMOS and CDFS separately. 
To avoid confusion, the data points were horizontal shifted by $+0.01$ and $-0.01$ dex for COSMOS and CDFS respectively. 
A clear bright end bump is seen in COSMOS but not in CDFS. 
{Middle panel:  the LF averaged over two fields (red). To illustrate the effect of ``selection incompleteness", we over-plot the ``selection incompleteness" uncorrected LF of this work in green open circles and green dot-dashed line. We also over-plot the data points of $z\sim$ 7.0 LAE LFs from  \citet{Ota2017} and \citet{Itoh2018} in green diamonds and green crosses, respectively, in both of which the ``selection incompleteness" was not corrected.
}
The left most LF bin (semi-transparent data points) in both the top and middle panels suffer strong incompleteness, and were  excluded from further analyses.
In both panels, the best-fit Schechter functions with $\alpha=-2.5$ are over-plotted (see text for details). 
Lower panel: the number  of LAEs in each field in each luminosity bin.
}
\end{figure*}

The resulting luminosity functions are plotted in Fig. \ref{fig:lf_z7}. 
The bright end luminosity bump in COSMOS, first reported by \citet{Zheng2017}, is confirmed with a {doubly large} sample size.
A detailed comparison with the LF in \citet{Zheng2017} is presented in \S \ref{ssc:bbd}.
 We interpret the bump in COSMOS as an evidence of ionized bubbles at $z\sim7$ (see Section \ref{ssc:beb}). 
Remarkably, while the faint end luminosity functions from two field agree with each other, the bright end bump is not seen in CDFS. 

\begin{table*}
\caption{Best-fit Schechter Parameters of \lya\ luminosity function and \lya\ luminosity density from $z\sim 5.7$ to $7.3$}
\label{tab:lf}
\centering
\begin{tabular}{l l l l l l l l l}
\hline
\hline
\tabincell{l}{$z$\\ \ } & \tabincell{l}{Field\\ \ } & \tabincell{l}{$L$ Fitted Range\\ (erg s$^{-1}$)} & \tabincell{l}{$\log_{10} L^\ast_{Ly\alpha}$ \\ (erg s$^{-1}$)} & \tabincell{l}{$\log_{10} \Phi^\ast$ \\ (Mpc$^{-3}$)} & \tabincell{l}{$\alpha$ \\ \ } & \tabincell{l}{$\log_{10} \rho_{Ly\alpha}$ \\ (erg s$^{-1}$ Mpc$^{-3}$)} & \tabincell{l}{Transmission \\ ($T^{IGM}_z/T^{IGM}_{5.7}$) } \\ 
\hline
\multicolumn{8}{c}{Selection incompleteness corrected}\\
\hline
6.9 & COSMOS & $42.65-43.4$ & $42.75^{+0.15}_{-0.11}$ & $-3.02^{+0.38}_{-0.44}$ & $-2.5$(fixed) & $39.57\pm0.13$ \\
6.9 & CDFS & $42.65-43.4$ & $42.93^{+0.21}_{-0.13}$ & $-3.62^{+0.37}_{-0.48}$ & $-2.5$(fixed) & $39.38\pm0.09$\\
6.9 & COSMOS+CDFS & $42.65-43.65$ & $42.94^{+0.11}_{-0.09}$ & $-3.60^{+0.25}_{-0.28}$ & $-2.5$(fixed) & $39.42\pm0.08$\\
\hline
\multicolumn{8}{c}{Selection incompleteness uncorrected}\\
\hline
5.7$^a$ & HSC SSP & $42.4-44.0$ & $43.21^{+0.36}_{-0.24}$ & $-4.07^{+0.51}_{-0.28}$ & $-2.56^{+0.53}_{-0.45}$ & $39.54$ & 1 \\ 
6.6$^a$ & HSC SSP & $42.4-44.0$ & $43.22^{+0.07}_{-0.23}$ & $-4.33^{+0.61}_{-1.27}$ & $-2.49^{+0.50}_{-0.50}$ & $39.26$ & $0.70\pm0.15$ \\ 
6.9& CDFS+COSMOS & $42.65-43.65$ & $43.08^{+0.14}_{-0.11}$ & $-4.19^{+0.26}_{-0.31}$ & $-2.5$(fixed) & $39.13\pm0.07$ & $0.63\pm0.12$ \\ 
 7.3$^b$ & SXDS+COSMOS & $42.4-43.0$ & $42.77^{+1.23}_{-0.34}$ & $-4.09^{+1.09}_{-1.91}$ & $-2.5$(fixed) & $38.55\pm0.17$ & $0.29\pm0.19$  \\ 
\hline
\end{tabular}
\begin{tablenotes}
\item[$^a$] $^a$ The LFs at $z\sim 5.7$ and $6.6$ are from \citet{Konno2018}.
\item[$^c$] $^b$ The best-fit Schechter parameters for LF at $z\sim7.3$ when assuming a fixed slope of $\alpha=-2.5$ and using the data points from \citet{Konno2014}.
\end{tablenotes}
\end{table*}

We fit a Schechter function to the luminosity functions as:  
\begin{align}
\Phi(L)dL=\Phi^{\ast}\left( \frac{L}{L^\ast}\right)^\alpha \exp \left( -\frac{L}{L^\ast} \right) d\left( \frac{L}{L^\ast} \right),
\end{align}
where $L^\ast$ and  $\Phi^\ast$ are the characteristic luminosity and number density, respectively. 
We fix the faint end LF slope $\alpha$ of the Schechter function to $-2.5$, consistent to those observed at $z\sim 5.7$ and $6.6$ \citep[e.g.][]{Konno2018, Santos2016, Matthee2015}.
The three LAEs in the lowest luminosity bin in COSMOS have selection completeness $f_{comp}$ (see equation \ref{equation_lf}) $<$ 0.1. This indicates this luminosity bin suffers {from} incompleteness too strong to be accurately estimated and corrected, {and} thus {we} exclude {it} from further analyses.
In Fig. \ref{fig:lf_z7} we fit the luminosity functions from both fields in the luminosity range of 10$^{42.65}$ -- 10$^{43.4}$ erg s$^{-1}$, i.e.,
excluding the two brightest luminosity bins for COSMOS as there are no LAEs in CDFS in these two bins. 
We use {the} Cash Statistics { \citep[a maximum likelihood-based statistics for Poisson data, i.e., low number of counts, ][]{Cash1979}} to estimate the best-fit value and error of $L^\ast$ and  $\Phi^\ast$.
The best-fit curves are plotted in the Fig. \ref{fig:lf_z7} and the best-fit Schechter parameters are listed in Tab. \ref{tab:lf}.
To better illustrate the bright end bump in COSMOS,  we plot the Schechter function elevated and truncated at the bright end to match the two brightest bins.

{We also present the LF averaged over two fields in the middle panel of Fig. \ref{fig:lf_z7}, together with the best-fit Schechter function (over the full luminosity range of 10$^{42.64}$ -- 10$^{43.65}$ erg s$^{-1}$). 
For comparison, we over-plot the ``selection incompleteness" uncorrected LF, e.g., with only ``detection incompleteness" corrected. Leaving the ``selection incompleteness" uncorrected clearly yields underestimated LF. 
The z $\sim$ 7.0 LFs from \cite{Itoh2018} and \cite{Ota2017} are also over-plotted,  in both of which the ``selection incompleteness" correction was unavailable thus not applied.   
}

\subsection{The Effect of the Underlying Broadband Depth 
\label{ssc:bbd}}

\begin{figure}
   \centering
   \includegraphics[width=3.5in]{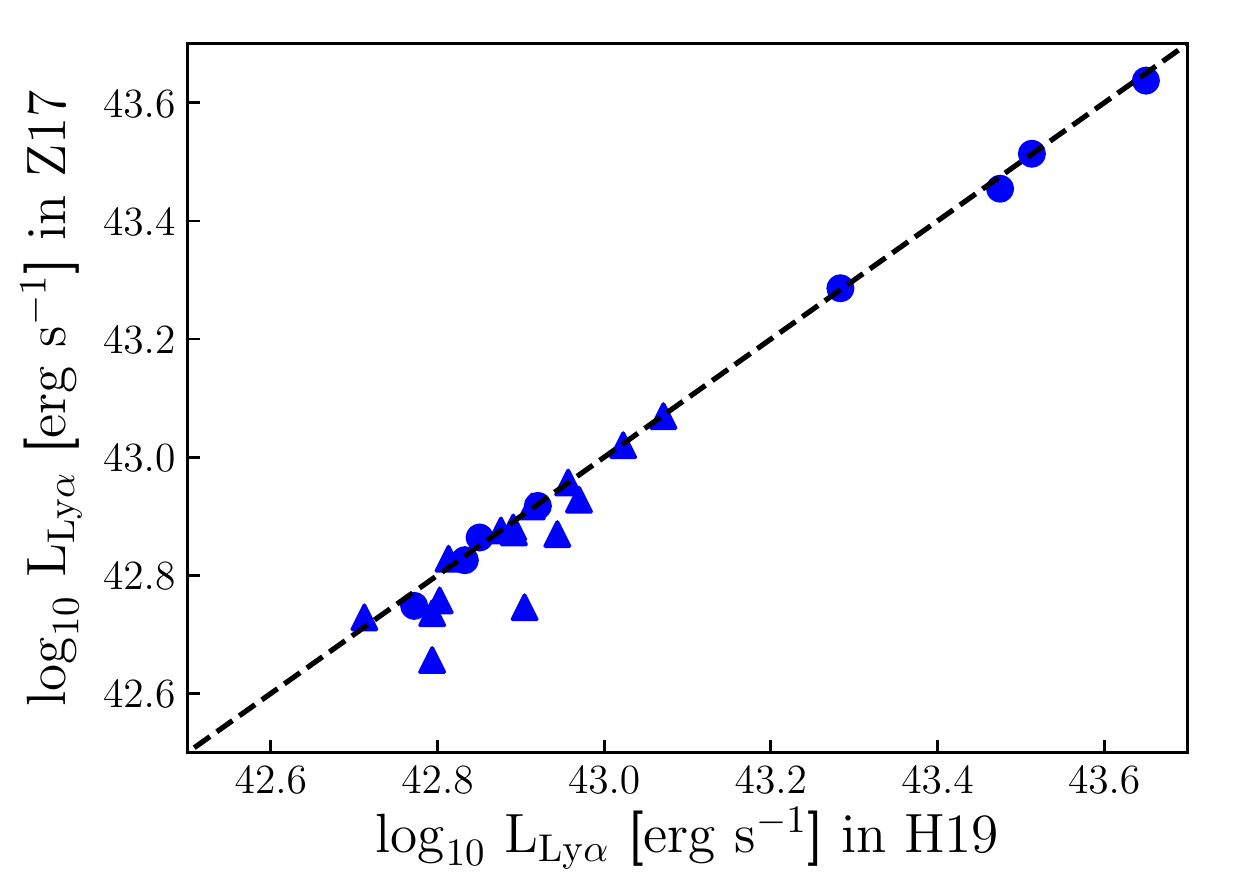}
   \includegraphics[width=3.5in]{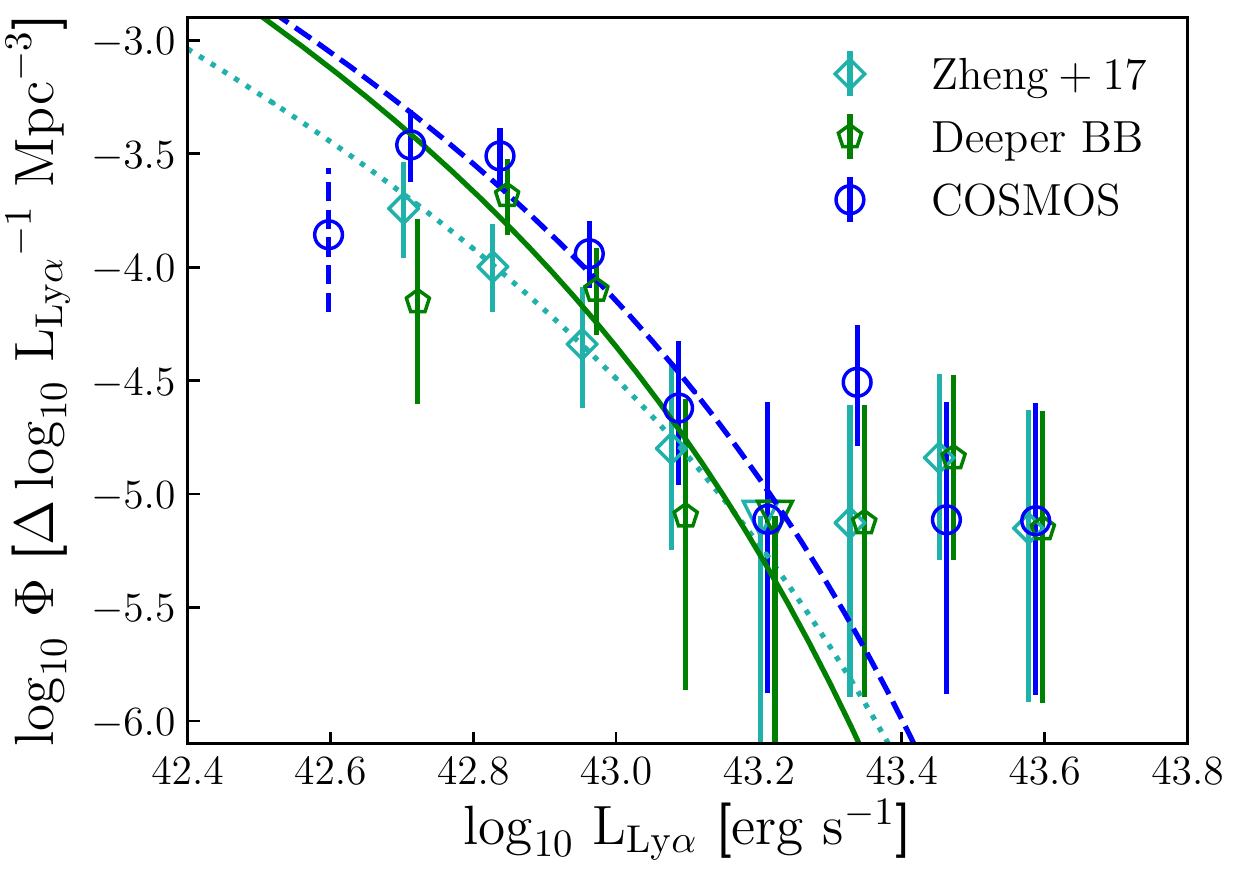}
   \caption{\label{fig:bbd} { Upper panel: The \lya\ line luminosities for the 23 LAEs selected by \cite{Zheng2017}, recalculated with the much deeper underlying HSC-$y$ photometry in this work (x-axis),
   versus those calculated with the shallower UltraVISTA Y band photometry (y-axis). The triangles plot the sources non-detected in the underlying broadband in Z17.}
   Lower Panel: Comparing the COSMOS field \lya\ LF obtained in this work (blue circles) with that from \citet[][cyan diamonds]{Zheng2017}.
   The green pentagons plot the recalculated LF based on the LAE sample of \citet{Zheng2017} but with much deeper underlying broadband photometry (HSC-$y$ band).
   To be consistent with \citet{Zheng2017}, we fit the Schechter function to the data points within $L$ range $42.65\leqslant \log_{10}L \leqslant 43.25$ in this figure.
   }
\end{figure}

In Fig. \ref{fig:bbd} we compare the Ly$\alpha$ luminosity function in COSMOS obtained in this work with that reported in \citet{Zheng2017}.
To enable a direct comparison, the selection incompleteness described in \S\ref{ssc:compp} is ignored in this plot, as it was uncorrected in \citet{Zheng2017}.
While at the highest luminosity bins both luminosity functions appear consistent, the new luminosity function obtained in this work is considerably higher
at fainter luminosity bins. {However, this is} only partly due to the fact that we select more candidates in this work.  

Another and dominant reason is the depth of the underlying broadband image.
As described in Section \ref{ss:cs}, for a LAE candidate which is not detected in the underlying broadband, a widely used approach is to place a 2$\sigma$ upper limit
to its broadband flux density and use such upper limit and narrowband flux density to estimate the \lya\ flux. As we demonstrate below, this step would yield significantly
under-estimated \lya\ flux and bias the LF if the underlying broadband is not sufficiently deep.
\citet{Zheng2017} adopted NB964 and the UltraVISTA $Y$ band image to calculate the \lya\ fluxes. For most of the faint candidates, UltraVISTA Y band detections are not available and 2$\sigma$ upper limits were given to their BB fluxes. In this work, the underlying broadband in COSMOS is HSC-$y$, which is considerably deeper than UltraVISTA $Y$  (by $\sim$ 1.1 -- 2.5 mag, as the depth of UltraVISTA $Y$ is not uniform). 
To simply illustrate such effect, we re-calculate the \lya\ line luminosity and the luminosity function of 23 LAE candidates selected by \citet{Zheng2017} with old NB964 photometry but new HSC-$y$ photometry. 
For sources  which are not detected in HSC-$y$, we adopt the 2$\sigma$ limiting magnitude of HSC-$y$.
{ The comparison of the resulted \lya\ luminosities is given in the upper panel of Fig. \ref{fig:bbd}, where we see that for
a significant fraction of faint LAEs, the \lya\ luminosities had been under-estimated with shallower underlying broadband photometry.}
As shown in the Fig. \ref{fig:bbd}, the deeper underlying broadband could significantly elevate the faint end LF to a level more consistent with this work. 
The bright end LF does not change much because most of the luminous LAEs were already detected in UltraVISTA $Y$. 

With simulations, \citet{Zheng2014} showed that the depth of the underlying broadband could significantly affect the LAE selection, and {a} broadband image $\sim$ 0.5 -- 1.0 mag deeper than the narrowband is most efficient in selecting emission line sources.
In this work, we show an additional effect of the underlying broadband depth, which would affect the calculation of line flux and thus LF. 
This effect could be particularly significant for faint LAEs with large line equivalent width, for which we expect very weak underlying broadband signal,
{and whose line flux measurements would be obviously biased by an upper limit from the BB if it is not sufficiently deep.}
To estimate the required BB depth which would eliminate {this} bias, 
we assume an extreme case with \lya\ line only. The line signal is detected in the narrowband with S/N $>$ 5, and we expect {to detect it} in the BB with S/N $>$ 2. 
The underlying broadband is thus required to be $2.5\log (W_{BB} / W_{NB}) - 2.5\log(5/2)$ deeper than the inside narrowband, where $W_{NB}$ and $W_{BB}$ are the width of the narrow- and  broadband filters. Practically, LAEs have finite line EW, thus we expect continuum signal in both narrowband and the underlying broadband, and BB $\sim$ 1.0 -- 1.5 mag deeper than the NB would be sufficient, depending on the bandpass of the filters.\footnote {Whether the bandpass of the BB and NB overlap also matters. For instance, the bandpass of UltraVISTA Y in fact does not overlap with that of NB964, thus in UltraVISTA Y we only expect
UV continuum signal but not the \lya\ line. In this sense, an overlapping BB (like HSC $y$) is preferred.}
 The underlying broadbands adopted in this work are indeed sufficiently deep, and the luminosity functions we obtained are free from significant bias.

\section{Discussion} \label{sc:ds}
\subsection{The Bright End Bump in the \lya\ Luminosity Function at $z\sim7.0$ \label{ssc:beb}}
\citet{Zheng2017} firstly detected a bright end bump in the LF of $z\sim7.0$ LAEs in COSMOS field with four luminous LAEs (L$_{Ly\alpha}$ $>10^{43.3}$ erg s$^{-1}$).
This suggests the existence of ionized bubbles at $z\sim7$ which {reduce} the opacity of neutral IGM around the luminous LAEs \citep[see \S 4.3 in][]{Zheng2017}.
Such a bright end LF bump is confirmed in this work, with six luminous LAEs (L$_{Ly\alpha}$ $>$ 10$^{43.3}$ erg s$^{-1}$) selected in COSMOS field. 
One of the newly selected luminous LAE did not pass the selection in \cite{Zheng2017} because it is close to a nearby bright source and its 2\arcsec\ aperture veto band photometry is 
significantly contaminated. It is selected in this work as we adopt a 1.35\arcsec\ aperture in our veto band photometry. 
Another newly selected luminous LAE was classified as a possible foreground source in \citet{Zheng2017}  due to visually identified marginal signal in one of the previously adopted veto bands.
With the deeper HSC ultra deep images used in this work (also with better seeing), we did not 
reveal any signal in any of the new veto bands. The marginal signal in the old veto band is indeed due to data processing flaw, and {was confirmed to be} artificial with improved reprocessing
of the old veto band images. 

Strikingly, while the faint end LF from a second LAGER field CDFS is {quite} consistent with that from COSMOS, 
the bright end LF bump is not seen in CDFS, in which only one luminous LAE is selected.  
Such field-to-field variation in the bright end LF is also visible {when} comparing with other $z\sim7.0$ LAE samples in literature. 
\citet{Ota2017} identified 20 $z\sim7.0$ LAE candidates using Subaru Suprime-Cam and NB973$_{SSC}$ filter. The sample was selected in 
a smaller volume (0.61$\times$10$^6$ Mpc$^3$) with no LAEs with L$_{Ly\alpha}$ $>$ 10$^{43.3}$ erg s$^{-1}$. 
\citet{Itoh2018} identified 34 $z\sim7.0$ LAE candidates using Subaru HSC  
and NB973$_{HSC}$ filter in two fields (COSMOS: 1.15$\times$10$^6$ Mpc$^3$, SXDS: 1.04$\times$10$^6$ Mpc$^3$).
While \citet{Itoh2018} claimed no evidence of bright end LF bump, they did identify 4 luminous LAEs in COSMOS, but zero in SXDS. Such field-to-field variation is similar to what we see in {the} two LAGER fields. 
Additional reasons that \citet{Itoh2018} {did not detect} the bright end LF bump include {that}: 1) \citet{Itoh2018} adopted larger luminosity bins in their LFs \citep[0.2 dex comparing with 0.125 dex adopted in this work 
and][]{Zheng2017};\footnote{{ We examine the effect of luminosity bin using our own dataset, and confirm that adopting 0.2 dex luminosity bin could weaken the bright end LF bump we seen with 0.125 dex bin in COSMOS.}}
2) the NB973$_{HSC}$ filter that \citet{Itoh2018} used has Gaussian-like transmission curve with clear wings \citep{Itoh2018}, while the transmission curve of our NB964 filter is more {box-car shaped} (see \citealt{Zheng2019} and Fig. \ref{fig:transmission}). 
A Gaussian-like transmission curve would yield large uncertainties in the \lya\ luminosity derived from narrowband photometry, and would significantly underestimate the luminosity and number of LAEs whose \lya\ lines fall on the wings of the bandpass. 
Such large uncertainties could likely smear out the bump feature in the LF. 

As the bandpass of NB973$_{HSC}$ and our NB964 overlap \citep{Zheng2019}, 
we compare our LAEs with that of \citet{Itoh2018} in COSMOS and find 7 common LAEs selected by both programs. 
Particularly 3 out of the 4 luminous LAEs selected by \citet{Itoh2018} are included in our sample. 
Two of them (HSC-z7LAE3 and HSC-z7LAE25) were classified as luminous LAEs by \citet{Itoh2018}, but only after they recalibrated their \lya\ luminosities using  our spectroscopic redshifts \citep{Hu2017}, i.e., they fall on the NB973$_{HSC}$ transmission curve wing.
We select another (HSC-z7LAE2) that coincides with the core of one LAE-overdense region in Fig. \ref{fig:2d}, but its NB964-based \lya\ luminosity (10$^{42.7}$ erg s$^{-1}$) is considerably lower than 10$^{43.4}$ from \citet{Itoh2018}.
The last luminous LAE (HSC-z7LAE1) selected by \cite{Itoh2018} is also significantly detected in our NB964 image. 
This source, however, did not pass our selection due to foreground contamination in the veto bands\footnote{{Furthermore, we also detected} HSC-z7LAE7 of \cite{Itoh2018} {in} our NB964 image but {it} did not pass our selection due to contamination by adjacent sources.
Its NB964-derived \lya\ luminosity is 10$^{43.02}$ erg s$^{-1}$, similar to 10$^{43.18}$ from \cite{Itoh2018}.}.
After excluding the contamination, the NB964-derived \lya\ luminosity of HSC-z7LAE1 is 10$^{43.0}$ erg s$^{-1}$, also considerably lower than 10$^{43.5}$ from \cite{Itoh2018}.
Both HSC-z7LAE1 and HSC-z7LAE2 
might fall on the transmission curve wing of our NB964 filter, i.e., have underestimated NB964-based \lya\ luminosity. 

If HSC-z7LAE1 and HSC-z7LAE2 are included as luminous LAEs in our sample, the number 
of luminous LAEs selected in LAGER COSMOS rises to 8, further strengthening the robustness of the bright end LF bump and the field-to-field variation. 
We also stress that the three luminous LAEs in COSMOS field and the single luminous LAE in CDFS have been spectroscopically confirmed \citep[][]{Hu2017, Yang2019}.
Meanwhile, as shown in Fig. \ref{fig:evl_lf}, our new \lya\ LF at $z\sim7.0$, averaged over two LAGER fields, is well consistent from those from \citet{Ota2017} and \citet{Itoh2018}.

All 6 luminous LAEs are located within the two large scale structures (Fig. \ref{fig:2d}). 
This indicates that large ionized bubbles at $z\sim$ 7.0 are closely associated with cosmic overdensities. 
{ Note that two $z\sim7$ LAEs with projected distance of $\sim90$ pkpc are confirmed by \citet{Castellano2018}, which are also selected in an overdense region identified with several Lyman Break Galaxy (LBG) candidates \citep{Castellano2016}.}
{These provide direct observational supports} to the inside-out reionization topology \citep[e.g.][]{Iliev2006, Choudhury2009,Friedrich2011}.
Further clustering analysis and follow-up observations of the overdense regions in COSMOS are essential to study the patchy reionization.
The clear field-to-field variation of the bright end LF manifests the need of LAE searches in even more fields to probe the reionization and large scale structures in the early universe.

\subsection{Evolution of \lya\ Luminosity Function and Constraint to Neutral Hydrogen Fraction}

\begin{figure*}
\centering
\includegraphics[width=6.7in]{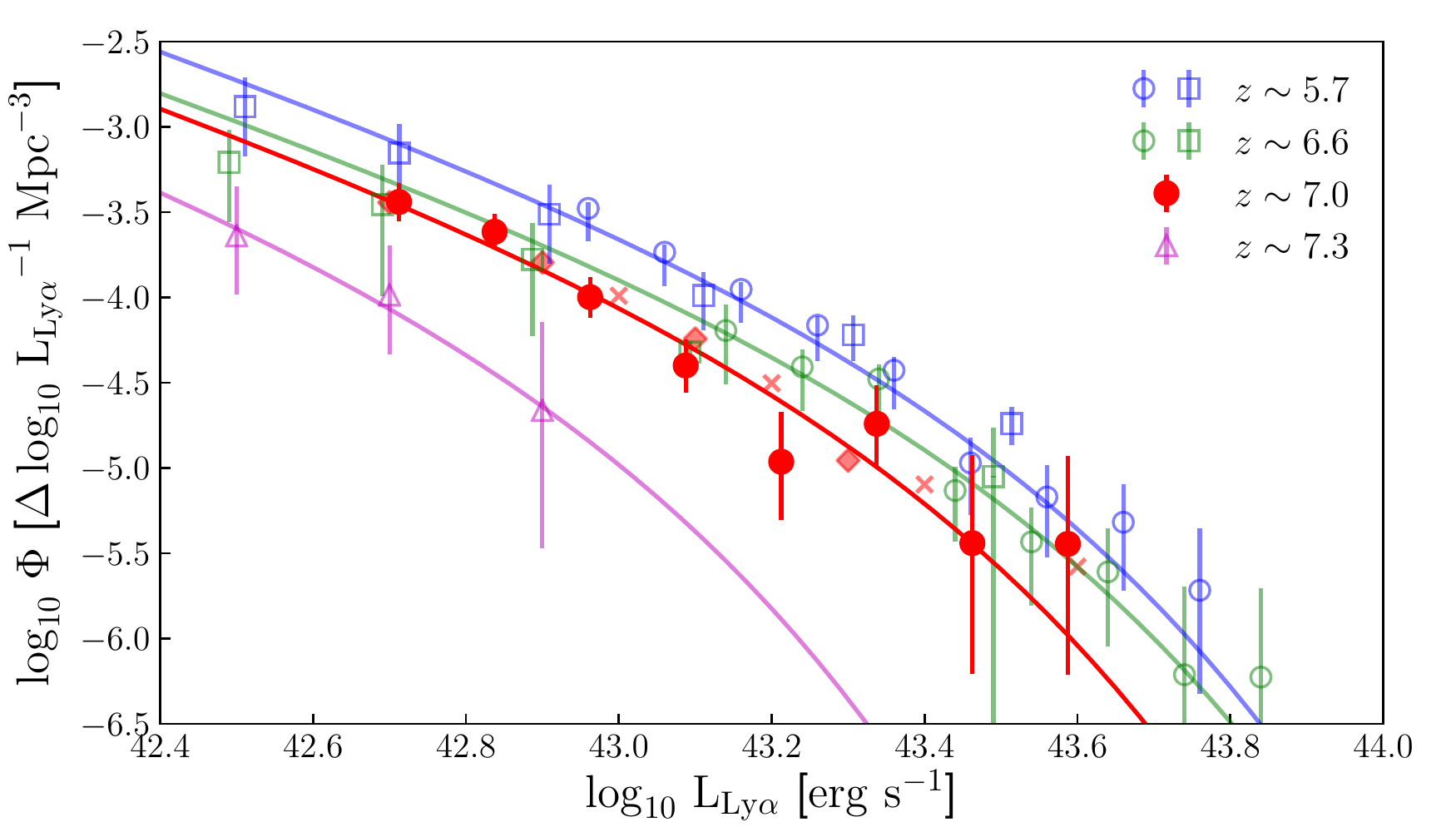}
\caption{\label{fig:evl_lf} Evolution of Ly$\alpha$ LFs from $z\sim 5.7$ to $7.3$. The red filled circles are the LF of our $z\sim7.0$ LAEs and the red solid line is best-fit Schechter function. 
The $z\sim7.0$ LFs from \cite{Itoh2018} and \cite{Ota2017} are plotted with red crosses and diamonds respectively, which {agrees well} with ours.   
The blue squares and circles are the Ly$\alpha$ LFs at $z\sim5.7$ from \citet{Ouchi2008} and \citet{Konno2018}, respectively. The green opened squares and circles are the Ly$\alpha$ LFs at $z\sim6.6$ from \citet{Ouchi2010} and \citet{Konno2018}, respectively. The purple opened triangle are the Ly$\alpha$ LF at $z\sim7.3$ from \citet{Konno2014}. The blue, green and purple solid lines are the corresponding best-fit Schechter function from $z\sim5.7$, $6.6$, $7.3$ LFs. }
\end{figure*}

\begin{figure}
\centering
\includegraphics[width=3.5in]{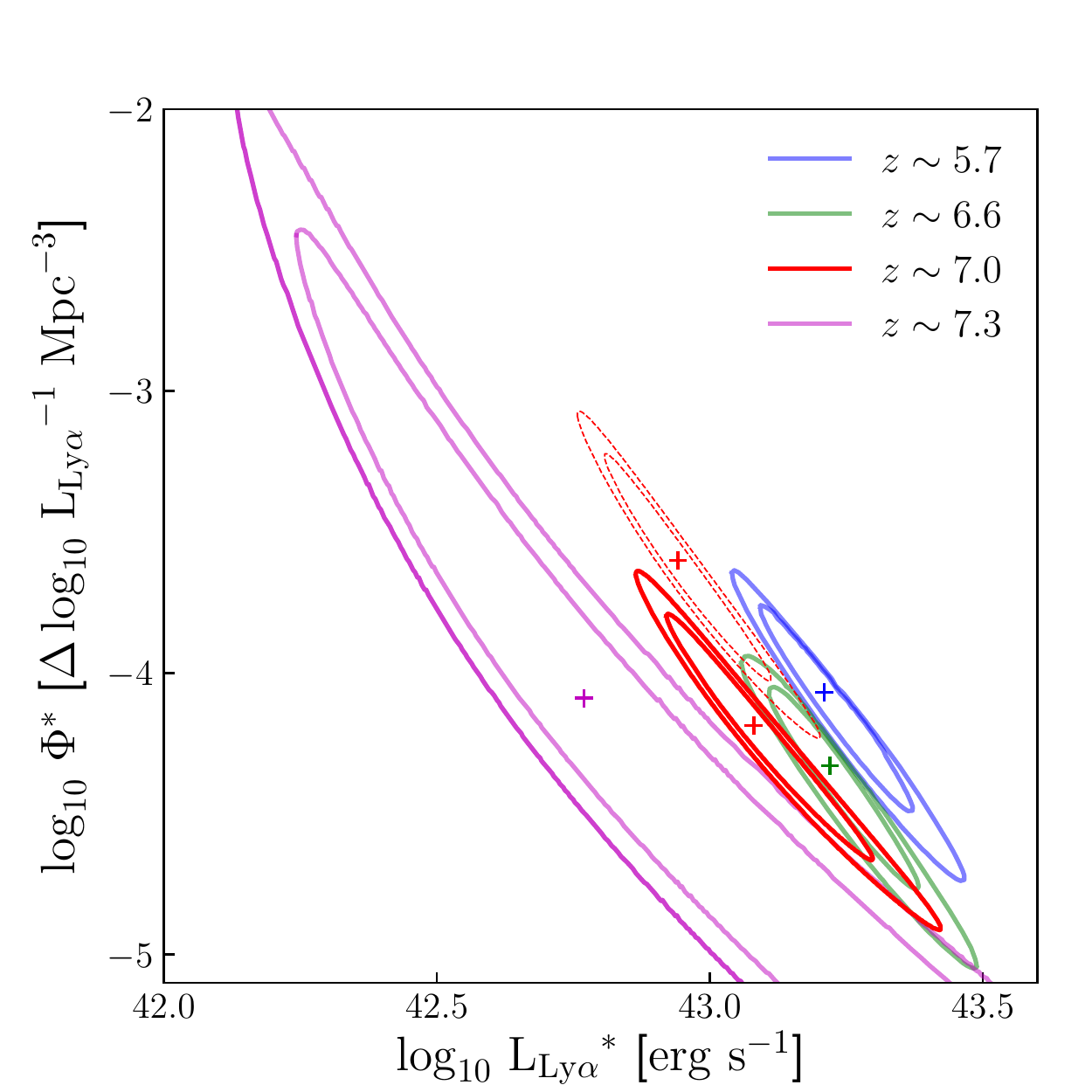}
\caption{\label{fig:lf_conf} $68\%$ and $90\%$ Confidence intervals of the best-fit Schechter parameters L$^\ast$ and $\Phi^\ast$ for our $z\sim7.0$ LF from two LAGER fields (solid red: {uncorrected} ``selection incompleteness").  The ``selection incompleteness" corrected version is plotted with dashed red lines to illustrate the effect of such incompleteness.  We also plot the confidence interval of best-fit Schechter parameters of LFs ({all with uncorrected} ``selection incompleteness") at $z\sim5.7$, $6.6$ \citep{Konno2018}, and $7.3$ \citep{Konno2014} }
\end{figure}

In Fig. \ref{fig:evl_lf}, we plot our \lya\ LF (averaged over two LAGER fields) at $z\sim$ 7.0 together with those at $z\sim5.7$ to $7.3$ \citep{Ouchi2008, Ouchi2010, Konno2014, Konno2018}.
We stress that the ``selection incompleteness" described in \S\ref{ssc:compp} was not corrected in this plot, as such incompleteness was not available for LFs given in literature. 
We assume those LAE samples at redshift 5.7 -- 7.3 suffer from similar ``selection incompleteness" when we compare the uncorrected LFs to demonstrate cosmic evolution in the luminosity function.
A gradual evolution between redshift of 5.7 and 7.3 is clearly seen in Fig. \ref{fig:evl_lf}. Note the LF at $z\sim$ 7.3 is based on a rather small photometric sample \citep[7 LAEs,][]{Konno2014}, thus the error bars  are considerably larger. 
To further quantify the evolution of LFs from $z\sim5.7$ to $7.3$, we plot the contours of best-fit Schechter function parameters (L$^*$ and $\phi^*$) for LFs at $z\sim5.7$, $6.6$, $7.0$, and $7.3$ in Fig. \ref{fig:lf_conf}. 
In this plot, for our LF at $z\sim$ 7.0, we plot both ``selection incompleteness" corrected (dashed red) and uncorrected (solid red) results to illustrate the effect of such incompleteness correction.

\begin{figure}
\centering
\includegraphics[width=3.5in]{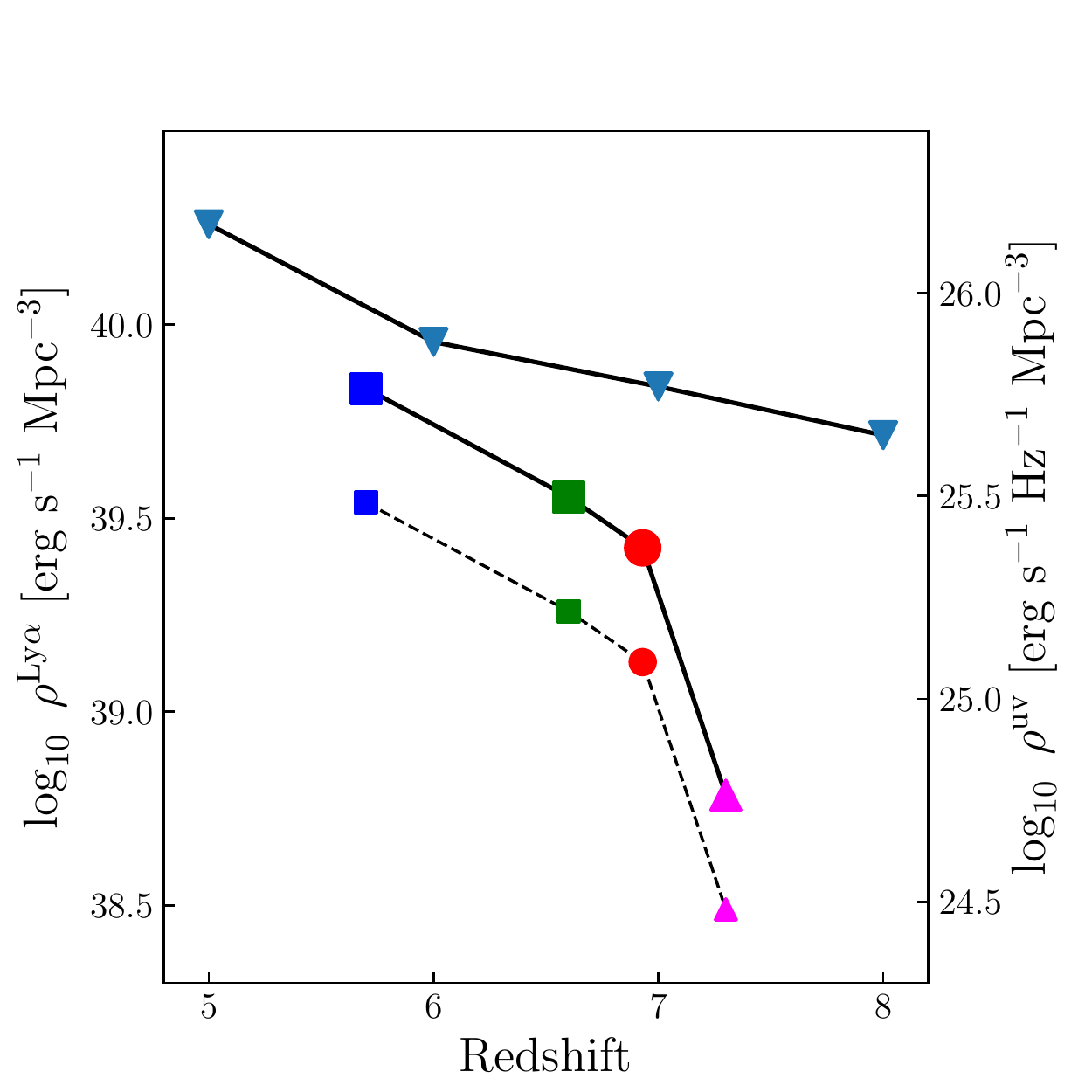}
\caption{\label{fig:evol_ld} Evolution of Ly$\alpha$ luminosity density (dashed line, uncorrected for sample selection incompleteness) from $z\sim5.7$ to $7.3$. The Ly$\alpha$ luminosity density at $z\sim5.7$ and $6.6$ are from \citet{Konno2018} and the Ly$\alpha$ luminosity density at $z\sim7.3$ are from \citet{Konno2014}. We also plot the evolution of UV luminosity density (inverse triangles) derived by \citet{Finkelstein2015}. 
The larger symbols and the solid line for Ly$\alpha$ luminosity density are corrected for LAE sample ``selection incompleteness", assuming a common correction factor of 0.3 dex as we derived for our $z\sim$ 7.0 LAE sample. }
\end{figure}

The luminosity density of \lya\ photons is derived by integrating our \lya\ LF in the luminosity range of $\log \mathrm{L_{Ly\alpha}}$ [erg s$^{-1}$] $= 42.4-44$.
The evolution of \lya\ luminosity density from $z\sim5.7$ to $7.3$ are plotted in the Fig. \ref{fig:evol_ld}. 
We also plot the UV luminosity density from \citet{Finkelstein2015}, 
based on the galaxy LFs from $z\sim4$ -- $8$ using galaxies selected by photometric redshift with \textit{Hubble Space Telescope} (HST) imaging data. 

Below we estimate the effective IGM transmission factor $T^{IGM}_z$  and neutral hydrogen fraction $\chi_{HI}$ 
following \citet{Ouchi2010}.
The observed \lya\ luminosity density can be simply converted from UV luminosity density:
\begin{equation}
\rho^{Ly\alpha}=\kappa T^{IGM} f_{esc} \rho^{UV},
\end{equation}
where $\kappa$ is the conversion factor from UV photons to \lya\ photons; $f_{esc}$ the \lya\ escape fraction through the ISM \citep{Dijkstra2007b,Dijkstra2010,Cai2014,Dayal2018}; and $T^{IGM}$ the transmission of the IGM.
Assuming the properties of ISM {and} stellar population are the same at $z=5.7$ and $7.0$, the IGM transmission at $z \sim 7.0$ can be calculated: 
\begin{equation}
\label{eq:trans}
\frac {T^{IGM}_{7.0}}{T^{IGM}_{5.7}} = \frac {\rho^{Ly\alpha}_{7.0}/\rho^{Ly\alpha}_{5.7}}{\rho^{UV}_{7.0}/\rho^{UV}_{5.7}}.
\end{equation}
We linearly interpolate the UV luminosity density in Fig. \ref{fig:evol_ld} \citep{Finkelstein2015} and estimate ${\rho^{UV}_{7.0}/\rho^{UV}_{5.7}} = 0.63\pm 0.09.$\footnote{
{As} \citet{Finkelstein2015} {did}, \citet{Bouwens2015} select thousands of galaxies from $z\sim4$ to $10$ with absolute magnitude down to $M_{\mathrm{UV}}=-17$ with HST data.
Using the results from \citet{Bouwens2015}, we obtain a rather similar ${\rho^{UV}_{7.0}/\rho^{UV}_{5.7}} = 0.61\pm 0.07$.}
We then obtain $T^{IGM}_{7.0}/T^{IGM}_{5.7} = 0.63\pm0.12$ with Eq. \ref{eq:trans},  which indicates statistically significant evolution in {the} IGM suppression to {the} \lya\ line.

It is model dependent to estimate the neutral hydrogen fraction $\chi_{\mathrm{HI}}$ based on the evolution of $T^{IGM}$. 
Below we present $\chi_{HI}$ inferred with several theoretical models.

With an analytical approach, \citet{Santos2004} calculated the \lya\ emission transmission through IGM as a function of neutral hydrogen fraction $\chi_{HI}$ in the early universe, considering the effects of IGM dynamics and galactic winds.
Though the \lya\ transmission through the IGM is highly sensitive to the \lya\ line velocity offset, we can estimate the IGM neutral hydrogen fraction by comparing the observed $T^{IGM}_{7.0}/T^{IGM}_{5.7}$ with Fig. 25 of \citet{Santos2004} while assuming no evolution in the  \lya\ line velocity offset between redshift 5.7 and 7.0. 
By doing so, we estimate a neutral hydrogen fraction of 0.25 -- 0.50 for a galactic wind model with \lya\ velocity offset of 360 km/s,
and 0.30 -- 0.50 for the case of no \lya\ velocity shift.

Secondly, the observed evolution of \lya\ LF can also be compared with radiative transfer simulations to constrain the reionization. 
\citet{McQuinn2007} calculated the effect of reionization on \lya\ LF at $z=6.6$ with 200-Mpc radiative transfer simulations. 
The expected suppression to \lya\ LF is given in Fig. 5 of \citet{McQuinn2007} for various global IGM neutral hydrogen fraction.
Comparing our observations with the simulation results, we obtain similar constraints to $\chi_{\mathrm{HI}}$ (0.2 -- 0.4).

We also compare our observations with the analytical calculations of \citet{Dijkstra2007a} and \citet{Furlanetto2006}, which assume complete neutral IGM outside of ionized bubbles.
Consistent with previous estimations, our observed $T^{IGM}_{7.0}/T^{IGM}_{5.7}$  corresponds to a globally averaged $0.2<\chi_{\mathrm{HI}}<0.5$.

\section{Conclusion}

Narrowband imaging surveys are powerful approaches to search for high redshift Ly$\alpha$ emitting galaxies and probe the cosmic reionization. 
We deploy a large area survey for $z\sim$ 7.0 LAEs (Lyman Alpha Galaxies in the Epoch of Reionization, abbr. LAGER) with a custom made 
narrowband filter installed on DECam onboard CTIO 4m Blanco telescope. 
In this paper, we present LAEs selected in the ultradeep LAGER-COSMOS field and a second deep field LAGER-CDFS. 
We present the Ly$\alpha$ luminosity function at $z\sim$ 7.0 and new knowledge inferred {about} cosmic reionization. 
Our major results are listed below:

\begin{enumerate}

\item 
We accumulate 47.25 hrs DECam NB964 exposure in COSMOS and 32.9 hrs in CDFS field.
We select 49 $z\sim$ 7.0 LAEs in COSMOS 
and 30 in CDFS, 
building a largest ever LAE sample at $z\sim$ 7.0.  
 
\item 
We find obvious LAE sample incompleteness due to 
foreground contamination {in} bluer veto broadband photometry.  
Such selection incompleteness (30\% -- 40\% in this work), depending on the confusion level of the broadband images (seeing and depth), could cause underestimation of the luminosity function of high redshift galaxies, and thus should be carefully corrected.

\item We show that while calculating the Ly$\alpha$ luminosity based on narrow- and underlying broadband photometry, placing an upper limit to the broadband flux for non-detection might significantly bias the calculation of \lya\ flux and luminosity function, if the broadband image is not sufficiently deep. We recommend the underlying BB {be} $\sim$ 1.0 -- 1.5 mag deeper than the NB to avoid such bias. 
 
\item Six luminous LAEs with L$_{Ly\alpha}$ $>$ 10$^{43.3}$ erg s$^{-1}$ constitute a bright end bump in the luminosity function in COSMOS, supporting the patch reionization scenario. 
The bump is however not seen in CDFS in which only one luminous LAE is selected. Except for the bright end bump, the luminosity functions from two fields agree with each other,
and with those at $z\sim$ 7.0 in literature.

\item Two clear LAE overdense regions are detected in COSMOS, making them the highest redshift protoclusters observed to date. 
All six luminous LAEs in COSMOS fall in the overdense regions, further supporting the inside-out reionization topology. 

\item We compare the LAGER LAE luminosity function at $z\sim$ 7.0 with those at $z\sim$ 5.7, 6.6, and 7.3 reported in literature,  assuming  they suffer similar ``selection incompleteness".
We infer an average neutral hydrogen fraction of $\chi_{\mathrm{H I}} = 0.2 - 0.4$ at $z\sim$ 7.0.

\end{enumerate}

\acknowledgments
{We thank the anonymous referee for the valuable comments and Zhen-Yi Cai and Edmund Christian Herenz for the informative discussions. }
This work is supported by National Science Foundation of China (grants No. 11421303 $\&$ 11890693)
and CAS Frontier Science Key Research Program (QYZDJ-SSW-SLH006).
Z.Y.Z. is sponsored by Shanghai Pujiang Program, the National Science Foundation of China (11773051), the China-Chile Joint Research Fund (CCJRF No. 1503) and the CAS Pioneer Hundred Talents Program.
The work of the US coauthors on this project has been supported in part by the US National Science Foundation through grant NSF AST-1518057, and by NASA through the WFIRST Science Investigation Team program, contract number NNG16PJ33C.
L.F.B was partially supported by Anillo ACT-1417 and by CONICYT Project BASAL AFB-170002

This project used the data obtained with the Dark Energy Camera (DECam), which was constructed by the Dark Energy Survey (DES), 
and public archival data from the Dark Energy Survey (DES). 

Based on observations at Cerro Tololo Inter-American Observatory, National Optical
Astronomy Observatory (NOAO Prop. ID: 2016A-0386, 2017A-0366, 2017B-0330; PI: Malhotra S.; CNTAC Prop. ID: 2015B-0603, 2016A-0610, 2016B-0924, 2017A-0920 ; PI: Infante L.), which is operated by the Association of
Universities for Research in Astronomy (AURA) under a cooperative agreement with the
National Science Foundation.

This paper makes use of software developed for the Large Synoptic Survey Telescope. We thank the LSST Project for making their code available as free software at  http://dm.lsst.org

Based [in part] on data collected at the Subaru Telescope and retrieved from the HSC data archive system, which is operated by Subaru Telescope and Astronomy Data Center at National Astronomical Observatory of Japan.

{\it Facilities:}\facility{$Blanco$ (DECam)}

\bibliography{ms.bbl}

\begin{thebibliography}{}
\providecommand\natexlab[1]{#1}
\providecommand\JournalTitle[1]{#1}

\bibitem[{{Abbott} {et~al.}(2018){Abbott}, {Abdalla}, {Allam}, {Amara},
  {Annis}, {Asorey}, {Avila}, \& {aaa}}]{Abbott2018}
{Abbott}, T.~M.~C., {Abdalla}, F.~B., {Allam}, S., {et~al.} 2018,
  \href{http://dx.doi.org/10.3847/1538-4365/aae9f0}{\JournalTitle{\apjs}, 239,
  18}

\bibitem[{{Aihara} {et~al.}(2018){Aihara}, {Armstrong}, {Bickerton}, {Bosch},
  {Coupon}, {Furusawa}, {Hayashi}, {Ikeda}, {Kamata}, {Karoji}, {Kawanomoto},
  {Koike}, {Komiyama}, {Lang}, {Lupton}, {Mineo}, {Miyatake}, {Miyazaki},
  {Morokuma}, {Obuchi}, {Oishi}, {Okura}, {Price}, {Takata}, {Tanaka},
  {Tanaka}, {Tanaka}, {Uchida}, {Uraguchi}, {Utsumi}, {Wang}, {Yamada},
  {Yamanoi}, {Yasuda}, {Arimoto}, {Chiba}, {Finet}, {Fujimori}, {Fujimoto},
  {Furusawa}, {Goto}, {Goulding}, {Gunn}, {Harikane}, {Hattori}, {Hayashi},
  {He{\l}miniak}, {Higuchi}, {Hikage}, {Ho}, {Hsieh}, {Huang}, {Huang},
  {Imanishi}, {Iwata}, {Jaelani}, {Jian}, {Kashikawa}, {Katayama}, {Kojima},
  {Konno}, {Koshida}, {Kusakabe}, {Leauthaud}, {Lee}, {Lin}, {Lin},
  {Mandelbaum}, {Matsuoka}, {Medezinski}, {Miyama}, {Momose}, {More}, {More},
  {Mukae}, {Murata}, {Murayama}, {Nagao}, {Nakata}, {Niida}, {Niikura},
  {Nishizawa}, {Oguri}, {Okabe}, {Ono}, {Onodera}, {Onoue}, {Ouchi}, {Pyo},
  {Shibuya}, {Shimasaku}, {Simet}, {Speagle}, {Spergel}, {Strauss}, {Sugahara},
  {Sugiyama}, {Suto}, {Suzuki}, {Tait}, {Takada}, {Terai}, {Toba}, {Turner},
  {Uchiyama}, {Umetsu}, {Urata}, {Usuda}, {Yeh}, \& {Yuma}}]{Aihara2018}
{Aihara}, H., {Armstrong}, R., {Bickerton}, S., {et~al.} 2018,
  \href{http://dx.doi.org/10.1093/pasj/psx081}{\JournalTitle{\pasj}, 70, S8}

\bibitem[{{Allen} {et~al.}(2017){Allen}, {Kacprzak}, {Glazebrook}, {Labb{\'e}},
  {Tran}, {Spitler}, {Cowley}, {Nanayakkara}, {Papovich}, {Quadri},
  {Straatman}, {Tilvi}, \& {van Dokkum}}]{Allen2017}
{Allen}, R.~J., {Kacprzak}, G.~G., {Glazebrook}, K., {et~al.} 2017,
  \href{http://dx.doi.org/10.3847/2041-8213/834/2/L11}{\JournalTitle{\apjl},
  834, L11}

\bibitem[{{Annis} {et~al.}(2014){Annis}, {Soares-Santos}, {Strauss}, {Becker},
  {Dodelson}, {Fan}, {Gunn}, {Hao}, {Ivezi{\'c}}, {Jester}, {Jiang},
  {Johnston}, {Kubo}, {Lampeitl}, {Lin}, {Lupton}, {Miknaitis}, {Seo}, {Simet},
  \& {Yanny}}]{Annis2014}
{Annis}, J., {Soares-Santos}, M., {Strauss}, M.~A., {et~al.} 2014,
  \href{http://dx.doi.org/10.1088/0004-637X/794/2/120}{\JournalTitle{\apj},
  794, 120}

\bibitem[{{Ba{\~n}ados} {et~al.}(2018){Ba{\~n}ados}, {Venemans},
  {Mazzucchelli}, {Farina}, {Walter}, {Wang}, {Decarli}, {Stern}, {Fan},
  {Davies}, {Hennawi}, {Simcoe}, {Turner}, {Rix}, {Yang}, {Kelson}, {Rudie}, \&
  {Winters}}]{Banados2018}
{Ba{\~n}ados}, E., {Venemans}, B.~P., {Mazzucchelli}, C., {et~al.} 2018,
  \href{http://dx.doi.org/10.1038/nature25180}{\JournalTitle{\nat}, 553, 473}

\bibitem[{{Bertin}(2011)}]{Bertin2011}
{Bertin}, E. 2011, in Astronomical Society of the Pacific Conference Series,
  Vol. 442, Astronomical Data Analysis Software and Systems XX, ed. I.~N.
  {Evans}, A.~{Accomazzi}, D.~J. {Mink}, \& A.~H. {Rots}, 435

\bibitem[{{Bertin} \& {Arnouts}(1996)}]{Bertin1996}
{Bertin}, E., \& {Arnouts}, S. 1996,
  \href{http://dx.doi.org/10.1051/aas:1996164}{\JournalTitle{\aaps}, 117, 393}

\bibitem[{{Bertin} {et~al.}(2002){Bertin}, {Mellier}, {Radovich}, {Missonnier},
  {Didelon}, \& {Morin}}]{Bertin2002}
{Bertin}, E., {Mellier}, Y., {Radovich}, M., {et~al.} 2002, in Astronomical
  Society of the Pacific Conference Series, Vol. 281, Astronomical Data
  Analysis Software and Systems XI, ed. D.~A. {Bohlender}, D.~{Durand}, \&
  T.~H. {Handley}, 228

\bibitem[{{Bouwens} {et~al.}(2015){Bouwens}, {Illingworth}, {Oesch}, {Trenti},
  {Labb{\'e}}, {Bradley}, {Carollo}, {van Dokkum}, {Gonzalez}, {Holwerda},
  {Franx}, {Spitler}, {Smit}, \& {Magee}}]{Bouwens2015}
{Bouwens}, R.~J., {Illingworth}, G.~D., {Oesch}, P.~A., {et~al.} 2015,
  \href{http://dx.doi.org/10.1088/0004-637X/803/1/34}{\JournalTitle{\apj}, 803,
  34}

\bibitem[{{Cai} {et~al.}(2014){Cai}, {Lapi}, {Bressan}, {De Zotti}, {Negrello},
  \& {Danese}}]{Cai2014}
{Cai}, Z.-Y., {Lapi}, A., {Bressan}, A., {et~al.} 2014,
  \href{http://dx.doi.org/10.1088/0004-637X/785/1/65}{\JournalTitle{\apj}, 785,
  65}

\bibitem[{{Cash}(1979)}]{Cash1979}
{Cash}, W. 1979, \href{http://dx.doi.org/10.1086/156922}{\JournalTitle{\apj},
  228, 939}

\bibitem[{{Castellano} {et~al.}(2016){Castellano}, {Dayal}, {Pentericci},
  {Fontana}, {Hutter}, {Brammer}, {Merlin}, {Grazian}, {Pilo}, {Amorin},
  {Cristiani}, {Dickinson}, {Ferrara}, {Gallerani}, {Giallongo}, {Giavalisco},
  {Guaita}, {Koekemoer}, {Maiolino}, {Paris}, {Santini}, {Vallini}, {Vanzella},
  \& {Wagg}}]{Castellano2016}
{Castellano}, M., {Dayal}, P., {Pentericci}, L., {et~al.} 2016,
  \href{http://dx.doi.org/10.3847/2041-8205/818/1/L3}{\JournalTitle{\apjl},
  818, L3}

\bibitem[{{Castellano} {et~al.}(2018){Castellano}, {Pentericci}, {Vanzella},
  {Marchi}, {Fontana}, {Dayal}, {Ferrara}, {Hutter}, {Carniani}, {Cristiani},
  {Dickinson}, {Gallerani}, {Giallongo}, {Giavalisco}, {Grazian}, {Maiolino},
  {Merlin}, {Paris}, {Pilo}, \& {Santini}}]{Castellano2018}
{Castellano}, M., {Pentericci}, L., {Vanzella}, E., {et~al.} 2018,
  \href{http://dx.doi.org/10.3847/2041-8213/aad59b}{\JournalTitle{\apjl}, 863,
  L3}

\bibitem[{{Castelli} \& {Kurucz}(2004)}]{Castelli2004}
{Castelli}, F., \& {Kurucz}, R.~L. 2004, \JournalTitle{ArXiv Astrophysics
  e-prints}, \href{http://arxiv.org/abs/astro-ph/0405087}{{\sffamily
  astro-ph/0405087}}

\bibitem[{{Choudhury} {et~al.}(2009){Choudhury}, {Haehnelt}, \&
  {Regan}}]{Choudhury2009}
{Choudhury}, T.~R., {Haehnelt}, M.~G., \& {Regan}, J. 2009,
  \href{http://dx.doi.org/10.1111/j.1365-2966.2008.14383.x}{\JournalTitle{\mnras},
  394, 960}

\bibitem[{{Dayal} \& {Ferrara}(2018)}]{Dayal2018}
{Dayal}, P., \& {Ferrara}, A. 2018,
  \href{http://dx.doi.org/10.1016/j.physrep.2018.10.002}{\JournalTitle{\physrep},
  780, 1}

\bibitem[{{Diehl} {et~al.}(2008){Diehl}, {Angstadt}, {Campa}, {Cease},
  {Derylo}, {Emes}, {Estrada}, {Kubik}, {Flaugher}, \& {Holland }}]{Diehl2008}
{Diehl}, H.~T., {Angstadt}, R., {Campa}, J., {et~al.} 2008,
  \href{http://dx.doi.org/10.1117/12.790053}{in Society of Photo-Optical
  Instrumentation Engineers (SPIE) Conference Series, Vol. 7021, High Energy,
  Optical, and Infrared Detectors for Astronomy III}, 702107

\bibitem[{{Dijkstra}(2014)}]{Dijkstra2014}
{Dijkstra}, M. 2014,
  \href{http://dx.doi.org/10.1017/pasa.2014.33}{\JournalTitle{\pasa}, 31, e040}

\bibitem[{{Dijkstra} {et~al.}(2007{\natexlab{a}}){Dijkstra}, {Lidz}, \&
  {Wyithe}}]{Dijkstra2007b}
{Dijkstra}, M., {Lidz}, A., \& {Wyithe}, J. S.~B. 2007{\natexlab{a}},
  \href{http://dx.doi.org/10.1111/j.1365-2966.2007.11666.x}{\JournalTitle{\mnras},
  377, 1175}

\bibitem[{{Dijkstra} \& {Wyithe}(2010)}]{Dijkstra2010}
{Dijkstra}, M., \& {Wyithe}, J. S.~B. 2010,
  \href{http://dx.doi.org/10.1111/j.1365-2966.2010.17112.x}{\JournalTitle{\mnras},
  408, 352}

\bibitem[{{Dijkstra} {et~al.}(2007{\natexlab{b}}){Dijkstra}, {Wyithe}, \&
  {Haiman}}]{Dijkstra2007a}
{Dijkstra}, M., {Wyithe}, J.~S.~B., \& {Haiman}, Z. 2007{\natexlab{b}},
  \href{http://dx.doi.org/10.1111/j.1365-2966.2007.11936.x}{\JournalTitle{\mnras},
  379, 253}

\bibitem[{{Fan} {et~al.}(2006){Fan}, {Strauss}, {Becker}, {White}, {Gunn},
  {Knapp}, {Richards}, {Schneider}, {Brinkmann}, \& {Fukugita}}]{Fan2006}
{Fan}, X., {Strauss}, M.~A., {Becker}, R.~H., {et~al.} 2006,
  \href{http://dx.doi.org/10.1086/504836}{\JournalTitle{\aj}, 132, 117}

\bibitem[{{Finkelstein} {et~al.}(2011){Finkelstein}, {Cohen}, {Windhorst},
  {Ryan}, {Hathi}, {Finkelstein}, {Anderson}, {Grogin}, {Koekemoer},
  {Malhotra}, {Mutchler}, {Rhoads}, {McCarthy}, {O'Connell}, {Balick}, {Bond},
  {Calzetti}, {Disney}, {Dopita}, {Frogel}, {Hall}, {Holtzman}, {Kimble},
  {Luppino}, {Paresce}, {Saha}, {Silk}, {Trauger}, {Walker}, {Whitmore}, \&
  {Young}}]{Finkelstein2011}
{Finkelstein}, S.~L., {Cohen}, S.~H., {Windhorst}, R.~A., {et~al.} 2011,
  \href{http://dx.doi.org/10.1088/0004-637X/735/1/5}{\JournalTitle{\apj}, 735,
  5}

\bibitem[{{Finkelstein} {et~al.}(2015){Finkelstein}, {Ryan}, {Papovich},
  {Dickinson}, {Song}, {Somerville}, {Ferguson}, {Salmon}, {Giavalisco},
  {Koekemoer}, {Ashby}, {Behroozi}, {Castellano}, {Dunlop}, {Faber}, {Fazio},
  {Fontana}, {Grogin}, {Hathi}, {Jaacks}, {Kocevski}, {Livermore}, {McLure},
  {Merlin}, {Mobasher}, {Newman}, {Rafelski}, {Tilvi}, \&
  {Willner}}]{Finkelstein2015}
{Finkelstein}, S.~L., {Ryan}, Jr., R.~E., {Papovich}, C., {et~al.} 2015,
  \href{http://dx.doi.org/10.1088/0004-637X/810/1/71}{\JournalTitle{\apj}, 810,
  71}

\bibitem[{{Forrest} {et~al.}(2017){Forrest}, {Tran}, {Broussard}, {Allen},
  {Apfel}, {Cowley}, {Glazebrook}, {Kacprzak}, {Labb{\'e}}, {Nanayakkara},
  {Papovich}, {Quadri}, {Spitler}, {Straatman}, \& {Tomczak}}]{Forrest2017}
{Forrest}, B., {Tran}, K.-V.~H., {Broussard}, A., {et~al.} 2017,
  \href{http://dx.doi.org/10.3847/2041-8213/aa653b}{\JournalTitle{\apjl}, 838,
  L12}

\bibitem[{{Friedrich} {et~al.}(2011){Friedrich}, {Mellema}, {Alvarez},
  {Shapiro}, \& {Iliev}}]{Friedrich2011}
{Friedrich}, M.~M., {Mellema}, G., {Alvarez}, M.~A., {Shapiro}, P.~R., \&
  {Iliev}, I.~T. 2011,
  \href{http://dx.doi.org/10.1111/j.1365-2966.2011.18219.x}{\JournalTitle{\mnras},
  413, 1353}

\bibitem[{{Furlanetto} {et~al.}(2006){Furlanetto}, {Zaldarriaga}, \&
  {Hernquist}}]{Furlanetto2006}
{Furlanetto}, S.~R., {Zaldarriaga}, M., \& {Hernquist}, L. 2006,
  \href{http://dx.doi.org/10.1111/j.1365-2966.2005.09785.x}{\JournalTitle{\mnras},
  365, 1012}

\bibitem[{{Greiner} {et~al.}(2009){Greiner}, {Kr{\"u}hler}, {Fynbo}, {Rossi},
  {Schwarz}, {Klose}, {Savaglio}, {Tanvir}, {McBreen}, {Totani}, {Zhang}, {Wu},
  {Watson}, {Barthelmy}, {Beardmore}, {Ferrero}, {Gehrels}, {Kann}, {Kawai},
  {Yolda{\c s}}, {M{\'e}sz{\'a}ros}, {Milvang-Jensen}, {Oates}, {Pierini},
  {Schady}, {Toma}, {Vreeswijk}, {Yolda{\c s}}, {Zhang}, {Afonso}, {Aoki},
  {Burrows}, {Clemens}, {Filgas}, {Haiman}, {Hartmann}, {Hasinger}, {Hjorth},
  {Jehin}, {Levan}, {Liang}, {Malesani}, {Pyo}, {Schulze}, {Szokoly}, {Terada},
  \& {Wiersema}}]{Greiner2009}
{Greiner}, J., {Kr{\"u}hler}, T., {Fynbo}, J.~P.~U., {et~al.} 2009,
  \href{http://dx.doi.org/10.1088/0004-637X/693/2/1610}{\JournalTitle{\apj},
  693, 1610}

\bibitem[{{Harikane} {et~al.}(2019){Harikane}, {Ouchi}, {Ono}, {Fujimoto},
  {Donevski}, {Shibuya}, {Faisst}, {Goto}, {Hatsukade}, {Kashikawa}, {Kohno},
  {Hashimoto}, {Higuchi}, {Inoue}, {Lin}, {Martin}, {Overzier}, {Smail},
  {Toshikawa}, {Umehata}, {Ao}, {Chapman}, {Clements}, {Im}, {Jing},
  {Kawaguchi}, {Lee}, {Lee}, {Lin}, {Matsuoka}, {Marinello}, {Nagao},
  {Onodera}, {Toft}, \& {Wang}}]{Harikane2019}
{Harikane}, Y., {Ouchi}, M., {Ono}, Y., {et~al.} 2019,
  \href{http://dx.doi.org/10.3847/1538-4357/ab2cd5}{\JournalTitle{\apj}, 883,
  142}

\bibitem[{{Hayashi} {et~al.}(2018){Hayashi}, {Tanaka}, {Shimakawa}, {Furusawa},
  {Momose}, {Koyama}, {Silverman}, {Kodama}, {Komiyama}, {Leauthaud}, {Lin},
  {Miyazaki}, {Nagao}, {Nishizawa}, {Ouchi}, {Shibuya}, {Tadaki}, \&
  {Yabe}}]{Hayashi2018}
{Hayashi}, M., {Tanaka}, M., {Shimakawa}, R., {et~al.} 2018,
  \href{http://dx.doi.org/10.1093/pasj/psx088}{\JournalTitle{Publications of
  the Astronomical Society of Japan}, 70, S17}

\bibitem[{{Hibon} {et~al.}(2012){Hibon}, {Kashikawa}, {Willott}, {Iye}, \&
  {Shibuya}}]{Hibon2012}
{Hibon}, P., {Kashikawa}, N., {Willott}, C., {Iye}, M., \& {Shibuya}, T. 2012,
  \href{http://dx.doi.org/10.1088/0004-637X/744/2/89}{\JournalTitle{\apj}, 744,
  89}

\bibitem[{{Hibon} {et~al.}(2011){Hibon}, {Malhotra}, {Rhoads}, \&
  {Willott}}]{Hibon2011}
{Hibon}, P., {Malhotra}, S., {Rhoads}, J., \& {Willott}, C. 2011,
  \href{http://dx.doi.org/10.1088/0004-637X/741/2/101}{\JournalTitle{\apj},
  741, 101}

\bibitem[{{Hibon} {et~al.}(2010){Hibon}, {Cuby}, {Willis}, {Cl{\'e}ment},
  {Lidman}, {Arnouts}, {Kneib}, {Willott}, {Marmo}, \& {McCracken}}]{Hibon2010}
{Hibon}, P., {Cuby}, J.-G., {Willis}, J., {et~al.} 2010,
  \href{http://dx.doi.org/10.1051/0004-6361/200912109}{\JournalTitle{\aap},
  515, A97}

\bibitem[{{Hu} {et~al.}(2010){Hu}, {Cowie}, {Barger}, {Capak}, {Kakazu}, \&
  {Trouille}}]{Hu2010}
{Hu}, E.~M., {Cowie}, L.~L., {Barger}, A.~J., {et~al.} 2010,
  \href{http://dx.doi.org/10.1088/0004-637X/725/1/394}{\JournalTitle{\apj},
  725, 394}

\bibitem[{{Hu} {et~al.}(2017){Hu}, {Wang}, {Zheng}, {Malhotra}, {Infante},
  {Rhoads}, {Gonzalez}, {Walker}, {Jiang}, {Jiang}, {Hibon}, {Barrientos},
  {Finkelstein}, {Galaz}, {Kang}, {Kong}, {Tilvi}, {Yang}, \& {Zheng}}]{Hu2017}
{Hu}, W., {Wang}, J., {Zheng}, Z.-Y., {et~al.} 2017,
  \href{http://dx.doi.org/10.3847/2041-8213/aa8401}{\JournalTitle{\apjl}, 845,
  L16}

\bibitem[{{Huang} {et~al.}(2015){Huang}, {Zheng}, {Wang}, {Ford}, {Lemze},
  {Moustakas}, {Shu}, {Van der Wel}, {Zitrin}, {Frye}, {Postman}, {Bartelmann},
  {Ben{\'{\i}}tez}, {Bradley}, {Broadhurst}, {Coe}, {Donahue}, {Infante},
  {Kelson}, {Koekemoer}, {Lahav}, {Medezinski}, {Moustakas}, {Rosati}, {Seitz},
  \& {Umetsu}}]{Huang2015}
{Huang}, X., {Zheng}, W., {Wang}, J., {et~al.} 2015,
  \href{http://dx.doi.org/10.1088/0004-637X/801/1/12}{\JournalTitle{\apj}, 801,
  12}

\bibitem[{{Iliev} {et~al.}(2006){Iliev}, {Mellema}, {Pen}, {Merz}, {Shapiro},
  \& {Alvarez}}]{Iliev2006}
{Iliev}, I.~T., {Mellema}, G., {Pen}, U.-L., {et~al.} 2006,
  \href{http://dx.doi.org/10.1111/j.1365-2966.2006.10502.x}{\JournalTitle{\mnras},
  369, 1625}

\bibitem[{{Itoh} {et~al.}(2018){Itoh}, {Ouchi}, {Zhang}, {Inoue}, {Mawatari},
  {Shibuya}, {Harikane}, {Ono}, {Kusakabe}, {Shimasaku}, {Fujimoto}, {Iwata},
  {Kajisawa}, {Kashikawa}, {Kawanomoto}, {Komiyama}, {Lee}, {Nagao}, \&
  {Taniguchi}}]{Itoh2018}
{Itoh}, R., {Ouchi}, M., {Zhang}, H., {et~al.} 2018,
  \href{http://dx.doi.org/10.3847/1538-4357/aadfe4}{\JournalTitle{\apj}, 867,
  46}

\bibitem[{{Jiang} {et~al.}(2013){Jiang}, {Egami}, {Fan}, {Windhorst}, {Cohen},
  {Dav{\'e}}, {Finlator}, {Kashikawa}, {Mechtley}, \& {Ouchi}}]{Jiang2013}
{Jiang}, L., {Egami}, E., {Fan}, X., {et~al.} 2013,
  \href{http://dx.doi.org/10.1088/0004-637X/773/2/153}{\JournalTitle{\apj},
  773, 153}

\bibitem[{{Jiang} {et~al.}(2014){Jiang}, {Fan}, {Bian}, {McGreer}, {Strauss},
  {Annis}, {Buck}, {Green}, {Hodge}, {Myers}, {Rafiee}, \&
  {Richards}}]{Jiang2014}
{Jiang}, L., {Fan}, X., {Bian}, F., {et~al.} 2014,
  \href{http://dx.doi.org/10.1088/0067-0049/213/1/12}{\JournalTitle{\apjs},
  213, 12}

\bibitem[{{Jiang} {et~al.}(2017){Jiang}, {Shen}, {Bian}, {Zheng}, {Wu},
  {Oyarz{\'u}n}, {Blanc}, {Fan}, {Ho}, {Infante}, {Wang}, {Wu}, {Mateo},
  {Bailey}, {Crane}, {Olszewski}, {Shectman}, {Thompson}, \&
  {Walker}}]{Jiang2017}
{Jiang}, L., {Shen}, Y., {Bian}, F., {et~al.} 2017,
  \href{http://dx.doi.org/10.3847/1538-4357/aa8561}{\JournalTitle{\apj}, 846,
  134}

\bibitem[{{Jiang} {et~al.}(2018){Jiang}, {Wu}, {Bian}, {Chiang}, {Ho}, {Shen},
  {Zheng}, {Bailey}, {Blanc}, {Crane}, {Fan}, {Mateo}, {Olszewski},
  {Oyarz{\'u}n}, {Wang}, \& {Wu}}]{Jiang2018}
{Jiang}, L., {Wu}, J., {Bian}, F., {et~al.} 2018,
  \href{http://dx.doi.org/10.1038/s41550-018-0587-9}{\JournalTitle{Nature
  Astronomy}, 2, 962}

\bibitem[{{Konno} {et~al.}(2014){Konno}, {Ouchi}, {Ono}, {Shimasaku},
  {Shibuya}, {Furusawa}, {Nakajima}, {Naito}, {Momose}, {Yuma}, \&
  {Iye}}]{Konno2014}
{Konno}, A., {Ouchi}, M., {Ono}, Y., {et~al.} 2014,
  \href{http://dx.doi.org/10.1088/0004-637X/797/1/16}{\JournalTitle{\apj}, 797,
  16}

\bibitem[{{Konno} {et~al.}(2018){Konno}, {Ouchi}, {Shibuya}, {Ono},
  {Shimasaku}, {Taniguchi}, {Nagao}, {Kobayashi}, {Kajisawa}, {Kashikawa},
  {Inoue}, {Oguri}, {Furusawa}, {Goto}, {Harikane}, {Higuchi}, {Komiyama},
  {Kusakabe}, {Miyazaki}, {Nakajima}, \& {Wang}}]{Konno2018}
{Konno}, A., {Ouchi}, M., {Shibuya}, T., {et~al.} 2018,
  \href{http://dx.doi.org/10.1093/pasj/psx131}{\JournalTitle{\pasj}, 70, S16}

\bibitem[{{Krug} {et~al.}(2012){Krug}, {Veilleux}, {Tilvi}, {Malhotra},
  {Rhoads}, {Hibon}, {Swaters}, {Probst}, {Dey}, {Dickinson}, \&
  {Jannuzi}}]{Krug2012}
{Krug}, H.~B., {Veilleux}, S., {Tilvi}, V., {et~al.} 2012,
  \href{http://dx.doi.org/10.1088/0004-637X/745/2/122}{\JournalTitle{\apj},
  745, 122}

\bibitem[{{Leclercq} {et~al.}(2017){Leclercq}, {Bacon}, {Wisotzki}, {Mitchell},
  {Garel}, {Verhamme}, {Blaizot}, {Hashimoto}, {Herenz}, \&
  {Conseil}}]{Leclercq2017}
{Leclercq}, F., {Bacon}, R., {Wisotzki}, L., {et~al.} 2017,
  \href{http://dx.doi.org/10.1051/0004-6361/201731480}{\JournalTitle{\aap},
  608, A8}

\bibitem[{{Madau}(1995)}]{Madau1995}
{Madau}, P. 1995, \href{http://dx.doi.org/10.1086/175332}{\JournalTitle{\apj},
  441, 18}

\bibitem[{{Malhotra} \& {Rhoads}(2004)}]{Malhotra2004}
{Malhotra}, S., \& {Rhoads}, J.~E. 2004,
  \href{http://dx.doi.org/10.1086/427182}{\JournalTitle{\apjl}, 617, L5}

\bibitem[{{Matthee} {et~al.}(2015){Matthee}, {Sobral}, {Santos},
  {R{\"o}ttgering}, {Darvish}, \& {Mobasher}}]{Matthee2015}
{Matthee}, J., {Sobral}, D., {Santos}, S., {et~al.} 2015,
  \href{http://dx.doi.org/10.1093/mnras/stv947}{\JournalTitle{\mnras}, 451,
  400}

\bibitem[{{McQuinn} {et~al.}(2007){McQuinn}, {Hernquist}, {Zaldarriaga}, \&
  {Dutta}}]{McQuinn2007}
{McQuinn}, M., {Hernquist}, L., {Zaldarriaga}, M., \& {Dutta}, S. 2007,
  \href{http://dx.doi.org/10.1111/j.1365-2966.2007.12085.x}{\JournalTitle{\mnras},
  381, 75}

\bibitem[{{Momose} {et~al.}(2016){Momose}, {Ouchi}, {Nakajima}, {Ono},
  {Shibuya}, {Shimasaku}, {Yuma}, {Mori}, \& {Umemura}}]{Momose2016}
{Momose}, R., {Ouchi}, M., {Nakajima}, K., {et~al.} 2016,
  \href{http://dx.doi.org/10.1093/mnras/stw021}{\JournalTitle{\mnras}, 457,
  2318}

\bibitem[{{Mortlock} {et~al.}(2011){Mortlock}, {Warren}, {Venemans}, {Patel},
  {Hewett}, {McMahon}, {Simpson}, {Theuns}, {Gonz{\'a}les-Solares}, {Adamson},
  {Dye}, {Hambly}, {Hirst}, {Irwin}, {Kuiper}, {Lawrence}, \&
  {R{\"o}ttgering}}]{Mortlock2011}
{Mortlock}, D.~J., {Warren}, S.~J., {Venemans}, B.~P., {et~al.} 2011,
  \href{http://dx.doi.org/10.1038/nature10159}{\JournalTitle{\nat}, 474, 616}

\bibitem[{{Muzzin} {et~al.}(2013){Muzzin}, {Marchesini}, {Stefanon}, {Franx},
  {Milvang-Jensen}, {Dunlop}, {Fynbo}, {Brammer}, {Labb{\'e}}, \& {van
  Dokkum}}]{Muzzin2013}
{Muzzin}, A., {Marchesini}, D., {Stefanon}, M., {et~al.} 2013,
  \href{http://dx.doi.org/10.1088/0067-0049/206/1/8}{\JournalTitle{\apjs}, 206,
  8}

\bibitem[{{Ota} {et~al.}(2017){Ota}, {Iye}, {Kashikawa}, {Konno}, {Nakata},
  {Totani}, {Kobayashi}, {Fudamoto}, {Seko}, {Toshikawa}, {Ichikawa},
  {Shibuya}, \& {Onoue}}]{Ota2017}
{Ota}, K., {Iye}, M., {Kashikawa}, N., {et~al.} 2017,
  \href{http://dx.doi.org/10.3847/1538-4357/aa7a0a}{\JournalTitle{\apj}, 844,
  85}

\bibitem[{{Ouchi} {et~al.}(2005){Ouchi}, {Shimasaku}, {Akiyama}, {Sekiguchi},
  {Furusawa}, {Okamura}, {Kashikawa}, {Iye}, {Kodama}, {Saito}, {Sasaki},
  {Simpson}, {Takata}, {Yamada}, {Yamanoi}, {Yoshida}, \&
  {Yoshida}}]{Ouchi2005}
{Ouchi}, M., {Shimasaku}, K., {Akiyama}, M., {et~al.} 2005,
  \href{http://dx.doi.org/10.1086/428499}{\JournalTitle{\apjl}, 620, L1}

\bibitem[{{Ouchi} {et~al.}(2008){Ouchi}, {Shimasaku}, {Akiyama}, {Simpson},
  {Saito}, {Ueda}, {Furusawa}, {Sekiguchi}, {Yamada}, {Kodama}, {Kashikawa},
  {Okamura}, {Iye}, {Takata}, {Yoshida}, \& {Yoshida}}]{Ouchi2008}
---. 2008, \href{http://dx.doi.org/10.1086/527673}{\JournalTitle{\apjs}, 176,
  301}

\bibitem[{{Ouchi} {et~al.}(2010){Ouchi}, {Shimasaku}, {Furusawa}, {Saito},
  {Yoshida}, {Akiyama}, {Ono}, {Yamada}, {Ota}, {Kashikawa}, {Iye}, {Kodama},
  {Okamura}, {Simpson}, \& {Yoshida}}]{Ouchi2010}
{Ouchi}, M., {Shimasaku}, K., {Furusawa}, H., {et~al.} 2010,
  \href{http://dx.doi.org/10.1088/0004-637X/723/1/869}{\JournalTitle{\apj},
  723, 869}

\bibitem[{{Pirzkal} {et~al.}(2013){Pirzkal}, {Rothberg}, {Ly}, {Malhotra},
  {Rhoads}, {Grogin}, {Dahlen}, {Noeske}, {Meurer}, {Walsh}, {Hathi}, {Cohen},
  {Bellini}, {Holwerda}, {Straughn}, {Mechtley}, \& {Windhorst}}]{Pirzkal2013}
{Pirzkal}, N., {Rothberg}, B., {Ly}, C., {et~al.} 2013,
  \href{http://dx.doi.org/10.1088/0004-637X/772/1/48}{\JournalTitle{\apj}, 772,
  48}

\bibitem[{{Planck Collaboration} {et~al.}(2018){Planck Collaboration},
  {Aghanim}, {Akrami}, {Ashdown}, {Aumont}, {Baccigalupi}, {Ballardini},
  {Banday}, {Barreiro}, {Bartolo}, {Basak}, {Battye}, {Benabed}, {Bernard},
  {Bersanelli}, {Bielewicz}, {Bock}, {Bond}, {Borrill}, {Bouchet}, {Boulanger},
  {Bucher}, {Burigana}, {Butler}, {Calabrese}, {Cardoso}, {Carron},
  {Challinor}, {Chiang}, {Chluba}, {Colombo}, {Combet}, {Contreras}, {Crill},
  {Cuttaia}, {de Bernardis}, {de Zotti}, {Delabrouille}, {Delouis}, {Di
  Valentino}, {Diego}, {Dor{\'e}}, {Douspis}, {Ducout}, {Dupac}, {Dusini},
  {Efstathiou}, {Elsner}, {En{\ss}lin}, {Eriksen}, {Fantaye}, {Farhang},
  {Fergusson}, {Fernandez-Cobos}, {Finelli}, {Forastieri}, {Frailis},
  {Franceschi}, {Frolov}, {Galeotta}, {Galli}, {Ganga}, {G{\'e}nova-Santos},
  {Gerbino}, {Ghosh}, {Gonz{\'a}lez-Nuevo}, {G{\'o}rski}, {Gratton},
  {Gruppuso}, {Gudmundsson}, {Hamann}, {Handley}, {Herranz}, {Hivon}, {Huang},
  {Jaffe}, {Jones}, {Karakci}, {Keih{\"a}nen}, {Keskitalo}, {Kiiveri}, {Kim},
  {Kisner}, {Knox}, {Krachmalnicoff}, {Kunz}, {Kurki-Suonio}, {Lagache},
  {Lamarre}, {Lasenby}, {Lattanzi}, {Lawrence}, {Le Jeune}, {Lemos},
  {Lesgourgues}, {Levrier}, {Lewis}, {Liguori}, {Lilje}, {Lilley}, {Lindholm},
  {L{\'o}pez-Caniego}, {Lubin}, {Ma}, {Mac{\'{\i}}as-P{\'e}rez}, {Maggio},
  {Maino}, {Mandolesi}, {Mangilli}, {Marcos-Caballero}, {Maris}, {Martin},
  {Martinelli}, {Mart{\'{\i}}nez-Gonz{\'a}lez}, {Matarrese}, {Mauri}, {McEwen},
  {Meinhold}, {Melchiorri}, {Mennella}, {Migliaccio}, {Millea}, {Mitra},
  {Miville-Desch{\^e}nes}, {Molinari}, {Montier}, {Morgante}, {Moss}, {Natoli},
  {N{\o}rgaard-Nielsen}, {Pagano}, {Paoletti}, {Partridge}, {Patanchon},
  {Peiris}, {Perrotta}, {Pettorino}, {Piacentini}, {Polastri}, {Polenta},
  {Puget}, {Rachen}, {Reinecke}, {Remazeilles}, {Renzi}, {Rocha}, {Rosset},
  {Roudier}, {Rubi{\~n}o-Mart{\'{\i}}n}, {Ruiz-Granados}, {Salvati}, {Sandri},
  {Savelainen}, {Scott}, {Shellard}, {Sirignano}, {Sirri}, {Spencer},
  {Sunyaev}, {Suur-Uski}, {Tauber}, {Tavagnacco}, {Tenti}, {Toffolatti},
  {Tomasi}, {Trombetti}, {Valenziano}, {Valiviita}, {Van Tent}, {Vibert},
  {Vielva}, {Villa}, {Vittorio}, {Wandelt}, {Wehus}, {White}, {White},
  {Zacchei}, \& {Zonca}}]{Planck2018VI}
{Planck Collaboration}, {Aghanim}, N., {Akrami}, Y., {et~al.} 2018,
  \JournalTitle{arXiv e-prints},
  \href{http://arxiv.org/abs/1807.06209}{{\sffamily arXiv:1807.06209}}

\bibitem[{{Robitaille} {et~al.}(2007){Robitaille}, {Whitney}, {Indebetouw}, \&
  {Wood}}]{Robitaille2007}
{Robitaille}, T.~P., {Whitney}, B.~A., {Indebetouw}, R., \& {Wood}, K. 2007,
  \href{http://dx.doi.org/10.1086/512039}{\JournalTitle{\apjs}, 169, 328}

\bibitem[{{Rowe} {et~al.}(2015){Rowe}, {Jarvis}, {Mandelbaum}, {Bernstein},
  {Bosch}, {Simet}, {Meyers}, {Kacprzak}, {Nakajima}, {Zuntz}, {Miyatake},
  {Dietrich}, {Armstrong}, {Melchior}, \& {Gill}}]{Rowe2014}
{Rowe}, B.~T.~P., {Jarvis}, M., {Mandelbaum}, R., {et~al.} 2015,
  \href{http://dx.doi.org/10.1016/j.ascom.2015.02.002}{\JournalTitle{Astronomy
  and Computing}, 10, 121}

\bibitem[{{Santos}(2004)}]{Santos2004}
{Santos}, M.~R. 2004,
  \href{http://dx.doi.org/10.1111/j.1365-2966.2004.07594.x}{\JournalTitle{\mnras},
  349, 1137}

\bibitem[{{Santos} {et~al.}(2016){Santos}, {Sobral}, \& {Matthee}}]{Santos2016}
{Santos}, S., {Sobral}, D., \& {Matthee}, J. 2016,
  \href{http://dx.doi.org/10.1093/mnras/stw2076}{\JournalTitle{\mnras}, 463,
  1678}

\bibitem[{{Shibuya} {et~al.}(2019){Shibuya}, {Ouchi}, {Harikane}, \&
  {Nakajima}}]{Shibuya2019}
{Shibuya}, T., {Ouchi}, M., {Harikane}, Y., \& {Nakajima}, K. 2019,
  \href{http://dx.doi.org/10.3847/1538-4357/aaf64b}{\JournalTitle{\apj}, 871,
  164}

\bibitem[{{Shimasaku} {et~al.}(2003){Shimasaku}, {Ouchi}, {Okamura},
  {Kashikawa}, {Doi}, {Furusawa}, {Hamabe}, {Hayashino}, {Kawabata}, {Kimura},
  {Kodaira}, {Komiyama}, {Matsuda}, {Miyazaki}, {Miyazaki}, {Nakata}, {Ohta},
  {Ohyama}, {Sekiguchi}, {Shioya}, {Tamura}, {Taniguchi}, {Yagi}, {Yamada}, \&
  {Yasuda}}]{Shimasaku2003}
{Shimasaku}, K., {Ouchi}, M., {Okamura}, S., {et~al.} 2003,
  \href{http://dx.doi.org/10.1086/374880}{\JournalTitle{\apjl}, 586, L111}

\bibitem[{{Suchyta} {et~al.}(2016){Suchyta}, {Huff}, {Aleksi{\'c}}, {Melchior},
  {Jouvel}, {MacCrann}, {Ross}, {Crocce}, {Gaztanaga}, {Honscheid}, {Leistedt},
  {Peiris}, {Rykoff}, {Sheldon}, {Abbott}, {Abdalla}, {Allam}, {Banerji},
  {Benoit-L{\'e}vy}, {Bertin}, {Brooks}, {Burke}, {Carnero Rosell}, {Carrasco
  Kind}, {Carretero}, {Cunha}, {D'Andrea}, {da Costa}, {DePoy}, {Desai},
  {Diehl}, {Dietrich}, {Doel}, {Eifler}, {Estrada}, {Evrard}, {Flaugher},
  {Fosalba}, {Frieman}, {Gerdes}, {Gruen}, {Gruendl}, {James}, {Jarvis},
  {Kuehn}, {Kuropatkin}, {Lahav}, {Lima}, {Maia}, {March}, {Marshall},
  {Miller}, {Miquel}, {Neilsen}, {Nichol}, {Nord}, {Ogando}, {Percival},
  {Reil}, {Roodman}, {Sako}, {Sanchez}, {Scarpine}, {Sevilla-Noarbe}, {Smith},
  {Soares-Santos}, {Sobreira}, {Swanson}, {Tarle}, {Thaler}, {Thomas},
  {Vikram}, {Walker}, {Wechsler}, {Zhang}, \& {DES
  Collaboration}}]{Suchyta2016}
{Suchyta}, E., {Huff}, E.~M., {Aleksi{\'c}}, J., {et~al.} 2016,
  \href{http://dx.doi.org/10.1093/mnras/stv2953}{\JournalTitle{\mnras}, 457,
  786}

\bibitem[{{Tanaka} {et~al.}(2017){Tanaka}, {Hasinger}, {Silverman},
  {Bickerton}, {Furusawa}, {Harikane}, {Hu}, {Ikeda}, {Li}, {McCracken},
  {Price}, {Strauss}, {Koike}, {Komiyama}, {Mineo}, {Miyazaki}, {Nishizawa},
  {Takata}, {Utsumi}, {Yamada}, \& {Yasuda}}]{Tanaka2017}
{Tanaka}, M., {Hasinger}, G., {Silverman}, J.~D., {et~al.} 2017,
  \JournalTitle{ArXiv e-prints},
  \href{http://arxiv.org/abs/1706.00566}{{\sffamily arXiv:1706.00566}}

\bibitem[{{Tilvi} {et~al.}(2010){Tilvi}, {Rhoads}, {Hibon}, {Malhotra}, {Wang},
  {Veilleux}, {Swaters}, {Probst}, {Krug}, {Finkelstein}, \&
  {Dickinson}}]{Tilvi2010}
{Tilvi}, V., {Rhoads}, J.~E., {Hibon}, P., {et~al.} 2010,
  \href{http://dx.doi.org/10.1088/0004-637X/721/2/1853}{\JournalTitle{\apj},
  721, 1853}

\bibitem[{{Valdes} {et~al.}(2014){Valdes}, {Gruendl}, \& {DES
  Project}}]{Valdes2014}
{Valdes}, F., {Gruendl}, R., \& {DES Project}. 2014, in Astronomical Society of
  the Pacific Conference Series, Vol. 485, Astronomical Data Analysis Software
  and Systems XXIII, ed. N.~{Manset} \& P.~{Forshay}, 379

\bibitem[{{Wang} {et~al.}(2005){Wang}, {Malhotra}, \& {Rhoads}}]{Wang2005}
{Wang}, J.~X., {Malhotra}, S., \& {Rhoads}, J.~E. 2005,
  \href{http://dx.doi.org/10.1086/429617}{\JournalTitle{\apjl}, 622, L77}

\bibitem[{{Wang} {et~al.}(2009){Wang}, {Malhotra}, {Rhoads}, {Zhang}, \&
  {Finkelstein}}]{Wang2009}
{Wang}, J.-X., {Malhotra}, S., {Rhoads}, J.~E., {Zhang}, H.-T., \&
  {Finkelstein}, S.~L. 2009,
  \href{http://dx.doi.org/10.1088/0004-637X/706/1/762}{\JournalTitle{\apj},
  706, 762}

\bibitem[{{Yang} {et~al.}(2017){Yang}, {Malhotra}, {Rhoads}, {Leitherer},
  {Wofford}, {Jiang}, \& {Wang}}]{Yang2017}
{Yang}, H., {Malhotra}, S., {Rhoads}, J.~E., {et~al.} 2017,
  \href{http://dx.doi.org/10.3847/1538-4357/aa6337}{\JournalTitle{\apj}, 838,
  4}

\bibitem[{{Yang} {et~al.}(2019){Yang}, {Infante}, {Rhoads}, {Hu}, {Zheng},
  {Malhotra}, {Wang}, {Barrientos}, {Kang}, \& {Jiang}}]{Yang2019}
{Yang}, H., {Infante}, L., {Rhoads}, J.~E., {et~al.} 2019,
  \href{http://dx.doi.org/10.3847/1538-4357/ab16ce}{\JournalTitle{\apj}, 876,
  123}

\bibitem[{{Zheng} {et~al.}(2016){Zheng}, {Malhotra}, {Rhoads}, {Finkelstein},
  {Wang}, {Jiang}, \& {Cai}}]{Zheng2016}
{Zheng}, Z.-Y., {Malhotra}, S., {Rhoads}, J.~E., {et~al.} 2016,
  \href{http://dx.doi.org/10.3847/0067-0049/226/2/23}{\JournalTitle{\apjs},
  226, 23}

\bibitem[{{Zheng} {et~al.}(2014){Zheng}, {Wang}, {Malhotra}, {Rhoads},
  {Finkelstein}, \& {Finkelstein}}]{Zheng2014}
{Zheng}, Z.-Y., {Wang}, J.-X., {Malhotra}, S., {et~al.} 2014,
  \href{http://dx.doi.org/10.1093/mnras/stu054}{\JournalTitle{\mnras}, 439,
  1101}

\bibitem[{{Zheng} {et~al.}(2017){Zheng}, {Wang}, {Rhoads}, {Infante},
  {Malhotra}, {Hu}, {Walker}, {Jiang}, {Jiang}, {Hibon}, {Gonzalez}, {Kong},
  {Zheng}, {Galaz}, \& {Barrientos}}]{Zheng2017}
{Zheng}, Z.-Y., {Wang}, J., {Rhoads}, J., {et~al.} 2017,
  \href{http://dx.doi.org/10.3847/2041-8213/aa794f}{\JournalTitle{\apjl}, 842,
  L22}

\bibitem[{{Zheng} {et~al.}(2019){Zheng}, {Rhoads}, {Wang}, {Malhotra},
  {Walker}, {Mooney}, {Jiang}, {Hu}, {Hibon}, \& {Jiang}}]{Zheng2019}
{Zheng}, Z.-Y., {Rhoads}, J.~E., {Wang}, J.-X., {et~al.} 2019,
  \href{http://dx.doi.org/10.1088/1538-3873/ab1c32}{\JournalTitle{\pasp}, 131,
  074502}

\bibitem[{{Zhu} {et~al.}(2015){Zhu}, {Comparat}, {Kneib}, {Delubac},
  {Raichoor}, {Dawson}, {Newman}, {Y{\`e}che}, {Zhou}, \&
  {Schneider}}]{Zhu2015}
{Zhu}, G.~B., {Comparat}, J., {Kneib}, J.-P., {et~al.} 2015,
  \href{http://dx.doi.org/10.1088/0004-637X/815/1/48}{\JournalTitle{\apj}, 815,
  48}

\end{thebibliography}

\begin{appendix}
{

\section{A:  On the color measurement}
\label{appd:a}

We examine the reliability of using SExtractor AUTO magnitudes to measure the narrowband to broadband color of LAEs, utilizing the injection and recovery simulations we introduced in \S\ref{ssc:compp}.
We use COSMOS field to present our analyses and results. 

Following \S\ref{ssc:compp}, we insert pseudo LAEs into the narrowband using a profile with S\'ersic index of 1.5 and half-light radius of 0.9 kpc. We also insert corresponding signals into the underlying broadband HSC-$y$ (assuming an intrinsic color excess of 1 mag) using identical source profile.
In the upper panel of Fig. \ref{fig:compcolor}, we plot the peak value and 1$\sigma$ scatter (measured through fitting the distribution with a Gaussian) of the output colors, as a function of the detected narrowband magnitude. 
We find while the AUTO-color could precisely recover the input value at the bright end, it slightly underestimate the color at the faint end. 
Meanwhile, the commonly used aperture color (using magnitudes measured on PSF-matched images within a fixed 2\arcsec\ aperture, e.g., \citealt{Ouchi2010,Ota2017}), though also underestimates the color at the faint end, behaves slightly better.

The color underestimation at the faint end is mainly due to contamination from foreground sources to the pseudo LAE photometry in the narrow and broadband, i.e., as most foreground sources show no color excess, the contamination to the broadband photometry is relatively more prominent than to the narrowband. As our AUTO magnitudes are generally measured in regions larger than a 2\arcsec\ aperture, the contamination effect is stronger for AUTO-color.
In the middle panel of Fig. \ref{fig:compcolor}, we exclude pseudo LAEs with S/N $>$ 2 in any veto band to minimize the effect of contamination. We then see negligible difference between two approaches, and both could reliably measure the input color at all magnitudes (though the AUTO-color shows slightly larger scatter).
Since we need to exclude sources with foreground contaminations anyway, both approaches are similarly reliable from this respect. 

However, it is known that the Ly$\alpha$ emission in LAEs at lower redshifts is more extended than the UV continuum \citep[e.g.][]{Momose2016,Yang2017,Leclercq2017}.
In this case the aperture-color would underestimate the intrinsic value of LAEs. 
To depict such effect, we perform injections adopting slightly larger Ly$\alpha$ emission size (1.2 kpc half-light radius in the narrowband, and 0.9 kpc in the broadband). The recovered color is presented in the lower panel of Fig. \ref{fig:compcolor}, in which we see that, in case of more extended Ly$\alpha$ emission, the AUTO-color behaves better than aperture-color, especially at the bright end. 
Note the sizes and the S\'ersic profiles here are adopted for illustration only. 
For instance, the half-light radius of 1.2 kpc we adopted is smaller than the typical size of the \lya\ profile measured by \citet{Leclercq2017} at lower redshifts (see also \S4.1). 
That is, if the \lya\ spatial profile of $z\sim$ 7.0 LAEs is similar to that seen at lower redshifts, the effect will be even more significant than the modest case we illustrate above. 

\begin{figure}[h]
   \centering
   \includegraphics[width=3.5in]{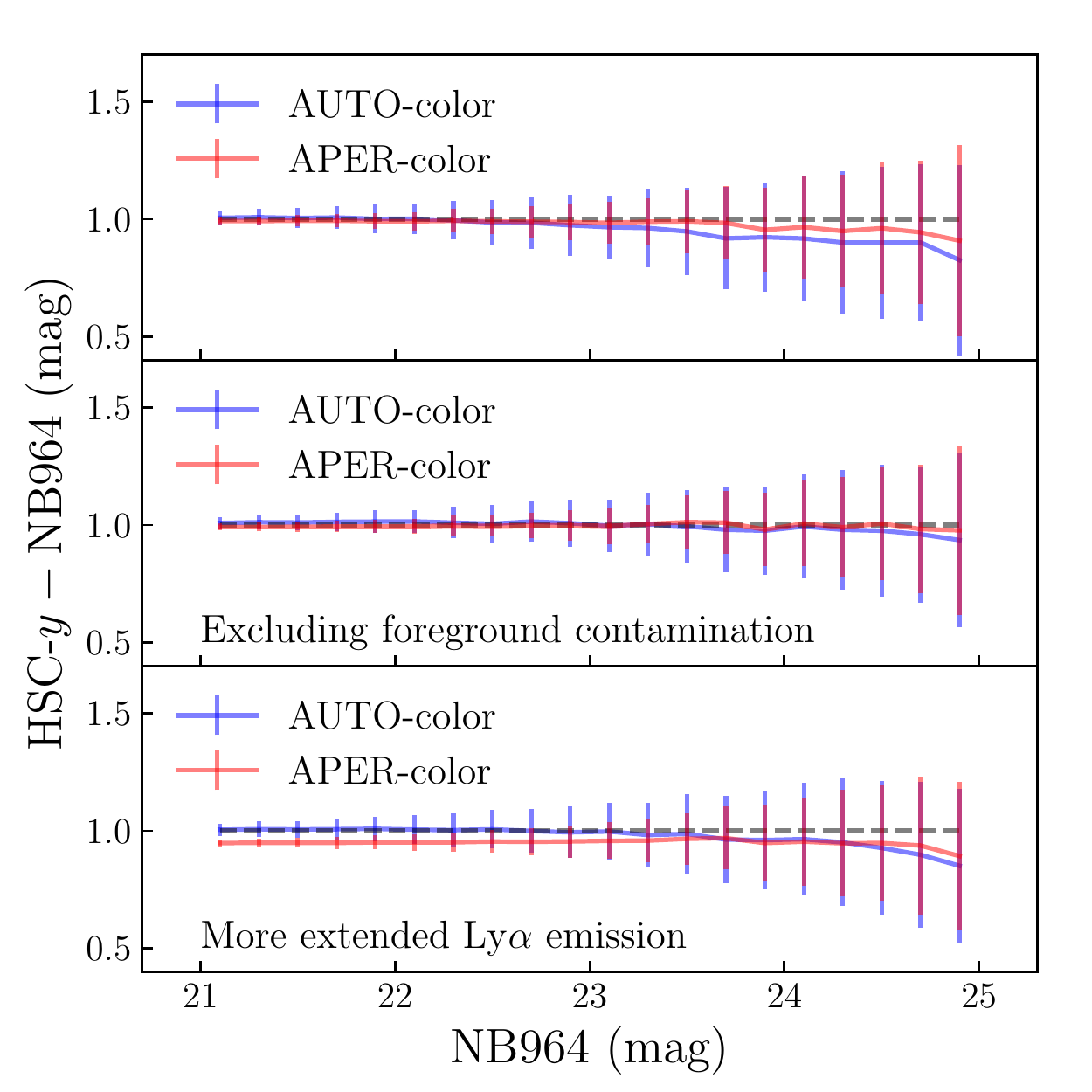}
   \caption{\label{fig:compcolor} The HSC-$y-\mathrm{NB964}$ color for simulated pseudo LAEs as a function of detected NB964 magnitude.  Upper panel: the recovered color for narrowband detected pseudo LAEs;
   Middle panel: for detected LAEs without foreground contamination in veto bands (S/N $<$ 2 in veto bands). In the upper and middle panel, the injected sources have identical  S\'ersic profile (with half-light radius of 0.9 kpc) in both the NB964 image and HSC-$y$ image. Lower panel: similar to the middle panel but the injected sources have slightly larger size in the narrowband (with half-light radius of 1.2 kpc in NB964 and 0.9 kpc in HSC-$y$.
   {The grey dashed lines indicate the input color.}
   }
\end{figure}

\section{B: Thumbnail Images of our LAE Candidates}
\label{appd}
We show the thumbnail images of our LAE candidates in COSMOS and CDFS field in Fig. \ref{fig:laethb}. 
We plot the veto broadband images, stacked veto broadband images (hereafter BB in the Fig. \ref{fig:laethb}), NB964 images and underlying broadband images for each LAE candidates. 
}

\begin{figure*}
\centering
\subfloat{\includegraphics[width=6.5in]{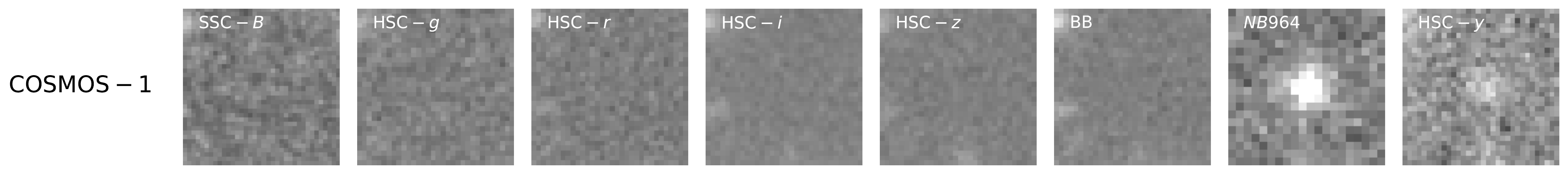}}\\
\subfloat{\includegraphics[width=6.5in]{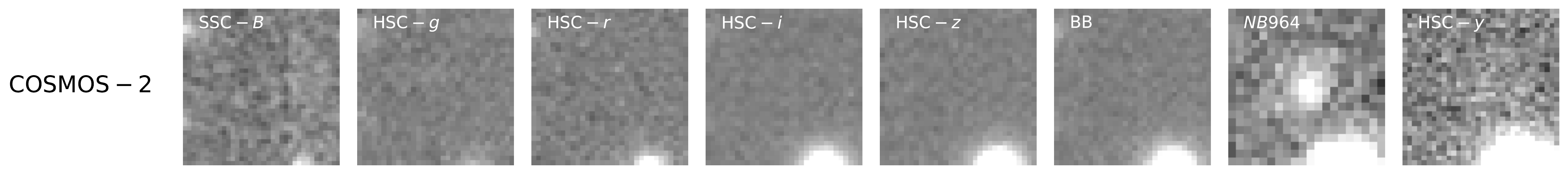}}\\
\subfloat{\includegraphics[width=6.5in]{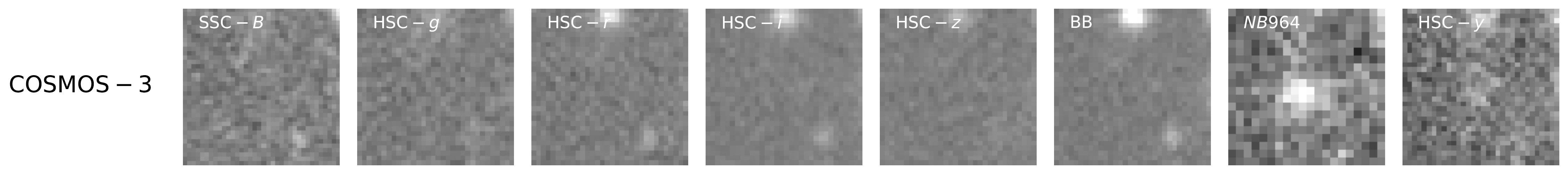}}\\
\subfloat{\includegraphics[width=6.5in]{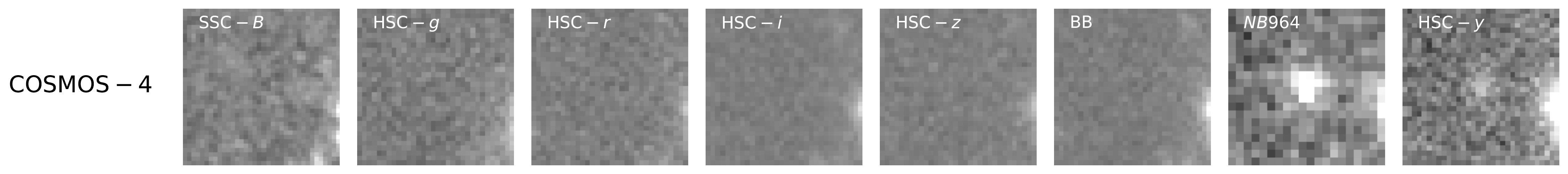}}\\
\subfloat{\includegraphics[width=6.5in]{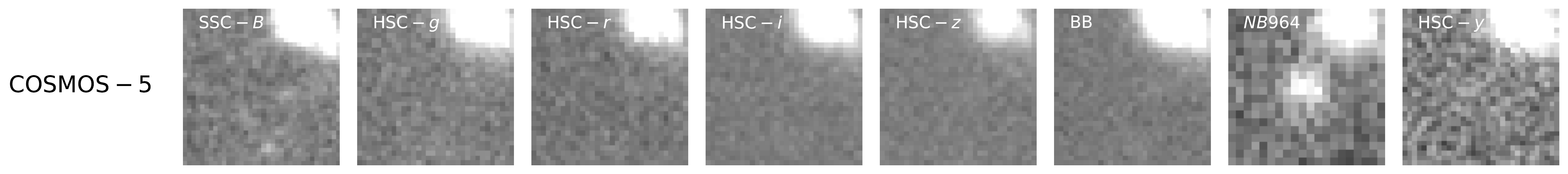}}\\
\subfloat{\includegraphics[width=6.5in]{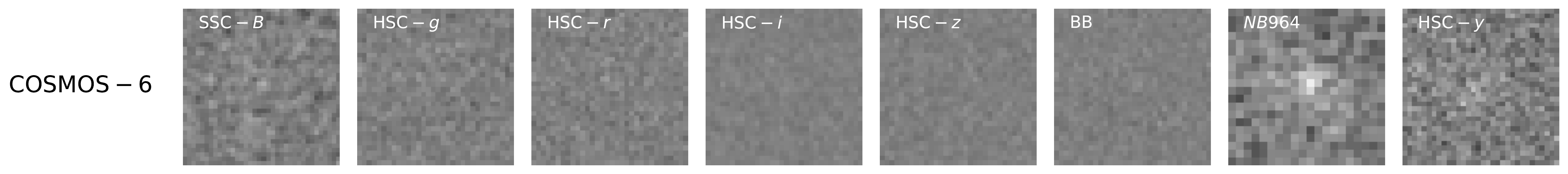}}\\
\subfloat{\includegraphics[width=6.5in]{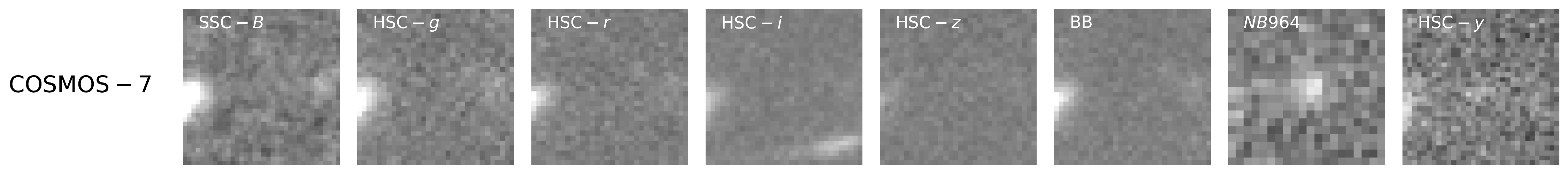}}\\
\subfloat{\includegraphics[width=6.5in]{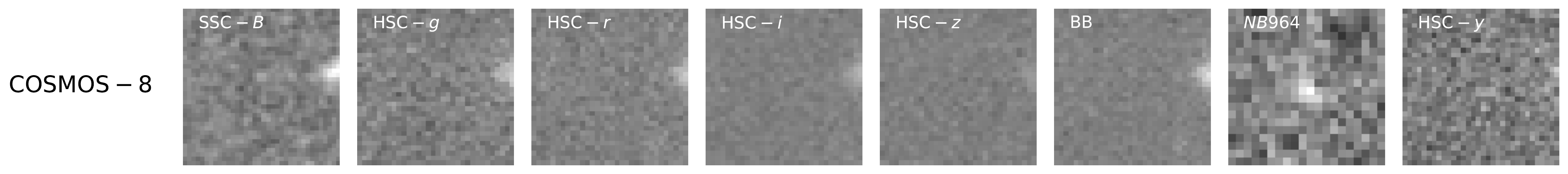}}\\
\subfloat{\includegraphics[width=6.5in]{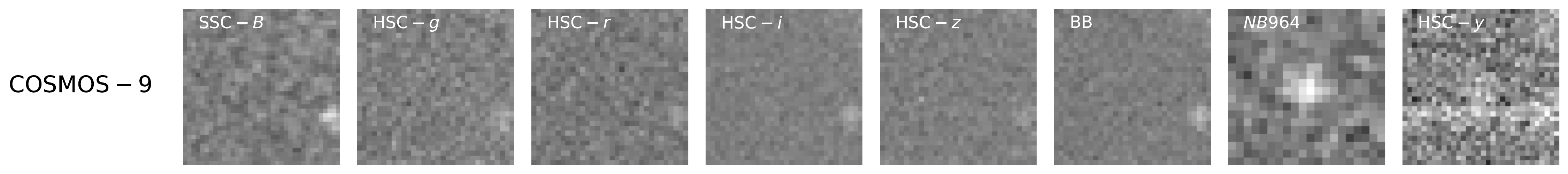}}\\
\subfloat{\includegraphics[width=6.5in]{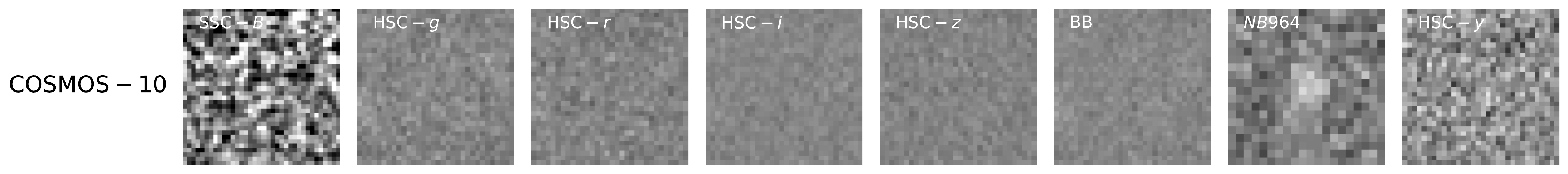}}\\
\subfloat{\includegraphics[width=6.5in]{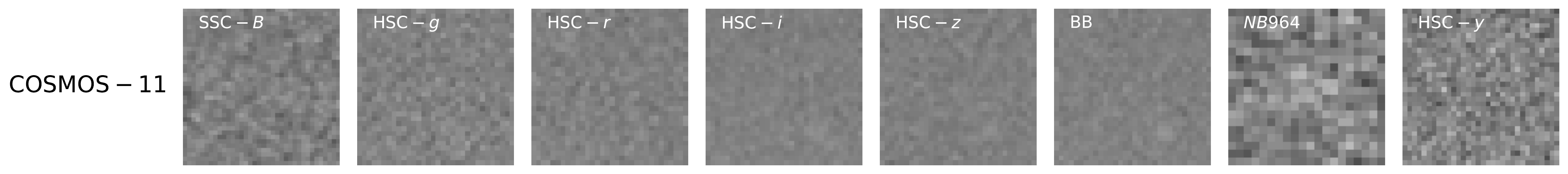}}\\
\subfloat{\includegraphics[width=6.5in]{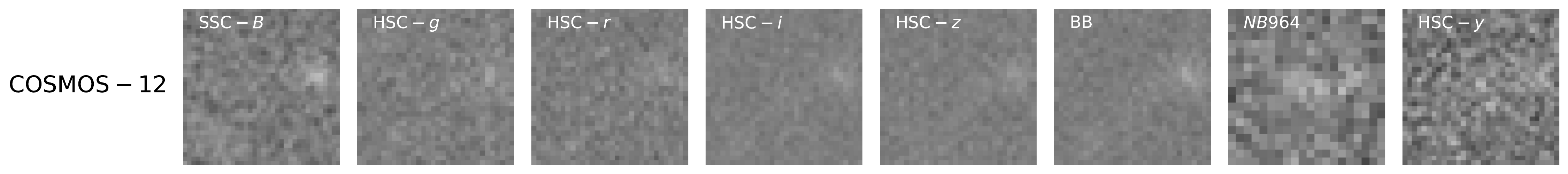}}
\caption{\label{fig:laethb} The veto broadband images, stacked veto broadband images (namely BB), NB964 images and the underlying broadband images for LAE candidates in the COSMOS and CDFS field. 
The size of each image is $\sim 5.4 \arcsec \times5.4 \arcsec$. The images are sorted by NB964 AUTO magnitude. Note the broadband images from  HSC plotted here were before resampling (for illustration only). }
\end{figure*}

\begin{figure*}
\centering
\ContinuedFloat
\subfloat{\includegraphics[width=6.5in]{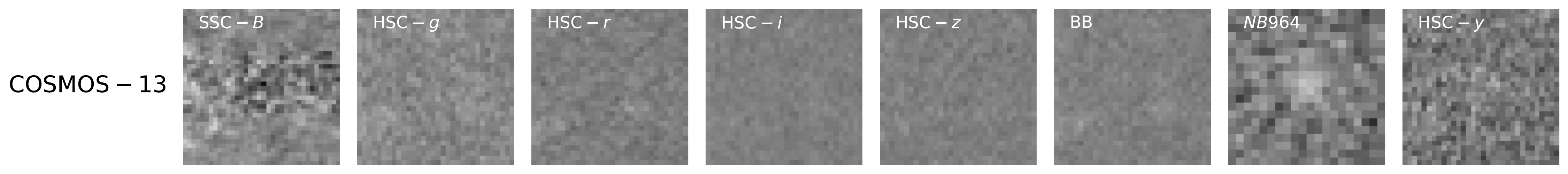}}\\
\subfloat{\includegraphics[width=6.5in]{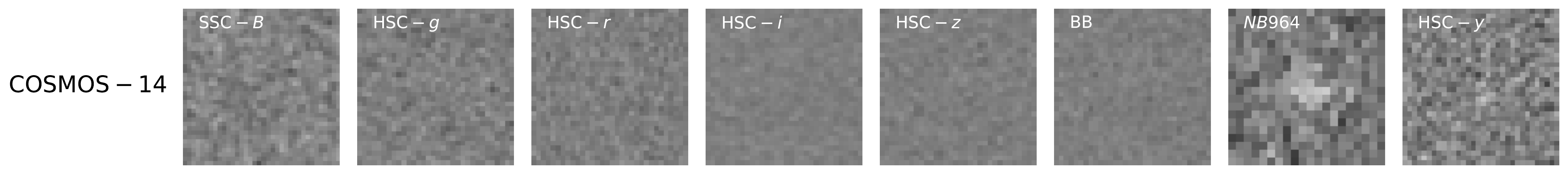}}\\
\subfloat{\includegraphics[width=6.5in]{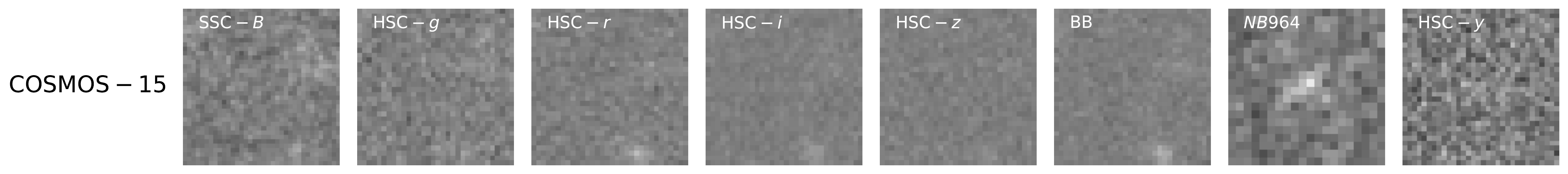}}\\
\subfloat{\includegraphics[width=6.5in]{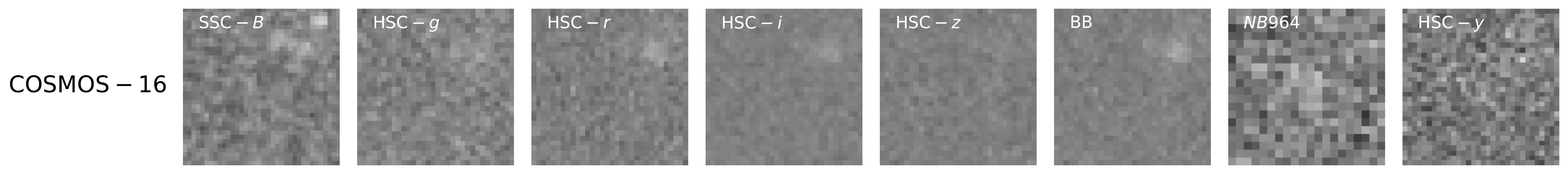}}\\
\subfloat{\includegraphics[width=6.5in]{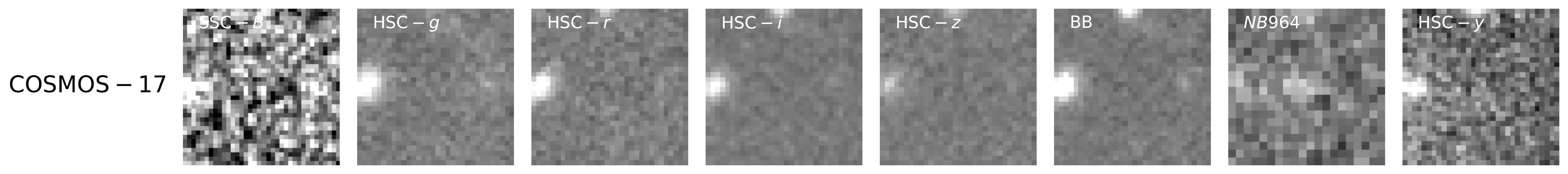}}\\
\subfloat{\includegraphics[width=6.5in]{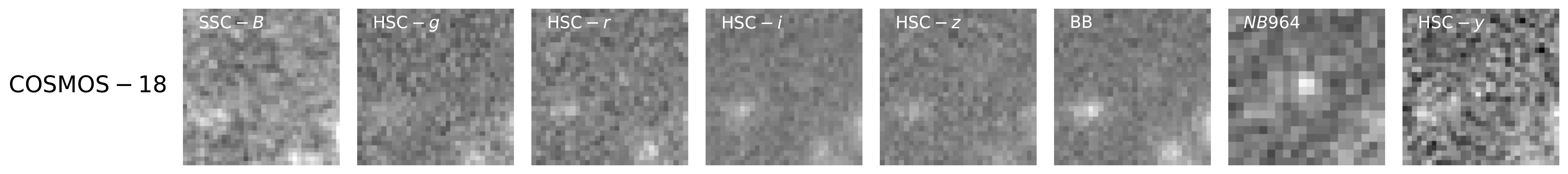}}\\
\subfloat{\includegraphics[width=6.5in]{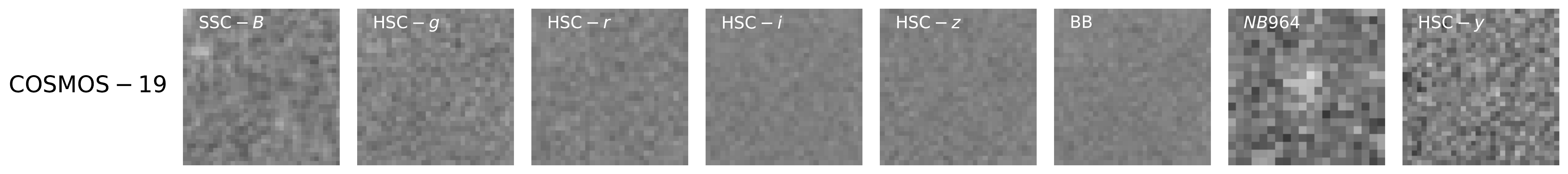}}\\
\subfloat{\includegraphics[width=6.5in]{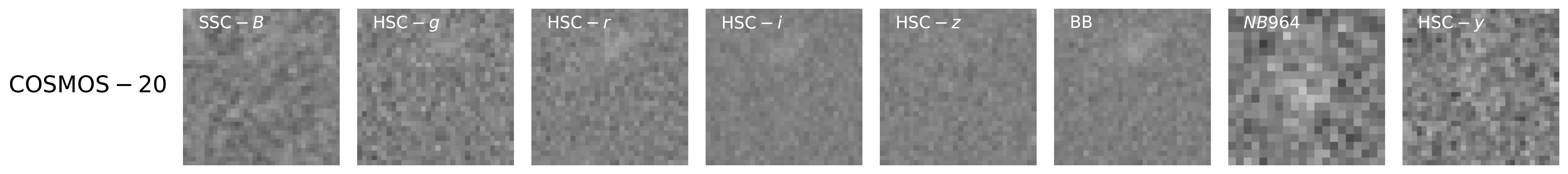}}\\
\subfloat{\includegraphics[width=6.5in]{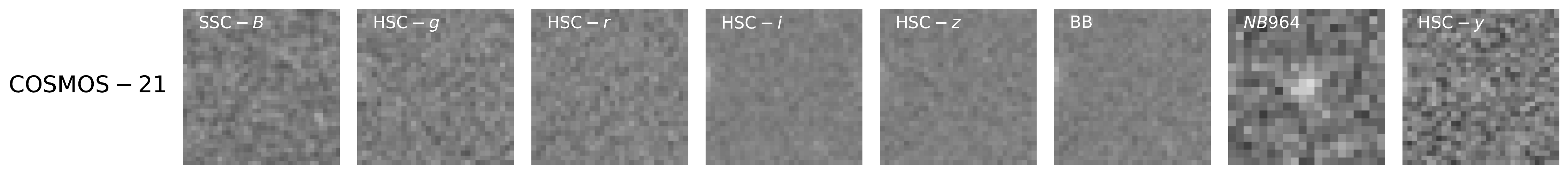}}\\
\subfloat{\includegraphics[width=6.5in]{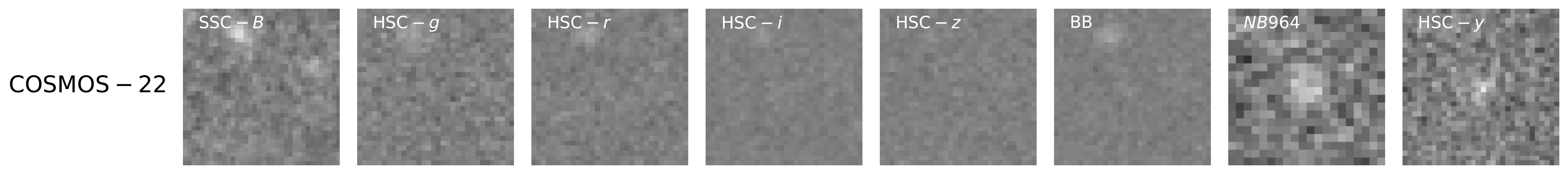}}\\
\subfloat{\includegraphics[width=6.5in]{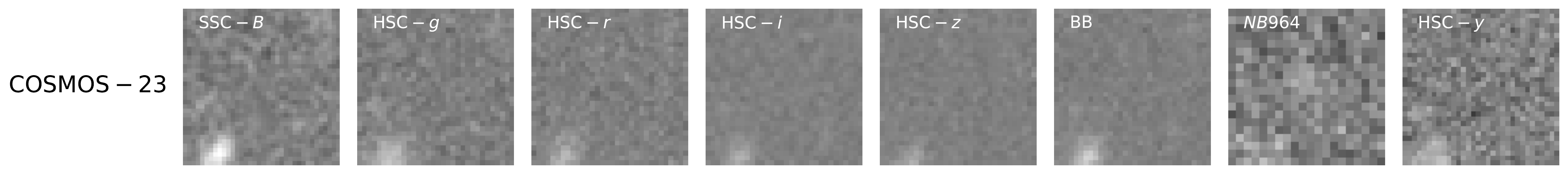}}\\
\subfloat{\includegraphics[width=6.5in]{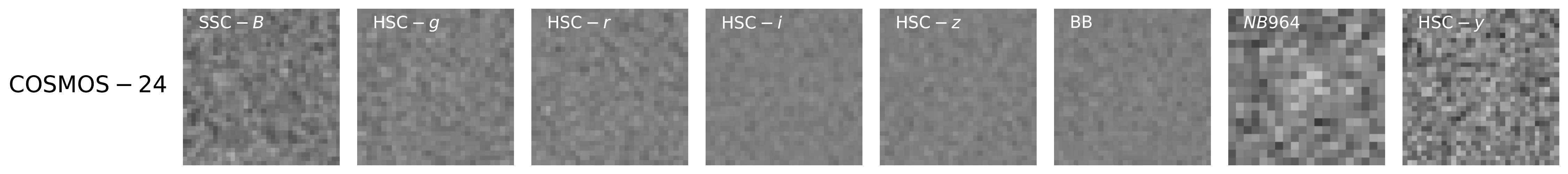}}
\caption{Continued}
\end{figure*}

\begin{figure*}
\centering
\ContinuedFloat
\subfloat{\includegraphics[width=6.5in]{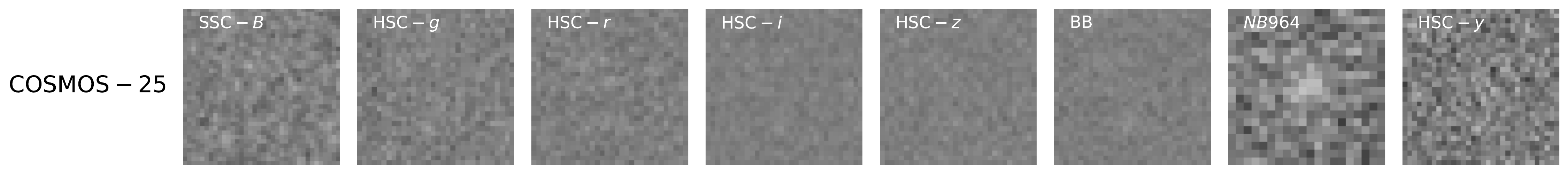}}\\
\subfloat{\includegraphics[width=6.5in]{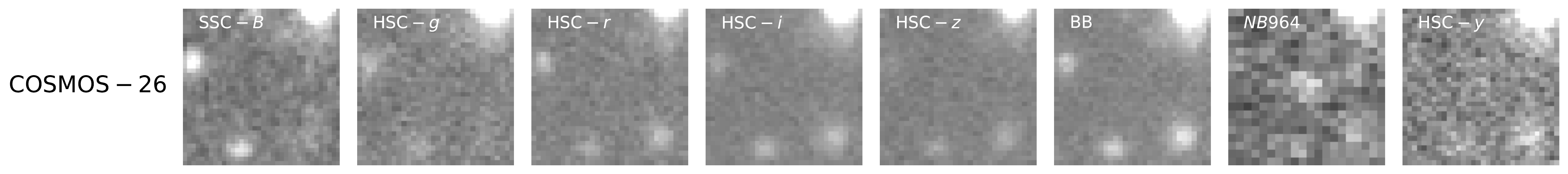}}\\
\subfloat{\includegraphics[width=6.5in]{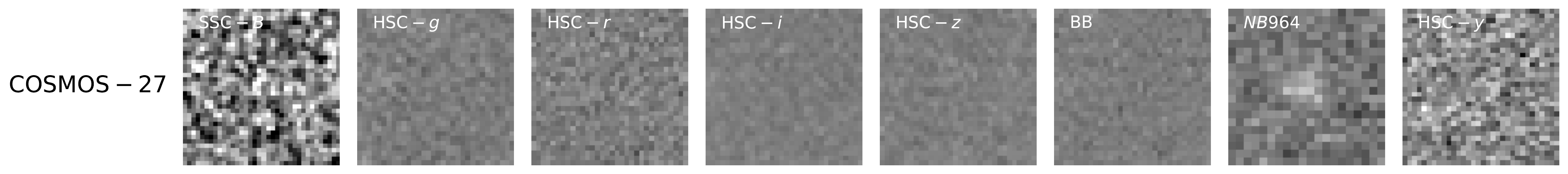}}\\
\subfloat{\includegraphics[width=6.5in]{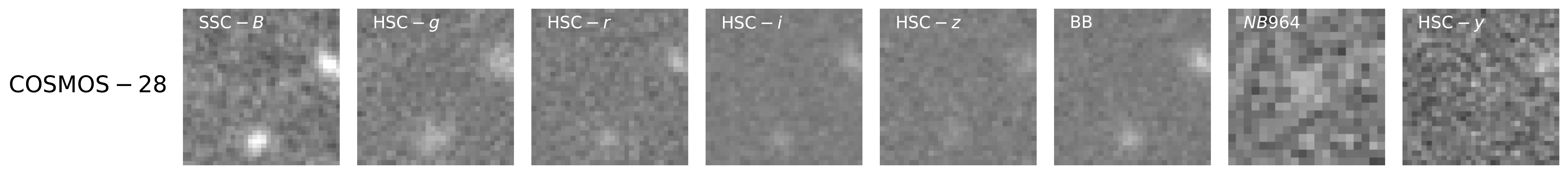}}\\
\subfloat{\includegraphics[width=6.5in]{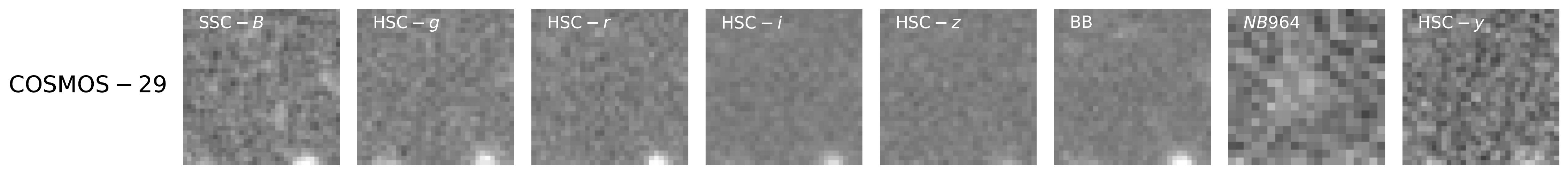}}\\
\subfloat{\includegraphics[width=6.5in]{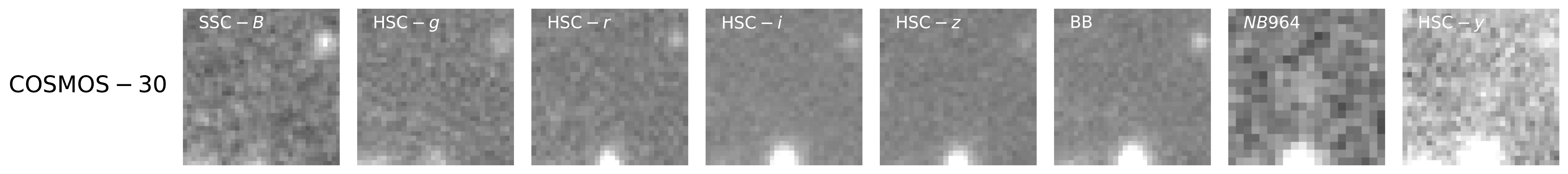}}\\
\subfloat{\includegraphics[width=6.5in]{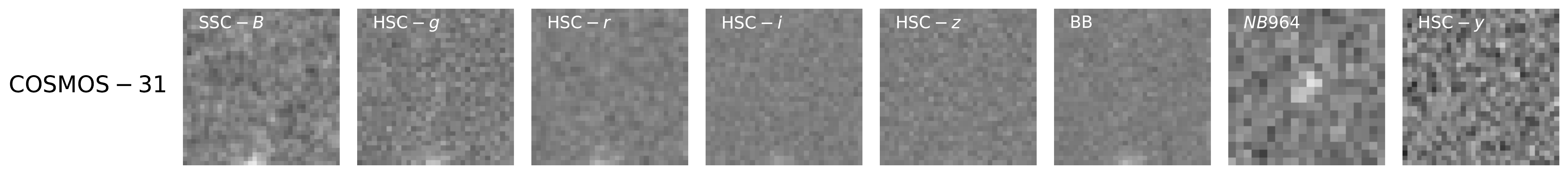}}\\
\subfloat{\includegraphics[width=6.5in]{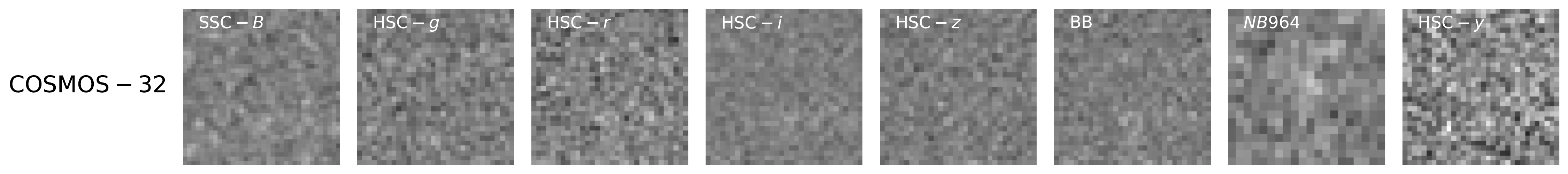}}\\
\subfloat{\includegraphics[width=6.5in]{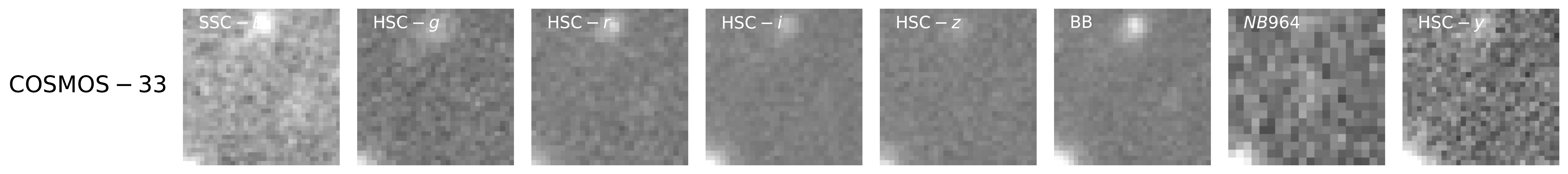}}\\
\subfloat{\includegraphics[width=6.5in]{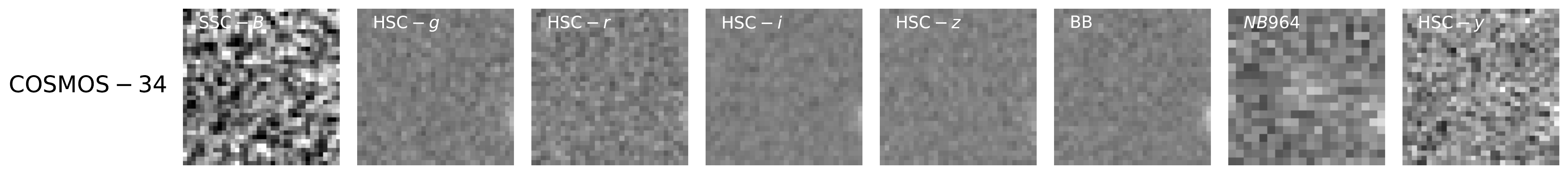}}\\
\subfloat{\includegraphics[width=6.5in]{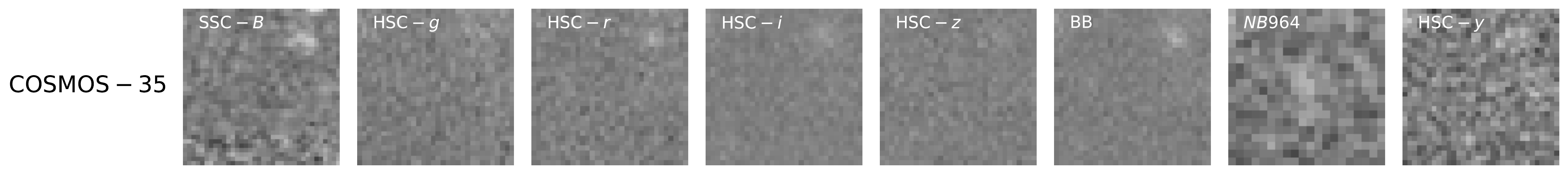}}\\
\subfloat{\includegraphics[width=6.5in]{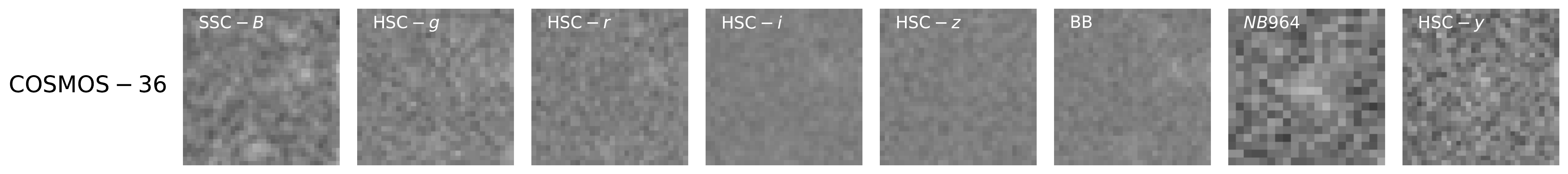}}
\caption{Continued.}
\end{figure*}

\begin{figure*}
\centering
\ContinuedFloat
\subfloat{\includegraphics[width=6.5in]{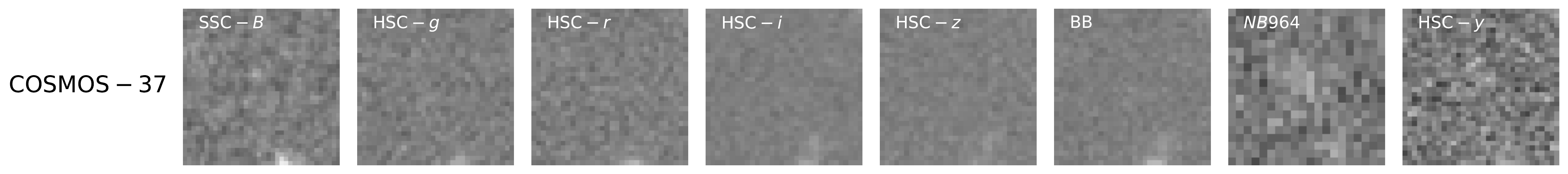}}\\
\subfloat{\includegraphics[width=6.5in]{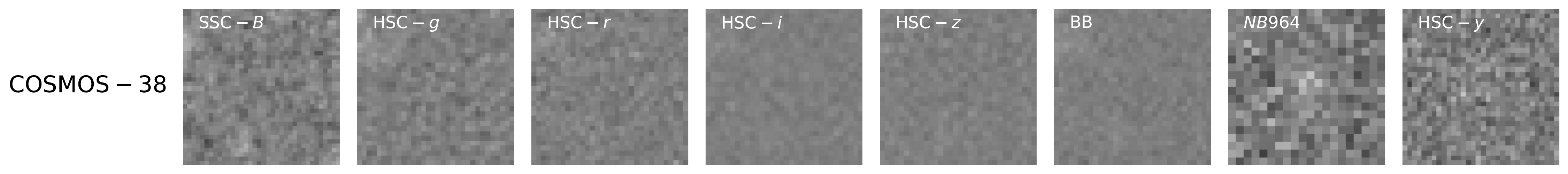}}\\
\subfloat{\includegraphics[width=6.5in]{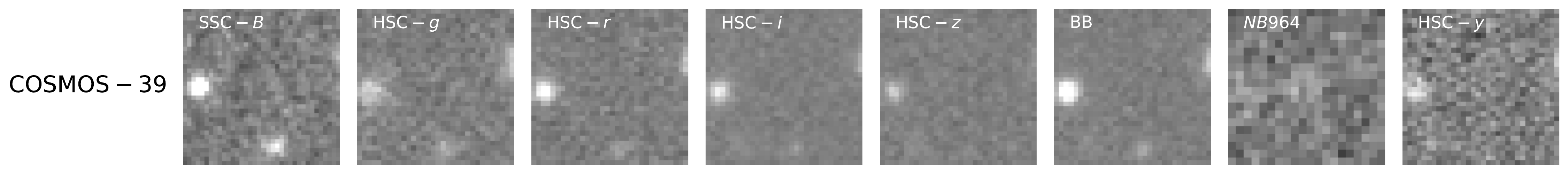}}\\
\subfloat{\includegraphics[width=6.5in]{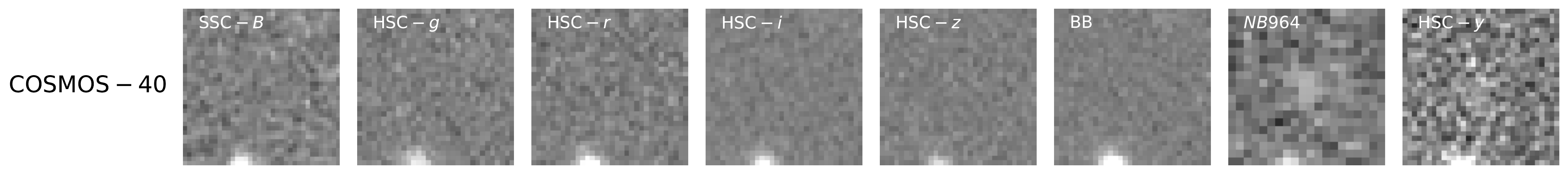}}\\
\subfloat{\includegraphics[width=6.5in]{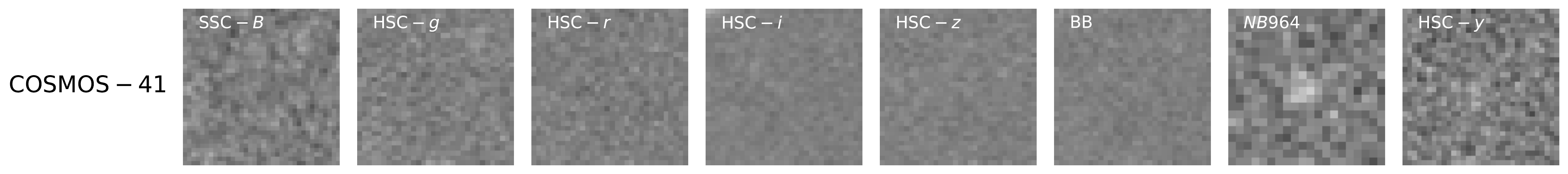}}\\
\subfloat{\includegraphics[width=6.5in]{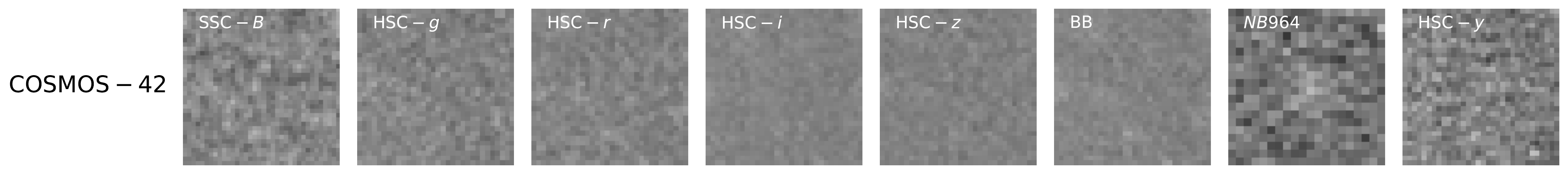}}\\
\subfloat{\includegraphics[width=6.5in]{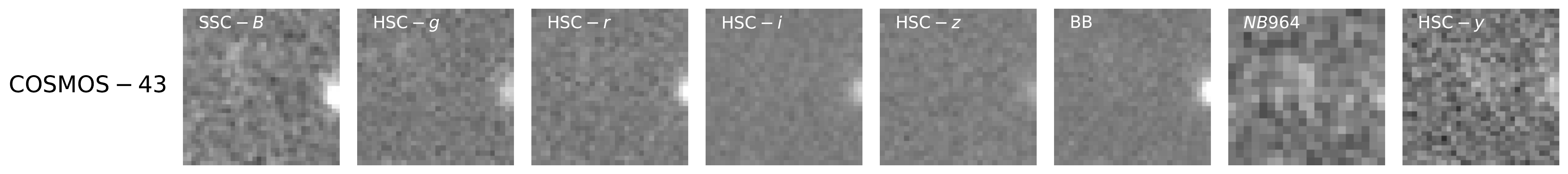}}\\
\subfloat{\includegraphics[width=6.5in]{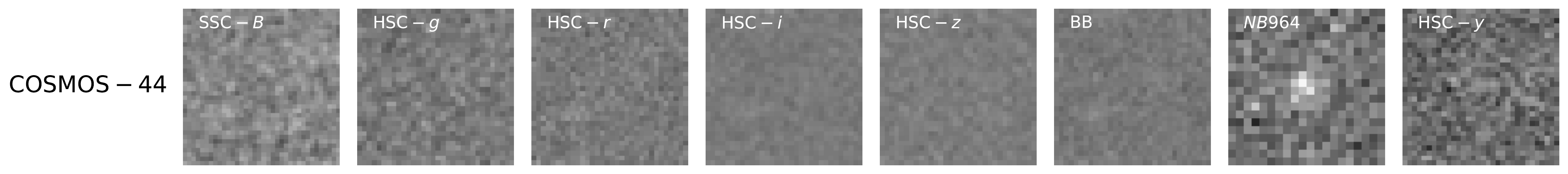}}\\
\subfloat{\includegraphics[width=6.5in]{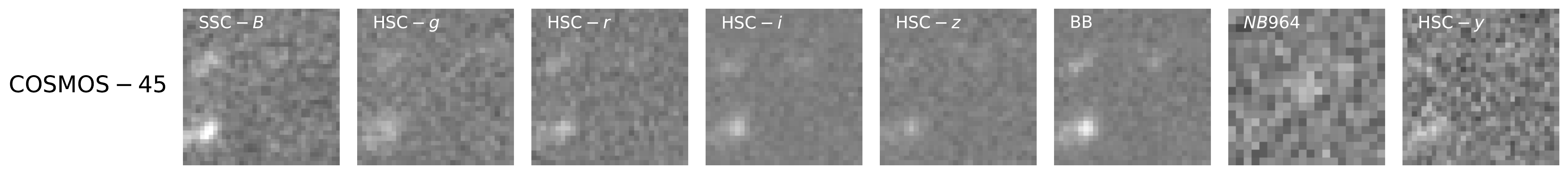}}\\
\subfloat{\includegraphics[width=6.5in]{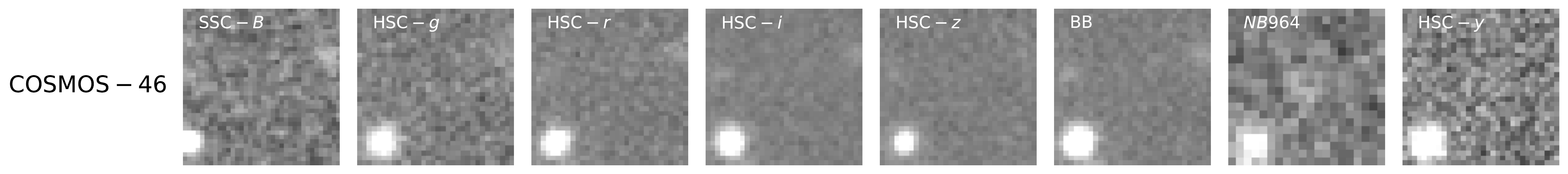}}\\
\subfloat{\includegraphics[width=6.5in]{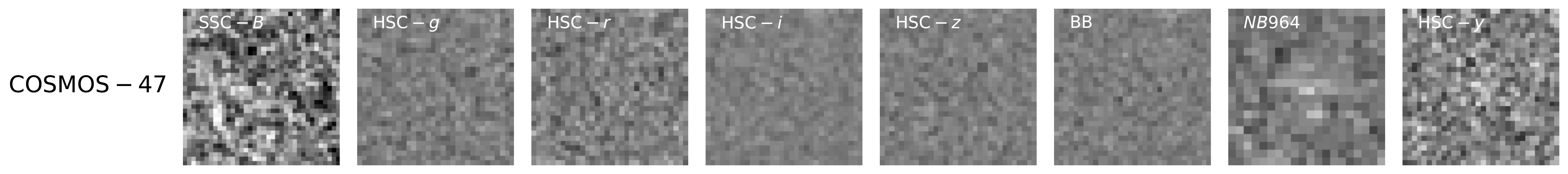}}\\
\subfloat{\includegraphics[width=6.5in]{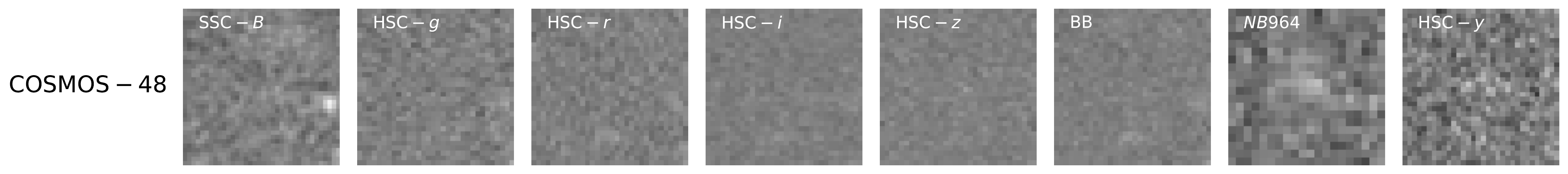}}
\caption{Continued.}
\end{figure*}

\begin{figure*}
\centering
\ContinuedFloat
\subfloat{\includegraphics[width=6.5in]{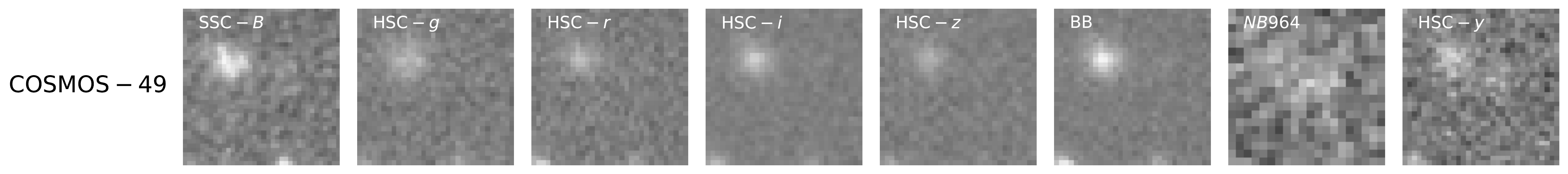}}\\
\subfloat{\includegraphics[width=5in]{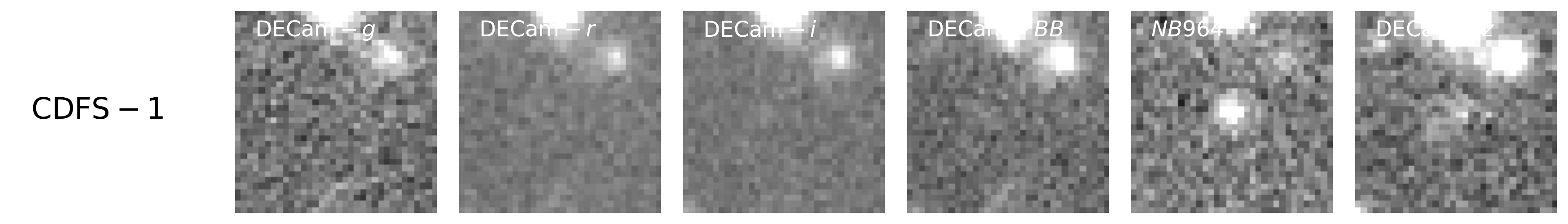}}\\
\subfloat{\includegraphics[width=5in]{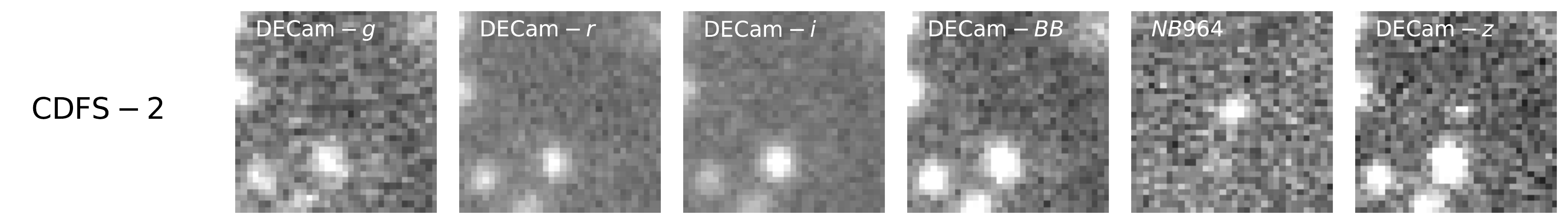}}\\
\subfloat{\includegraphics[width=5in]{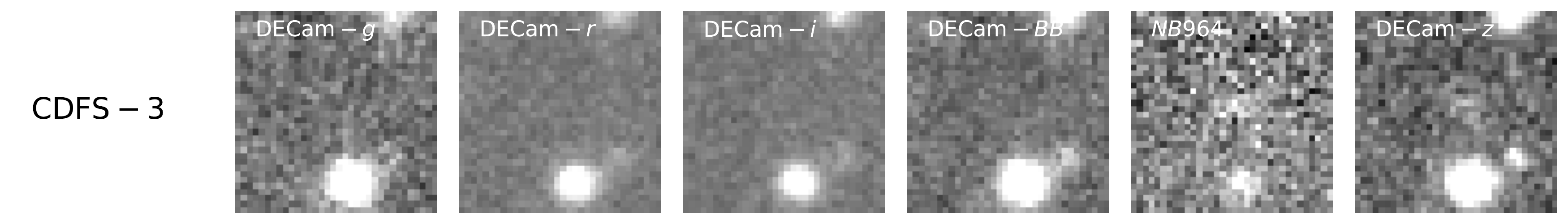}}\\
\subfloat{\includegraphics[width=5in]{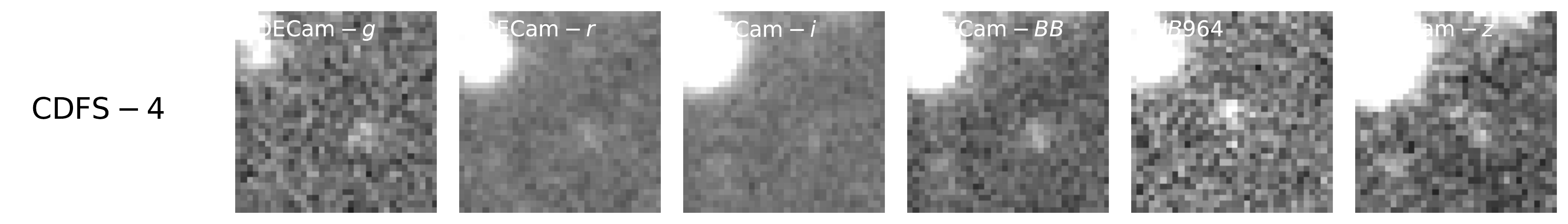}}\\
\subfloat{\includegraphics[width=5in]{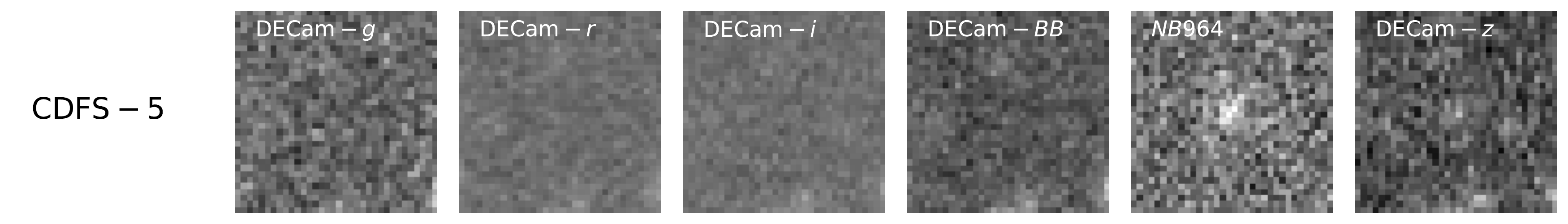}}\\
\subfloat{\includegraphics[width=5in]{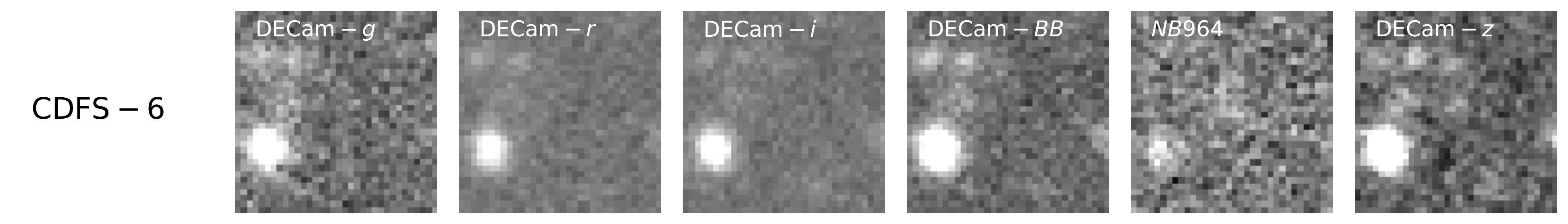}}\\
\subfloat{\includegraphics[width=5in]{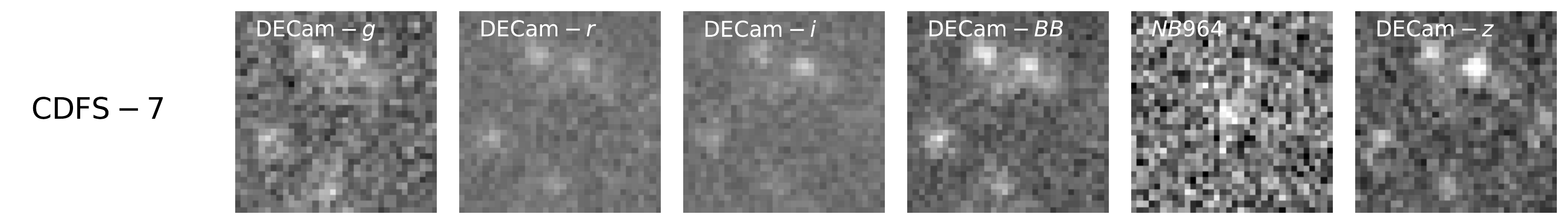}}\\
\subfloat{\includegraphics[width=5in]{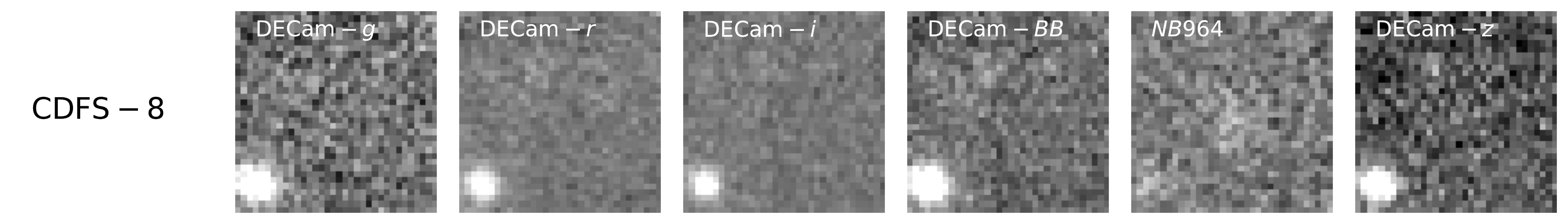}}\\
\subfloat{\includegraphics[width=5in]{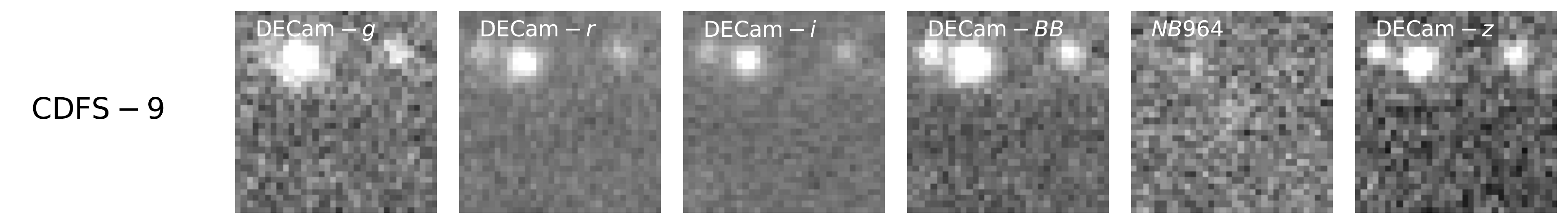}}\\
\subfloat{\includegraphics[width=5in]{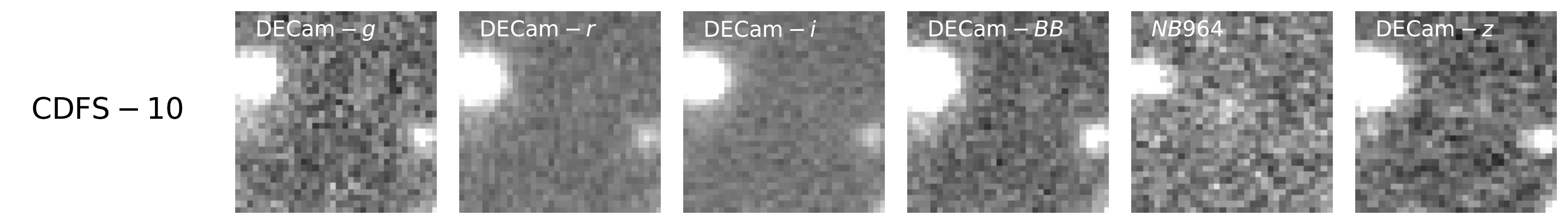}}\\
\subfloat{\includegraphics[width=5in]{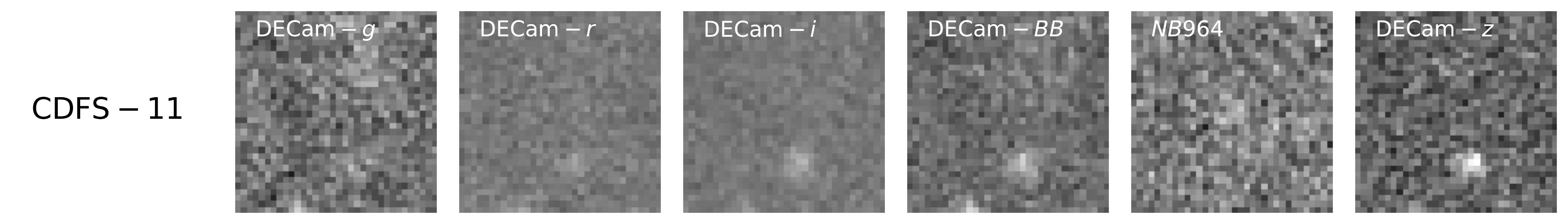}}
\caption{Continued.}
\end{figure*}

\begin{figure*}
\centering
\ContinuedFloat
\subfloat{\includegraphics[width=5in]{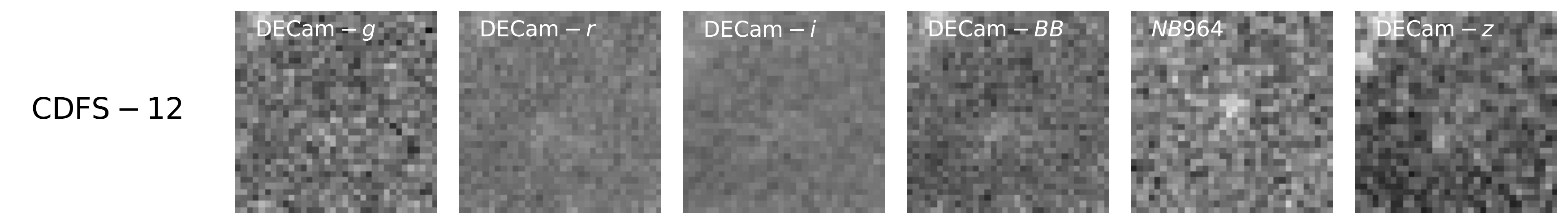}}\\
\subfloat{\includegraphics[width=5in]{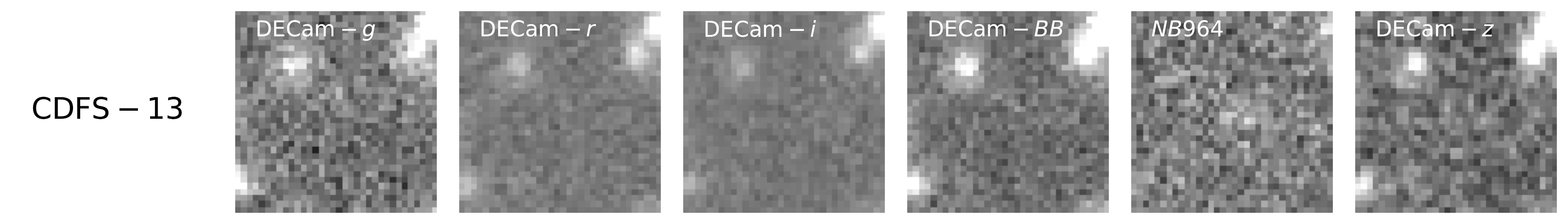}}\\
\subfloat{\includegraphics[width=5in]{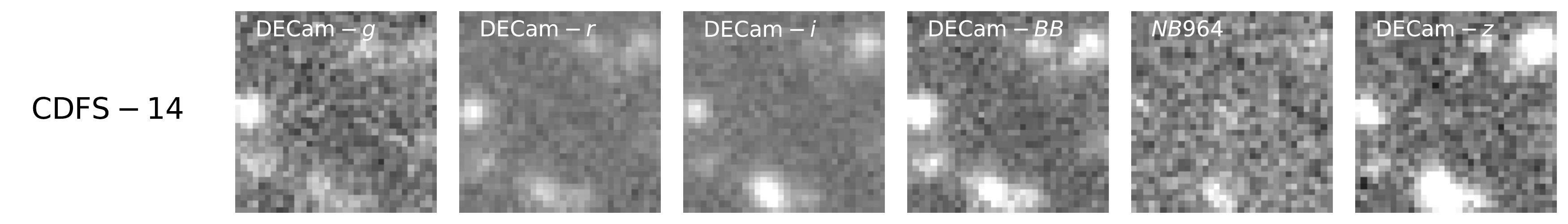}}\\
\subfloat{\includegraphics[width=5in]{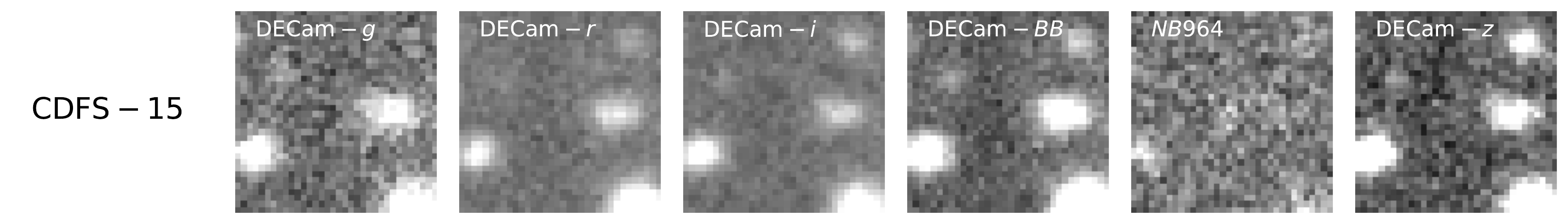}}\\
\subfloat{\includegraphics[width=5in]{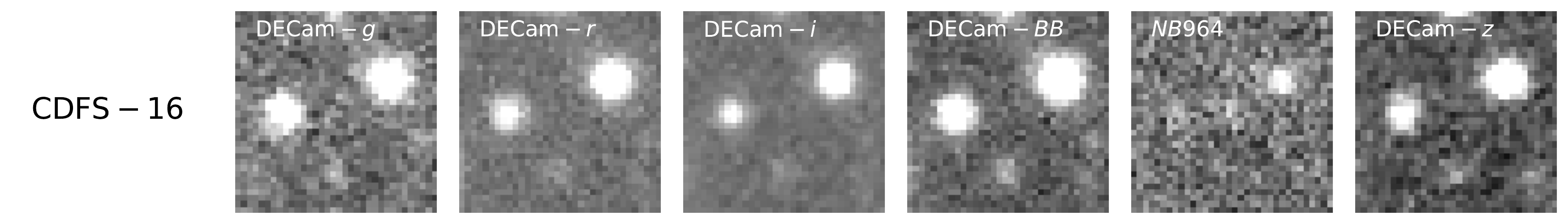}}\\
\subfloat{\includegraphics[width=5in]{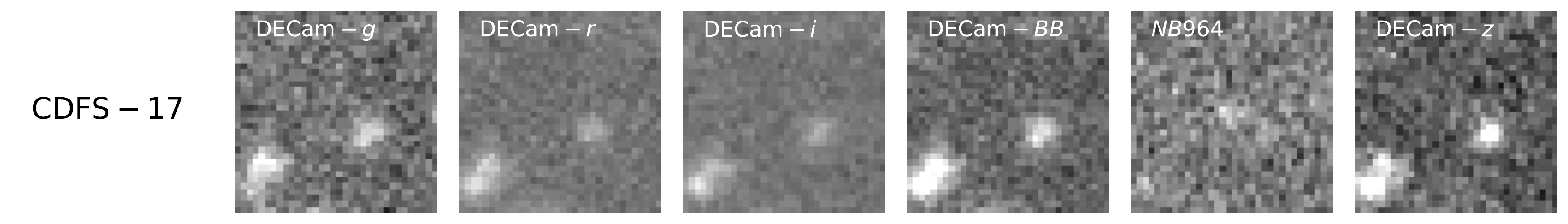}}\\
\subfloat{\includegraphics[width=5in]{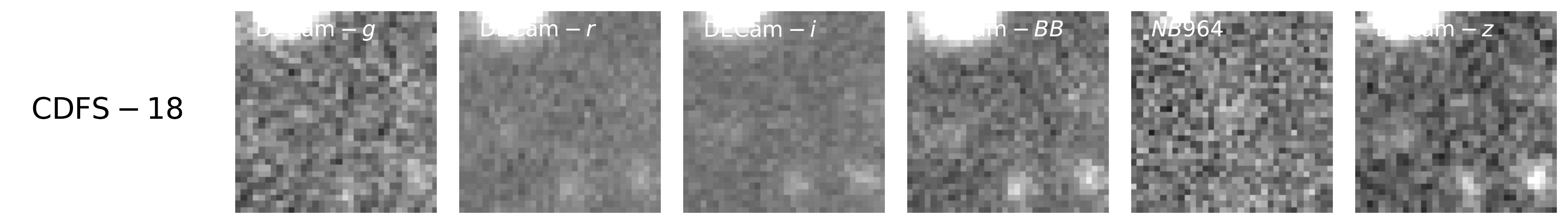}}\\
\subfloat{\includegraphics[width=5in]{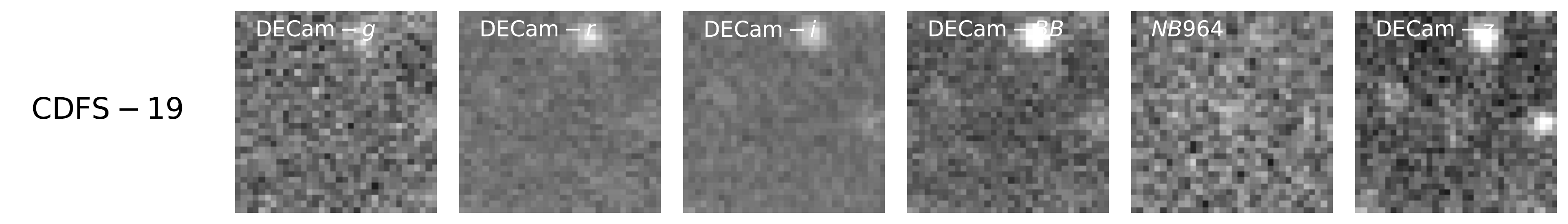}}\\
\subfloat{\includegraphics[width=5in]{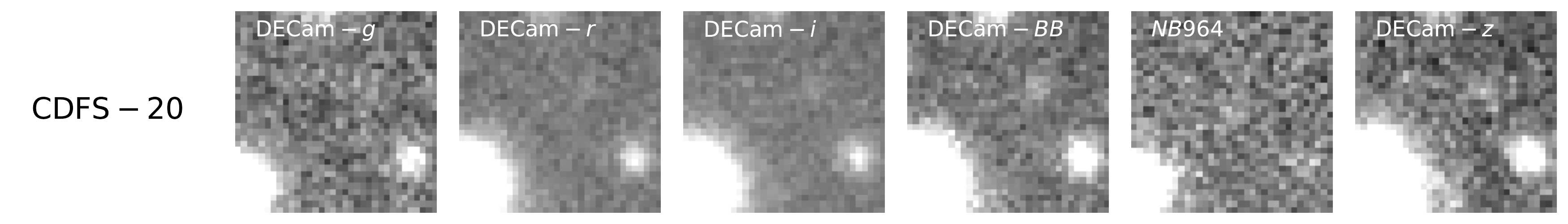}}\\
\subfloat{\includegraphics[width=5in]{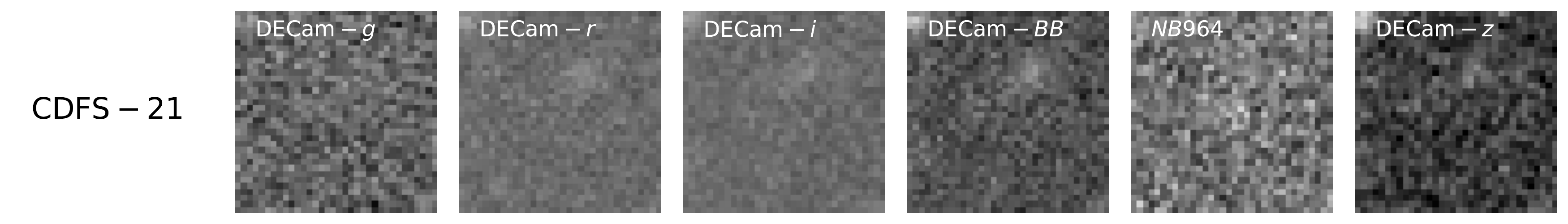}}\\
\subfloat{\includegraphics[width=5in]{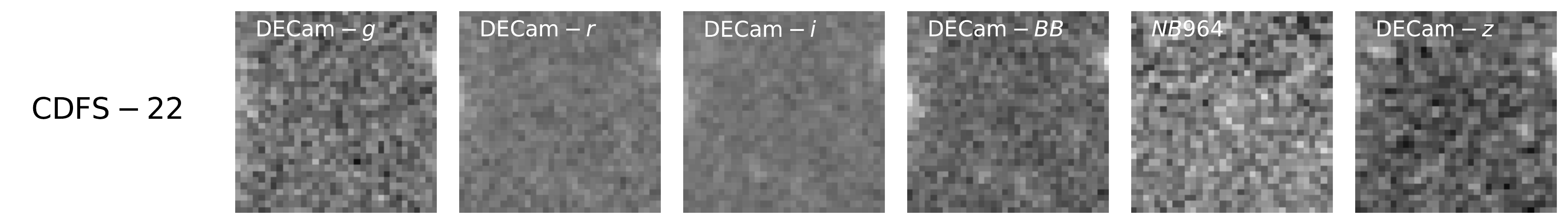}}\\
\subfloat{\includegraphics[width=5in]{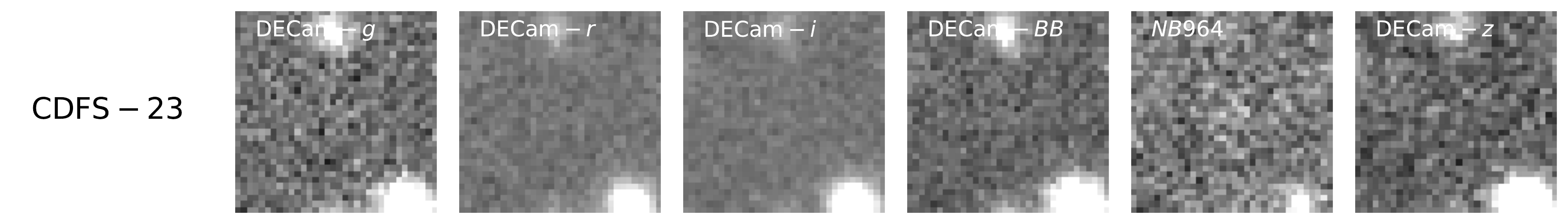}}
\caption{Continued.}
\end{figure*}

\begin{figure*}
\centering
\ContinuedFloat
\subfloat{\includegraphics[width=5in]{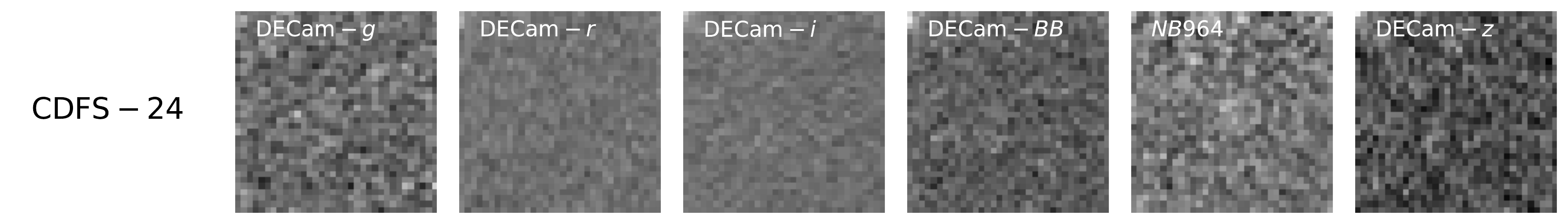}}\\
\subfloat{\includegraphics[width=5in]{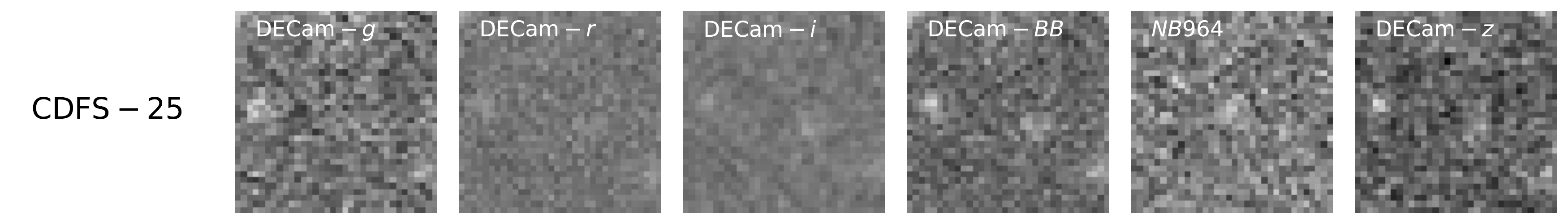}}\\
\subfloat{\includegraphics[width=5in]{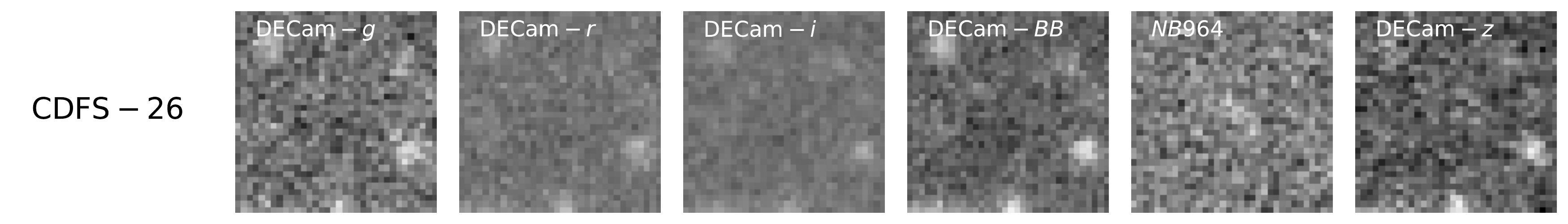}}\\
\subfloat{\includegraphics[width=5in]{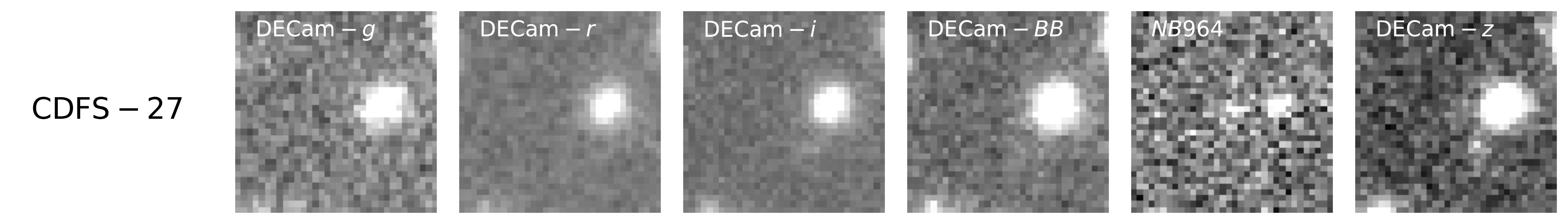}}\\
\subfloat{\includegraphics[width=5in]{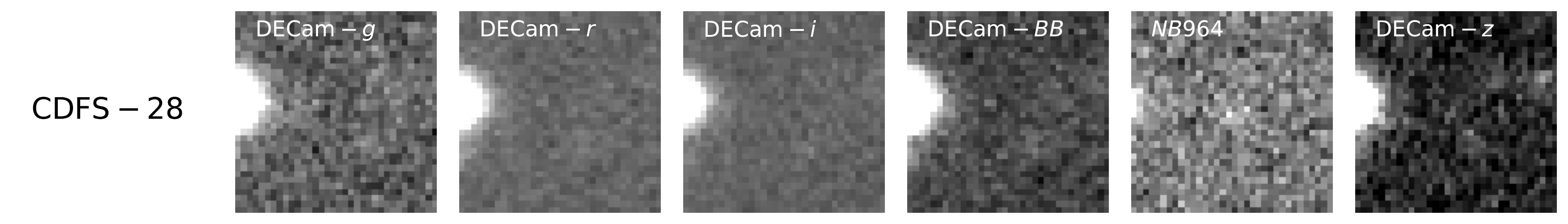}}\\
\subfloat{\includegraphics[width=5in]{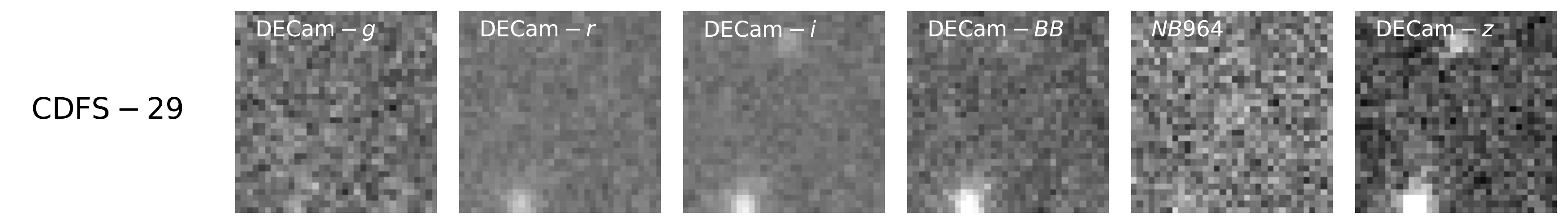}}\\
\subfloat{\includegraphics[width=5in]{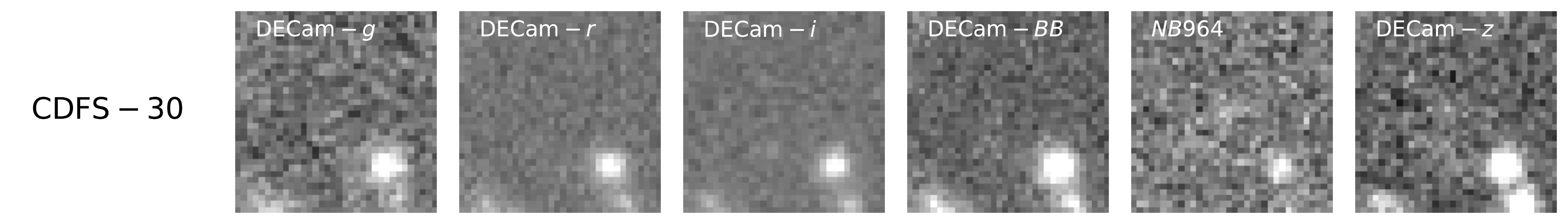}}
\caption{Continued.}
\end{figure*}

\end{appendix}

\end{document}